\documentclass[%
 reprint,
 superscriptaddress,
 noeprint,
 amsmath,amssymb,
 aps,
 prb,
 floatfix,
]{revtex4-2}

\usepackage[version=3]{mhchem} 
\usepackage{physics}
\usepackage{graphicx}
\usepackage{dcolumn}
\usepackage{bm}
\usepackage{hyperref}
\usepackage[caption=false]{subfig}

 \makeatletter
\NewDocumentCommand{\MakeTitleInner}{ +m +m +m }{
    \newpage%
    \null%
    \vskip 2em%
    \begin{center}%
        \let \footnote \thanks
        {\LARGE #1 \par}
        \vskip 1.5em%
        {%
            \large
            \lineskip .5em%
            \begin{tabular}[t]{c}%
                #2
            \end{tabular}\par%
        }%
        \vskip 1em%
        {\large #3}
    \end{center}%
    \par
    \vskip 1.5em%
}
\NewDocumentCommand{\MakeTitle}{ +m +m +m }{%
    \begingroup
        \renewcommand\thefootnote{\@fnsymbol\c@footnote}%
        \def\@makefnmark{\rlap{\@textsuperscript{\normalfont\@thefnmark}}}%
        \long\def\@makefntext##1{\parindent 1em\noindent
            \hb@xt@1.8em{%
                \hss\@textsuperscript{\normalfont\@thefnmark}%
            }##1%
        }%
        \if@twocolumn
            \ifnum \col@number=\@ne
                \MakeTitleInner{#1}{#2}{#3}
            \else
                \twocolumn[\MakeTitleInner{#1}{#2}{#3}]%
            \fi
        \else
            \newpage
            \global\@topnum\z@   
            \MakeTitleInner{#1}{#2}{#3}
        \fi
        \thispagestyle{plain}\@thanks
    \endgroup
    \setcounter{footnote}{0}%
}
\makeatother

\begin{document}

\preprint{APS/123-QED}

\title{Nonlocal Machine-Learned Exchange Functional for Molecules and Solids}

\author{Kyle Bystrom}
\email{kylebystrom@g.harvard.edu}
\affiliation{Harvard John A. Paulson School of Engineering and Applied Sciences}
\author{Boris Kozinsky}
\email{bkoz@g.harvard.edu}
\affiliation{Harvard John A. Paulson School of Engineering and Applied Sciences}
\affiliation{Robert Bosch LLC Research and Technology Center, Cambridge, MA, USA}

\date{\today}

\begin{abstract}
The design of better exchange-correlation functionals for Density Functional Theory (DFT) is a central challenge of modern electronic structure theory. However, current developments are limited by the mathematical form of the functional, with efficient semilocal functionals being inaccurate for many technologically important systems and the more accurate hybrid functionals being too expensive for large solid-state systems due to the use of the exact exchange operator. In this work, we use machine learning combined with exact physical constraints to design an exchange functional that is both orbital-dependent and nonlocal, but which can be evaluated at roughly the cost of semilocal functionals and is significantly faster than hybrid DFT in plane-wave codes. By training functionals with several different feature sets, we elucidate the roles of orbital-dependent and nonlocal features in learning the exchange energy and determine that both types of features provide vital and independently important information to the model. Having trained our new exchange functional with an expressive, nonlocal feature set, we substitute it into existing hybrid functionals to achieve hybrid-DFT accuracy on thermochemical benchmark sets and improve the accuracy of band gap predictions over semilocal DFT. To demonstrate the scalability of our approach as well as the practical benefits of improved band gap prediction, we compute charged defect transition levels in silicon using large supercells. Due to its transferability and computational efficiency for both molecular and extended systems, our model overcomes the cost-accuracy trade-off between semilocal and hybrid DFT, and our general approach provides a feasible path toward a universal exchange-correlation functional with post-hybrid DFT accuracy and semilocal DFT cost. 
\end{abstract}

\maketitle

\section{Introduction}

One of the central challenges of density functional theory (DFT) research is to design accurate approximations to the exchange-correlation (XC) functional~\cite{Mardirossian2017}. As the one unknown term in Kohn-Sham DFT~\cite{Kohn1965}, errors in the XC functional limit the predictive power of DFT for a wide variety of technological problems, including catalysis~\cite{Wellendorff2015}, battery materials~\cite{He2019}, and semiconductor physics~\cite{Lany2008,Freysoldt2014}. Increasing the complexity of the XC functional can improve its accuracy, but often at drastically increased computational cost, which can be prohibitive for large condensed matter simulations. In this work, we use machine learning (ML) to overcome this cost-accuracy trade-off by training a functional that obeys exact physical constraints and is accurate, transferable across a broad range of chemistries, and scalable to hundreds of atoms.

The key cost-accuracy trade-off we address is that between semilocal and hybrid XC functionals. Semilocal functionals, which include the local density approximation (LDA), generalized gradient approximation (GGA), and meta-generalized gradient approximation (meta-GGA)~\cite{Perdew2003}, are computationally efficient but insufficiently accurate for many applications. For example, LDA and GGA functionals are incapable of correctly predicting band gaps due to their lack of derivative discontinuity~\cite{Perdew1983,Godby1986}, and they are of limited accuracy for many molecular systems~\cite{Goerigk2017,Mardirossian2017}. Meta-GGAs introduce derivative discontinuities into the XC functional and can therefore improve band gap predictions, but they typically account for only 20-50\% of the GGA band gap error~\cite{Yang2016}. 
Meta-GGAs can be designed to further improve band gap prediction~\cite{Aschebrock2019,Kovacs2022}, but modifying a given meta-GGA form to predict larger, more accurate band gaps typically reduces the accuracy of that functional for other chemical and material properties, such as bond dissociation energies~\cite{Lebeda2022,Kovacs2022}.
On the other hand, hybrid functionals, which mix a fraction of the nonlocal, orbital-dependent exact exchange energy into an XC functional~\cite{10.1063/1.464913}, provide significantly improved predictions of band gaps~\cite{Yang2016,Janesko2009} and thermochemical properties~\cite{Goerigk2017,Mardirossian2017} over semilocal DFT, but this accuracy comes at a steep additional computational cost, especially within plane-wave DFT. In spite of the remarkable recent progress of efficient implementations for solids~\cite{Lin2016,Hu2017,Carnimeo_2019,Vinson2020,Ko2020,Lee2022}, hybrid DFT is not as efficient or practical as semilocal DFT for these calculations.

Several approaches to design improved XC functionals have been pursued in the literature. Many popular functionals, like PBE~\cite{Perdew1996} and SCAN~\cite{Sun2015}, are designed primarily or entirely based on mathematical constraints obeyed by the exact functional. Constraint-based functionals are often considered trustworthy because they can predict molecular properties without being explicitly tuned to match them, reducing the risk of bias toward a particular property or set of systems~\cite{Perdew2005,Kaplan2023}. However, some of the most popular and accurate XC functionals for main-group chemistry are empirically fit to chemical data~\cite{Stephens1994,Mardirossian2014}. Recently, significant progress has been made toward combining constraint and data-based design approaches using machine learning~\cite{PhysRevB.104.L161109,Nagai2020,Kasim2021,Nagai2022,Trepte2022,Sparrow2022,Bystrom2022}, culminating in the development of the DM21 functional~\cite{Kirkpatrick2021}, which achieves state-of-the-art accuracy on main-group chemistry benchmarks and also improves the description of static correlation by fitting fractional charge constraints.

While promising, these developments have a few key shortcomings, the most notable of which is their computational cost. DM21 uses a local range-separated hybrid formalism, which is both computationally expensive (even more so than global hybrid DFT) and not widely available for the plane-wave DFT codes typically used for investigating large extended systems. As a result, DM21 has not been applied to solids. To the best of our knowledge, no nonlocal machine-learned functional has yet been implemented for both molecular (Gaussian-type orbital) and extended (plane-wave) DFT calculations, with the exception of the DeePKS approach~\cite{doi:10.1021/acs.jctc.0c00872,Li2022}. However, DeePKS uses geometry-dependent (rather than universal) functionals and is trained for specific systems, so it does not fulfill the role of a broadly chemically transferable model. All other ML XC functionals that have been applied to solids have been limited to the semilocal form~\cite{Wellendorff2012,Wellendorff2014,Lundgaard2016,Trepte2022,Kovacs2022,Nagai2022} (with the exception of minimally parameterized van der Waals functionals and one concurrent work by Riemelmoser \emph{et al.}~\cite{Riemelmoser2023} that introduced a nonlocal functional for \ce{C}, \ce{H}, and \ce{O}-containing systems). This limitation restricts the application space of ML functionals by preventing their use in applications where hybrid DFT accuracy is needed. For example, hybrid DFT has proven useful for studying problems in catalysis within both Gaussian-type orbital and plane-wave codes~\cite{Mavros2014,Shukla2022} and for addressing the ``band-gap problem'' in DFT~\cite{Borlido2019}. In addition to this technical limitation, applications of ML functionals to complex and heterogeneous systems are limited by their lack of transferability and universality, as they have previously only been trained on molecular systems.

These limitations raise the question of whether machine learning models can leverage more efficient features to balance accuracy and computational cost. One possible approach in this direction is to generate nonlocal features by taking integrals of the density over some set of kernel functions~\cite{Lei2019descriptors,Nagai2020}. This way, an ML model of the XC functional can capture nonlocal information about the density without evaluating the computationally demanding exact exchange operator required for hybrid DFT. In line with this strategy, we recently introduced the Compressed Scale-Invariant Density Representation (CIDER) formalism for learning the exchange functional~\cite{Bystrom2022}, in which the nonlocal features were designed to enforce the uniform scaling rule~\cite{Levy1985} on the learned exchange functional. We demonstrated that the exchange functional can be accurately learned with the CIDER approach. However, the high computational cost of computing the features, lack of implementation for plane-wave DFT, and need to compute and fit the exchange energy density (rather than the total exchange energy) limited the practicality of our model and restricted its application to small molecules.

In this work, we overcome all of these limitations by revising the form of the features, implementing efficient algorithms to evaluate the new features for both molecular and extended systems, and redesigning our Gaussian process model to fit the total exchange energy. These improvements enable the construction of the first nonlocal, universal ML exchange functional for solids, which can be applied to large systems with similar computational cost to semilocal DFT. We use this exchange functional to implement efficient and accurate hybrid DFT surrogates that can be used in a variety of applications. Because our new functional form is both transferable and numerically stable, we believe that our approach will also be useful for learning the full XC functional, allowing us to address problems where DFT accuracy is limited by both exchange and correlation effects.

By training and testing exchange functionals with different feature sets, we find that to learn the exchange functional well enough to reproduce hybrid DFT accuracy, it is necessary to include both a semilocal orbital-dependent feature (i.e. the kinetic energy density) and nonlocal density-dependent features in the model input. Having developed this ``nonlocal meta-GGA,'' we show that it accurately predicts molecular properties and improves solid-state band gap predictions over standard meta-GGAs, which is a central problem in DFT. The performance for solid-state band gaps is particularly promising because only 9 systems in the training set are solids, with the others being isolated molecules and complexes. We also encounter only a few convergence problems in our self-consistent field (SCF) calculations of 2462 molecular and 453 periodic systems, indicating that the functional has good numerical stability. To demonstrate the transferability and efficiency of the model, we simulate systems with over 500 atoms to predict charged defect transition levels in silicon.

The paper is organized as follows. Section~\ref{sec:theory} describes the features and models used in this work.  Section~\ref{sec:results} presents our results and discusses the accuracy of the CIDER functionals as well as their computational efficiency within plane-wave DFT. Section~\ref{sec:conclusion} contains our conclusion. Appendix~\ref{sec:methods} provides a detailed description of the computational methods used in this study, and Appendix~\ref{app:feat_proj} provides additional mathematical details for the plane-wave implementation of CIDER.

\section{Theory}\label{sec:theory}

In Kohn-Sham DFT~\cite{Kohn1965}, it is often convenient to separate the XC energy into exchange and correlation parts, i.e. $E_\text{xc}[n]=E_\text{x}[n]+E_\text{c}[n]$. The exchange energy $E_\text{x}[n]$ accounts for the antisymmetric nature of the fermionic wave function, while the correlation energy $E_\text{c}[n]$ accounts for additional many-body correlation effects between the electrons. The exchange energy is the main focus of this work and can be trivially generalized from non-spin-polarized DFT to spin-polarized DFT as~\cite{Oliver1979}
\begin{equation}
    E_\text{x}[n_\uparrow,n_\downarrow]=\frac{1}{2}E_\text{x}[2n_\uparrow]+\frac{1}{2}E_\text{x}[2n_\downarrow],
\end{equation}
so all formulas in the theory section are written for the non-spin-polarized case for simplicity.

The exact exchange energy can be expressed in terms of the occupied Kohn-Sham orbitals $\{\phi_i(\mathbf{r})\}$ as
\begin{equation}
    E_\text{x}^\text{exact}[n] = - \sum_{ij} \int \dd[3]\mathbf{r} \int \dd[3]\mathbf{r}' \frac{\phi_i^*(\mathbf{r})\phi_j(\mathbf{r})\phi_i(\mathbf{r}')\phi_j^*(\mathbf{r}')}{|\mathbf{r}-\mathbf{r}'|}.\label{eq:exact_exchange_energy}
\end{equation}
Evaluating Eq.\ \ref{eq:exact_exchange_energy} is computationally expensive. However, the improved accuracy of hybrid DFT arises from mixing a fraction of $E_\text{x}^\text{exact}[n]$ into an otherwise semilocal XC functional. These considerations explain the cost-accuracy trade-off between semilocal and hybrid DFT. Our goal is to fit Eq.\ \ref{eq:exact_exchange_energy} as accurately as possible with a computationally efficient surrogate model.

The remainder of this section sets up our machine learning-based approach to this problem. Specifically, Sections~\ref{sec:theory_sldft}-\ref{sec:gp} cover the design of the exchange functional form, nonlocal features, and Gaussian process models, and Sections \ref{sec:rps}-\ref{sec:theory_ciderpaw} cover the computationally efficient implementation of the nonlocal features. The feature implementation requires different algorithms for Gaussian-type orbital and plane-wave DFT codes due to the use of different basis sets and integration grids within these codes.

\subsection{Uniform Scaling and Semilocal Exchange Functionals}\label{sec:theory_sldft}

The simplest approximate exchange functional is the local density approximation (LDA), in which the exchange energy is written as an integral over a function of the density $n(\mathbf{r})$ at each point in space
\begin{equation}
    E_\text{x}[n]=\int \dd[3]\mathbf{r}\, e_\text{x}^\text{LDA}\left(n(\mathbf{r})\right), \label{eq:lda_ex}
\end{equation}
where in the Kohn-Sham formalism~\cite{Kohn1965}, the density is expressed in terms of the occupied orbitals as
\begin{equation}
    n(\mathbf{r})=\sum_i \left|\phi_i(\mathbf{r})\right|^2.
\end{equation}
Typically, $e_\text{x}^\text{LDA}\left(n\right)$ is defined as the exchange energy density of the uniform electron gas~\cite{Dirac1930}
\begin{equation}
    e_\text{x}^\text{LDA}(n) = -\frac{3}{4}\left(\frac{3}{\pi}\right)^{1/3} n^{4/3}. \label{eq:ueg_xed}
\end{equation}
In addition to making Eq.\ \ref{eq:lda_ex} exact for the uniform electron gas, Eq.\ \ref{eq:ueg_xed} also satisfies an important property of the exchange functional known as uniform scaling~\cite{Levy1985,Dreizler1990}, which states that for a scaled density distribution $n_{\lambda}(\mathbf{r})=\lambda^3 n(\lambda \mathbf{r})$,
\begin{equation}
    E_\text{x}[n_\lambda]=\lambda E_\text{x}[n], \label{eq:unif_scaling}
\end{equation}
where $\lambda$ is a positive scalar.

More sophisticated functionals that obey Eq.\ \ref{eq:unif_scaling} can be designed by writing the exchange energy density as the product of Eq.\ \ref{eq:ueg_xed} and an exchange enhancement factor $F_\text{x}$:
\begin{equation}
    E_\text{x}[n] = \int \dd[3]\mathbf{r}\, e_\text{x}^\text{LDA}\left(n(\mathbf{r})\right) F_\text{x}(\mathbf{x}(\mathbf{r})). \label{eq:general_xf}
\end{equation}
In the above equation, $\mathbf{x}(\mathbf{r})$ is a vector of scale-invariant features $x_i$, where scale-invariance is defined as
\begin{equation}
    x_i[n_{\lambda}](\mathbf{r})=x_i[n](\lambda\mathbf{r}). \label{eq:scale_invariance}
\end{equation}
For example, a generalized gradient approximation (GGA) exchange functional is constructed by making $F_\text{x}$ a function of the reduced gradient $s$ of the density
\begin{equation}
    s = \frac{|\nabla n|}{2(3\pi^2)^{1/3}n^{4/3}}. \label{eq:reduced_grad_s}
\end{equation}
Because $s$ satisfies Eq.\ \ref{eq:scale_invariance}, the GGA form
\begin{equation}
    E_\text{x}[n] = \int \dd[3]\mathbf{r}\, e_\text{x}^\text{LDA}\left(n(\mathbf{r})\right) F_\text{x}(s(\mathbf{r}))
\end{equation}
satisfies Eq.\ \ref{eq:unif_scaling}. Likewise, a meta-GGA can be constructed by making the exchange enhancement factor a function of two ingredients $F_\text{x}(s,\alpha)$, where
\begin{equation}
    \alpha = \frac{\tau-\tau_W}{\tau_0} \label{eq:iso_orbital_alpha}
\end{equation}
is a scale-invariant quantity that depends on the kinetic energy density
\begin{equation}
    \tau(\mathbf{r}) = \frac{1}{2} \sum_i \left|\nabla \phi_i(\mathbf{r})\right|^2,
\end{equation}
and $\tau_W = \frac{|\nabla n|^2}{8n}$ and $\tau_0 = \frac{3}{10}(3\pi^2)^{2/3} n^{5/3}$ are the single-orbital and uniform electron gas kinetic energies, respectively.

\subsection{Updated Nonlocal Features for CIDER} \label{sec:theory_nldf}

In our initial construction of the CIDER formalism~\cite{Bystrom2022}, the exchange functional is defined as
\begin{align}
    E_\text{x}[n] = \int \dd[3]\mathbf{r}\,& e_\text{x}^\text{LDA}\left(n(\mathbf{r})\right) \notag\\ &\times F_\text{x}\left(s(\mathbf{r}), \alpha(\mathbf{r}), \{G_{nlm}(\mathbf{r})\}\right).\label{eq:exc_old}
\end{align}
In Eq.\ \ref{eq:exc_old}, the functions $\{G_{nlm}(\mathbf{r})\}$ are \emph{nonlocal} features of the density defined by
\begin{align}
    G_{nlm}(\mathbf{r}) &= \int \dd[3]\mathbf{r}'\,\, g_{nlm}(\mathbf{r}';\mathbf{r})\, n(\mathbf{r}+\mathbf{r}\,'),\label{eq:gnl_old}
\end{align}
where $g_{nlm}(\mathbf{r}';\mathbf{r})$ is the kernel function
\begin{align}
    g_{nlm}(\mathbf{r}';\mathbf{r}) &= M_l Y_{lm}(\mathbf{\hat{r}}') \left(a(\mathbf{r})|\mathbf{r}'|^2\right)^{n+l/2} \mathrm{e}^{-a(\mathbf{r})|\mathbf{r}'|^2},\label{eq:gkernel_old}
\end{align}
with $M_l = B_0^{3/2} \sqrt{4\pi^{l-1}} \left( \frac{8\pi}{3} \right)^{\frac{l}{3}}$.
The position-dependent exponent $a(\mathbf{r})$ is defined as a semilocal quantity
\begin{align}
    a[n](\mathbf{r}) &= \pi\left(\frac{n(\mathbf{r})}{2}\right)^{2/3} \left[B_0+C_0\left(\frac{\tau(\mathbf{r})}{\tau_0(\mathbf{r})}-1\right)\right]\label{eq:anl_old},
\end{align}
where $B_0$ and $C_0$ are adjustable constants that control how quickly the width of the Gaussian-type function in Eq.\ \ref{eq:gkernel_old} decreases as the density and kinetic energy density increase. Equation \ref{eq:anl_old} behaves quadratically under uniform scaling, i.e.
\begin{equation}
    a[n_\lambda](\mathbf{r})=\lambda^2 a[n](\lambda\mathbf{r}).\label{eq:quad_scaling}
\end{equation}
This property ensures that $G_{nlm}(\mathbf{r})$ is scale-invariant (Eq.\ \ref{eq:scale_invariance}) and therefore that Eq.\ \ref{eq:exc_old} satisfies the uniform scaling rule (Eq.\ \ref{eq:unif_scaling}).

Using these scale-invariant nonlocal features, we were able to train an ML exchange functional that accurately reproduced hybrid DFT atomization energies~\cite{Bystrom2022}, but there is room for improvement in making the features more physically intuitive and easier to implement with production-level performance. In this work, we simplify the features by removing the spherical harmonics and instead adding a second position-dependent exponent to the kernel function, resulting in the new features
\begin{align}
    G_i(\mathbf{r}_1) =& \left(\frac{B_i+B_0}{2}\right)^{3/2} \notag\\ &\times \int \dd[3]\mathbf{r}_2 \, \Phi\left(a(\mathbf{r}_2), b_i(\mathbf{r}_1), r_{12}\right) n(\mathbf{r}_2)\label{eq:cider_feature}\\
    \Phi\left(a, b, r\right) =& \exp\left[-\left(a+b\right)r^2\right],\label{eq:cider_kernel}
\end{align}
where $r_{12}=|\mathbf{r_1}-\mathbf{r_2}|$. Multiple features $G_1,G_2,...$ are generated by evaluating Eq.\ \ref{eq:cider_feature} with different choices of exponent $b_i(\mathbf{r})$, which in turn are tuned by constants $B_i$ and $C_i$ as discussed below. Because the position-dependent exponents increase as the density increases, the $a$ exponent effectively damps the value of the integrand when $\mathbf{r}_2$ is near the atomic core, which we intend to limit the interaction between the valence region and core region. The prefactor $\left(\frac{B_i+B_0}{2}\right)^{3/2}$ is chosen so that $G_i=2$ for the non-spin-polarized uniform electron gas.

To retain the scale-invariance of the features, any choice of $a[n](\mathbf{r})$ and $b_i[n](\mathbf{r})$ that satisfies Eq. \ref{eq:quad_scaling} is sufficient. For example, the exponents can be an orbital-independent function of the density and its gradient
\begin{align}
    a[n](\mathbf{r}) &= \pi\left(\frac{n}{2}\right)^{2/3} \left[B_0+C_0\left(\frac{|\nabla n|^2}{8n\tau_0}\right)\right]\label{eq:anl_new}\\
    b_i[n](\mathbf{r}) &= \pi\left(\frac{n}{2}\right)^{2/3} \left[B_i+C_i\left(\frac{|\nabla n|^2}{8n\tau_0}\right)\right]\label{eq:abnl_new}
\end{align}
or can be made kinetic-energy dependent, as in our previous work:
\begin{align}
    a[n](\mathbf{r}) &= \pi\left(\frac{n}{2}\right)^{2/3} \left[B_0+C_0\left(\frac{\tau}{\tau_0}-1\right)\right]\label{eq:anl_mgga_new}\\
    b_i[n](\mathbf{r}) &= \pi\left(\frac{n}{2}\right)^{2/3} \left[B_i+C_i\left(\frac{\tau}{\tau_0}-1\right)\right].\label{eq:abnl_mgga_new}
\end{align}
Equations \ref{eq:anl_new} and \ref{eq:abnl_new} are used for the nonlocal GGA presented in this work (c.f. Section~\ref{sec:gp}), while Eqs.\ \ref{eq:anl_mgga_new} and \ref{eq:abnl_mgga_new} are used for the nonlocal meta-GGA. Throughout Sections \ref{sec:rps}--\ref{sec:theory_ciderpaw}, the $\left(\frac{B_i+B_0}{2}\right)^{3/2}$ factor and the feature index $i$ will be dropped for simplicity, as the inclusion of the scaling factor and the introduction of additional features by introducing new $B_i$ and $C_i$ is trivial. The motivation for the form of Eqs.~\ref{eq:anl_new}-\ref{eq:abnl_mgga_new} is discussed in our previous work introducing CIDER~\cite{Bystrom2022}, and the heuristic selection of the $B_i$ and $C_i$ coefficients is discussed in Appendix~\ref{sec:hparam_sweep}.

Before continuing, we briefly discuss the construction of the XC potential when nonlocal features are employed. A typical semilocal functional has a multiplicative XC potential
\begin{equation}
    \fdv{E_\text{xc}}{n(\mathbf{r})} = \pdv{e_\text{xc}(\mathbf{r})}{n(\mathbf{r})}.
\end{equation}
A meta-GGA also has a non-multiplicative potential
\begin{equation}
    \fdv{E_\text{xc}}{\tau(\mathbf{r})} = \pdv{e_\text{xc}(\mathbf{r})}{\tau(\mathbf{r})}.
\end{equation}
This above term is non-multiplicative because it depends on the orbitals through $\tau(\mathbf{r})$, rather than depending exclusively on the density $n(\mathbf{r})$. For a given basis set $\left\{\chi_\mu(\mathbf{r})\right\}$, the contributions of the XC potential to the effective Kohn-Sham Hamiltonian matrix are
\begin{align}
    V_{\mu\nu}^{\text{xc}} =& \int \dd[3]\mathbf{r}\, \chi_\mu^*(\mathbf{r}) \chi_\nu(\mathbf{r}) \fdv{E_\text{xc}}{n(\mathbf{r})} \notag\\
    &+ \frac{1}{2} \int \dd[3]\mathbf{r}\, \nabla\chi_\mu^*(\mathbf{r}) \cdot \nabla\chi_\nu(\mathbf{r}) \fdv{E_\text{xc}}{\tau(\mathbf{r})}. \label{eq:vxc_matrix_elems}
\end{align}

When nonlocal features are included in the XC energy density, e.g.
\begin{equation}
    E_\text{xc} = \int \dd[3]\mathbf{r}\, e_\text{xc}\left(n(\mathbf{r}), |\nabla n(\mathbf{r})|, \tau(\mathbf{r}), G(\mathbf{r})\right),
\end{equation}
both $\fdv{E_\text{xc}}{n(\mathbf{r})}$ and $\fdv{E_\text{xc}}{\tau(\mathbf{r})}$ include terms that depend on other coordinates $\mathbf{r}'$:
\begin{align}
    \fdv{E_\text{xc}}{n(\mathbf{r})} =& \pdv{e_\text{xc}(\mathbf{r})}{n(\mathbf{r})} + \pdv{e_\text{xc}(\mathbf{r})}{G(\mathbf{r})} \pdv{G(\mathbf{r})}{n(\mathbf{r})} \notag\\
    & + \int \dd[3]\mathbf{r}'\, \pdv{e_\text{xc}(\mathbf{r}')}{G(\mathbf{r}')} \fdv{G(\mathbf{r}')}{n(\mathbf{r})} \label{eq:vxc_nlfeat} \\
    \fdv{E_\text{xc}}{\tau(\mathbf{r})} =& \pdv{e_\text{xc}(\mathbf{r})}{\tau(\mathbf{r})} + \pdv{e_\text{xc}(\mathbf{r})}{G(\mathbf{r})} \pdv{G(\mathbf{r})}{\tau(\mathbf{r})} \notag\\
    &+ \int \dd[3]\mathbf{r}'\, \pdv{e_\text{xc}(\mathbf{r}')}{G(\mathbf{r}')} \fdv{G(\mathbf{r}')}{\tau(\mathbf{r})} \label{eq:vxc_nlfeat_tau}
\end{align}
In Eq. \ref{eq:vxc_nlfeat}, $\pdv{G(\mathbf{r})}{n(\mathbf{r})}$ describes the local dependence of $G(\mathbf{r})$ on the density, and $\fdv{G(\mathbf{r}')}{n(\mathbf{r})}$ describes the nonlocal dependence on the density. While the expressions for $\fdv{E_\text{xc}}{n(\mathbf{r})}$ and $\fdv{E_\text{xc}}{\tau(\mathbf{r})}$ are more complicated than for a meta-GGA, they still contribute to the XC potential matrix exclusively via Eq. \ref{eq:vxc_matrix_elems}. Therefore, any DFT code supporting meta-GGAs can evaluate this type of functional if a routine is added to evaluate the nonlocal contributions to Eqs. \ref{eq:vxc_nlfeat} and \ref{eq:vxc_nlfeat_tau}.

\subsection{Gaussian Process Models for the Exchange Functional}\label{sec:gp}

\subsubsection{Overview}

To study the impact of different features on the accuracy of an ML exchange functional, we implement four functional types that use different sets of features.\par
\textbf{SL-GGA} is a standard, semilocal GGA like PBE~\cite{Perdew1996}, except trained via machine learning. The exchange enhancement factor ($F_\text{x}$ in Eq.\ \ref{eq:general_xf}) is a function of $s$ only.\par
\textbf{NL-GGA} is a \emph{nonlocal} GGA. The exchange enhancement factor is a function of $s$ and also three nonlocal features of the form in Eq.\ \ref{eq:cider_feature}, with Eqs.\ \ref{eq:anl_new} and \ref{eq:abnl_new} for the exponents.\par
\textbf{SL-MGGA} is a standard, semilocal meta-GGA like SCAN~\cite{Sun2015}, except trained via machine learning. The exchange enhancement factor is a function of $s$ and $\alpha$.\par
\textbf{NL-MGGA} is a \emph{nonlocal} meta-GGA. The exchange enhancement factor is a function of $s$, $\alpha$, and three nonlocal features of the form in Eq.\ \ref{eq:cider_feature}, with Eqs.\ \ref{eq:anl_mgga_new} and \ref{eq:abnl_mgga_new} for the exponents.\par
In short, the SL-GGA and SL-MGGA are semilocal functionals, while the NL-GGA and NL-MGGA augment the semilocal functional form with nonlocal features of the density. As described below, Gaussian process models for the exchange energy are trained using each of these four feature sets. Section~\ref{sec:gp_total_exx} explains how a Gaussian process can be fit to total exchange energies of systems, Section~\ref{sec:feature_defs} lists the specific mathematical form of each feature vector that is used as input to the Gaussian processes, and Section~\ref{sec:gp_kernels} lists the covariance kernels for each feature set.

Gaussian process regression is a Bayesian machine learning method in which the values of the predictive function $f(x)$ are treated as dependent random variables. The key ingredient in a Gaussian process is the covariance between values of the function $\text{Cov}(f(x), f(x'))=k(x, x')$, where the user-defined covariance kernel function $k(x, x')$ determines the relationship between the values of $f$ for different inputs $x$. Suppose one has training data of the form $(x_i, y_i)$, where $y_i$ is assumed to be the sum of $f(x_i)$ and Gaussian-distributed random noise of variance $\sigma_\text{noise}^2$. Then the predictive mean, or expected value, of the function $f$ is given by
\begin{equation}
    f(x_*) = \sum_i k(x_*, x_i) \alpha_i,
\end{equation}
with the weight vector $\boldsymbol{\alpha}$ defined as
\begin{equation}
    \boldsymbol{\alpha} = \left(\mathbf{K} + \sigma_\text{noise}^2\mathbf{I}\right)^{-1} \mathbf{y}.
\end{equation}
In the above equation $K_{ij}=k(x_i,x_j)$ is called the covariance matrix, and $\mathbf{I}$ is the identity matrix. The reader is referred to the textbook by Rasmussen and Williams~\cite{Rasmussen2006} for a more thorough introduction to Gaussian processes. The following subsection introduces a modified Gaussian process regression approach for training exchange-correlation functionals.

\subsubsection{Fitting the Total Exchange Energy}\label{sec:gp_total_exx}

In this section, we introduce a method to train a Gaussian process to the total exchange energy, as opposed to the nonunique exchange energy density used in our previous work~\cite{Bystrom2022}. While we only learn the exchange energy in this work, the approach in this section also applies to the XC energy, so we write the formulas for $E_\text{xc}$ for the sake of generality. We index the chemical systems in the training set by $i$ and $j$; while the feature implementations are different for Gaussian-type orbital and plane-wave DFT, the role of systems computed within these two frameworks is identical with regard to model training. The XC energy of a system $i$ is given by a numerical integral of the energy density
\begin{equation}
    E_{\text{xc}}^i = \sum_{g \in i} w_g^i e_{\text{xc}}(\mathbf{x}_g^i),
\end{equation}
where $g$ indexes grid points, $w_g^i$ are the quadrature weights, and $\mathbf{x}_g^i$ is the feature vector of molecule $i$ at point $g$.

Constructing a Gaussian process for $E_\text{xc}$ requires defining the covariance between the XC energies for two different systems, i.e. $\text{Cov}(E_{\text{xc}}^i, E_{\text{xc}}^j)$ (c.f. Chapter 2.2 of Ref. \cite{Rasmussen2006}). The covariance of sums of random variables $a,b,c$, and $d$ is
\begin{align}
    \text{Cov}(a+b,c+d)=&\text{Cov}(a,c)+\text{Cov}(b,c)\notag\\&+\text{Cov}(a,d)+\text{Cov}(b,d). \label{eq:cov_of_sum}
\end{align}
Using Eq.\ \ref{eq:cov_of_sum}, $\text{Cov}(E_{\text{xc}}^i, E_{\text{xc}}^j)$ can be expressed in terms of the covariances between the exchange energy densities as
\begin{align}
    \text{Cov}(E_{\text{xc}}^i, E_{\text{xc}}^j) =& \sum_{g \in i} \sum_{h \in j} w_g^i w_h^j \notag\\ &\times\text{Cov}(e_{\text{xc}}(\mathbf{x}_g^i), e_{\text{xc}}(\mathbf{x}_h^j)) \\
    =& \sum_{g \in i} \sum_{h \in j} w_g^i w_h^j k_{e_{\text{xc}}}(\mathbf{x}_g^i, \mathbf{x}_h^j),\label{eq:exc_cov_exact}
\end{align}
where $k_{e_{\text{xc}}}(\mathbf{x}, \mathbf{x}')$ is the kernel function for the covariance between the exchange energy densities for feature vectors $\mathbf{x}$ and $\mathbf{x}'$. Unfortunately, this expression can be quite expensive to compute because there are tens of thousands of grid points for each small molecule in the training database, and there are several hundred training molecules. Therefore, even for a database of this size, at least $10^{12}$ evaluations of the kernel are required. To circumvent this problem, a set of control points $\mathbf{\tilde{x}}_a$ can be sampled from the training database to use as a basis set for the feature space (as described in Appendix \ref{sec:control_points}), effectively creating a sparse Gaussian process~\cite{Rasmussen2006}. Equation \ref{eq:exc_cov_exact} is then approximated as
\begin{align}
    \text{Cov}(E_{\text{xc}}^i, E_{\text{xc}}^j) \approx&  \left(\mathbf{\tilde{k}}_{e_{\text{xc}}}^i\right)^T \mathbf{\widetilde{K}}^{-1} \mathbf{\tilde{k}}_{e_{\text{xc}}}^j\label{eq:nystrom_approx}\\
    \left(\mathbf{\widetilde{K}}\right)_{ab} =& k_{e_{\text{xc}}}(\mathbf{\tilde{x}}_a, \mathbf{\tilde{x}}_b) \label{eq:sparse_cov}\\
    \left(\mathbf{\tilde{k}}_{e_{\text{xc}}}^i\right)_a =& \sum_{g\in i} w_g^i k_{e_{\text{xc}}}(\mathbf{x}_g^i, \mathbf{\tilde{x}}_a).\label{eq:sparse_cross_cov}
\end{align}
The approximation of Eq.~\ref{eq:exc_cov_exact} by Eq.~\ref{eq:nystrom_approx} is called the Nystr\"om approximation, and it is covered in more detail in Chapter 8.1 of Ref.~\cite{Rasmussen2006}. The approximation can be thought of as interpolating the kernel function over the control points $\mathbf{\tilde{x}}_a$. Supplemental Material Section S9 investigates the accuracy of this approximation with respect to the number of control points used.

In this work, the exchange functional alone is trained, and the kernel function is expressed in terms of the exchange enhancement factor $F_\text{x}=e_\text{x}/e_\text{x}^\text{LDA}$. This leads to the adjusted expressions
\begin{align}
    \text{Cov}(E_\text{x}^i, E_\text{x}^j) &\approx \left(\mathbf{\tilde{k}}^i\right)^T \mathbf{\widetilde{K}}^{-1} \mathbf{\tilde{k}}^j = K_{ij} \\
    \left(\mathbf{\widetilde{K}}\right)_{ab} &= k_{F_\text{x}}(\mathbf{\tilde{x}}_a, \mathbf{\tilde{x}}_b)\\
    \left(\mathbf{\tilde{k}}^i\right)_a &= \sum_{g\in i} w_g^i \left(e_\text{x}^\text{LDA}\right)_g^i k_{F_\text{x}}(\mathbf{x}_g^i, \mathbf{\tilde{x}}_a),
\end{align}
where $k_{F_\text{x}}(\mathbf{x}, \mathbf{x}')$ is the covariance kernel for the exchange enhancement factor and $(e_\text{x}^\text{LDA})_g^i=e_\text{x}^\text{LDA}\left(n_i(\mathbf{r}_g)\right)$ is the LDA exchange energy density at coordinate $g$ for molecule $i$ (see Eq.\ \ref{eq:ueg_xed}). Having obtained these covariance expressions, they can be plugged into the predictive function for a Gaussian process to obtain the predicted exchange energy density~\cite{Rasmussen2006}
\begin{align}
    e_\text{x}^{\text{CIDER}}(\mathbf{x}_*) &= e_\text{x}^\text{LDA}(n_*) \sum_a k_{F_\text{x}}(\mathbf{x}_*, \mathbf{\tilde{x}}_a) \alpha_a \\
    \boldsymbol{\alpha} &= \sum_i \mathbf{\tilde{k}}^i \left\{\left[\mathbf{K} + \boldsymbol{\Sigma}_{\text{noise}}\right]^{-1} \mathbf{y}\right\}_i,\label{eq:gp_predictive}
\end{align}
where $n_*$ and $\mathbf{x}_*$ are the density and feature vector at a test point and $\boldsymbol{\Sigma}_{\text{noise}}$ is the noise matrix for the training points.

\subsubsection{Model Types and Feature vectors} \label{sec:feature_defs}

This section lists the feature vectors for each model type in terms of $s$ (Eq.\ \ref{eq:reduced_grad_s}), $\alpha$ (Eq.\ \ref{eq:iso_orbital_alpha}), and the nonlocal features $G_i$ (Eq.~\ref{eq:cider_feature}). All of the feature vectors below are designed such that for the uniform electron gas, $\mathbf{x}=\mathbf{0}$. Since the exchange enhancement factor is $F_\text{x}=1$ for the uniform electron gas~\cite{Dirac1930}, the uniform electron gas constraint is enforced with a single noiseless training point $F_\text{x}^\text{CIDER}(\mathbf{0})=1$. All functionals in this work satisfy the uniform electron gas constraint.

The primary purpose of the transformations performed on the features below is to constrain the ML feature vector elements to fall in a finite range (e.g.\ $[0,1]$), which makes it easier to map the resulting models to cubic splines as discussed in Section~\ref{sec:gp_kernels}. Aside from that, the exact form of the feature transformation is a flexible choice, informed by physical considerations, but should not matter much as long as the target is a reasonably smooth function of the transformed features. However, more systematically fine-tuning these transformations in the future could potentially improve model performance slightly.

For the SL-GGA, the feature vector $\mathbf{x}$ (Eq.\ \ref{eq:general_xf}) is of dimension 1:
\begin{equation}
    x_1 = \frac{\gamma s^2}{1 + \gamma s^2},\label{eq:gga_feature}
\end{equation}
where $\gamma=0.243$ was empirically chosen by Becke~\cite{Becke1986} to fit the exact exchange energies of noble gas atoms. This value of $\gamma$ was also used in our previous work~\cite{Bystrom2022} and in the B97-type functionals of the Head-Gordon group~\cite{Mardirossian2014,Mardirossian2015,Mardirossian2016}. For the SL-MGGA, the feature vector is of dimension 2:
\begin{align}
    x_1 &= \frac{\gamma s^2}{1 + \gamma s^2} \\
    x_2 &= \frac{2}{1 + \alpha^2} - 1.
\end{align}
The NL-GGA feature vector is of dimension 4:
\begin{align}
    x_1 &= \frac{\gamma s^2}{1 + \gamma s^2} \\
    x_2 &= \frac{G_1}{2+G_1} - \frac{1}{2} \\
    x_3 &= \frac{G_2}{2+G_2} - \frac{1}{2} \\
    x_4 &= \frac{G_3}{2+G_3} - \frac{1}{2}.
\end{align}
Lastly, the NL-MGGA feature vector is of dimension 5:
\begin{align}
    x_1 &= \frac{\gamma s^2}{1 + \gamma s^2} \\
    x_2 &= \frac{2}{1 + \alpha^2} - 1 \\
    x_3 &= \frac{G_1}{2+G_1} - \frac{1}{2} \\
    x_4 &= \frac{G_2}{2+G_2} - \frac{1}{2} \\
    x_5 &= \frac{G_3}{2+G_3} - \frac{1}{2}.
\end{align}
The three nonlocal features $G_1,G_2,G_3$ are given by Eq.\ \ref{eq:cider_feature}, and the parameterization of the kernel exponents (i.e. the constants $B_i$ and $C_i$ in Eqs.\ \ref{eq:anl_new}-\ref{eq:abnl_mgga_new}) is described in detail in Appendix~\ref{sec:hparam_sweep}.

\subsubsection{Kernels}\label{sec:gp_kernels}

To fully define the Gaussian process model for the exchange energy, we need to specify the kernel function $k_{F_\text{x}}(\mathbf{x}, \mathbf{x}')$ and the noise matrix $\boldsymbol{\Sigma}_{\text{noise}}$. The noise matrix is constructed to heuristically approximate the model uncertainty for each data point as described in Appendix~\ref{sec:noise_hparam}. To reflect the different number of features in each functional type, a different kernel is used for $k_{F_\text{x}}(\mathbf{x}, \mathbf{x}')$ for each, as specified below:
\begin{align}
    k_{\text{SL-GGA}}(\mathbf{x},\mathbf{x}') &= \Sigma_{\text{Cov}} k_1(x_1, x_1') \label{eq:slgga_kernel} \\
    k_{\text{SL-MGGA}}(\mathbf{x},\mathbf{x}') &= \Sigma_{\text{Cov}} k_1(x_1, x_1') k_2(x_2, x_2') \label{eq:slmgga_kernel} \\
    k_{\text{NL-GGA}}(\mathbf{x},\mathbf{x}') &= \Sigma_{\text{Cov}} k_1(x_1, x_1') \notag\\ &\times \sum_{i=2}^4 \sum_{j=i+1}^4 k_i(x_i, x_i') k_j(x_j, x_j') \label{eq:nlgga_kernel} \\
    k_{\text{NL-MGGA}}(\mathbf{x},\mathbf{x}') &= \Sigma_{\text{Cov}} k_1(x_1, x_1') \notag\\ &\times \sum_{i=2}^5 \sum_{j=i+1}^5 k_i(x_i, x_i') k_j(x_j, x_j'), \label{eq:nlmgga_kernel}
\end{align}
where $\Sigma_{\text{Cov}}$ is a covariance hyperparamter that can be tuned for each model, and the base kernel $k_i(x,x')$ for all of the above is the squared-exponential kernel
\begin{equation}
    k_i(x,x')=\exp[-\frac{1}{2}\left(\frac{x-x'}{l_i}\right)^2].\label{eq:basekernel}
\end{equation}
Equations \ref{eq:nlgga_kernel} and \ref{eq:nlmgga_kernel} use the additive Gaussian process approach of Duvenaud \emph{et al.}~\cite{Duvenaud2011}.

Because the cost of evaluating Gaussian processes scales linearly with the number of training points, we map the models to cubic splines after training so they can be evaluated efficiently. The semilocal kernels have 1 and 2 features, so they can be mapped to 1 and 2-dimensional cubic splines, respectively. The nonlocal kernels are a sum of 3-dimensional kernels, so they can be mapped to a sum of 3-dimensional cubic splines~\cite{Bystrom2022,Glielmo2018,Vandermause2019,Xie2021BayesianStanene}. For a description of the hyperparameter selection process for these models, see Appendix~\ref{sec:hparam_sweep}.

\subsection{Auxiliary Expansion of the CIDER Features}\label{sec:rps}

The rest of this section focuses on the efficient implementation of nonlocal features within Gaussian-type orbital and plane-wave DFT, which is vital for the practicality of CIDER as well as any nonlocal density functional. Evaluating Eq.\ \ref{eq:cider_feature} via simple numerical integration is computationally expensive due to quadratic scaling with the number of grid points. The key insight to designing a faster approach comes from Rom\'an-P\'erez and Soler~\cite{Roman-Perez2009}, who developed a method to expand van der Waals density functionals as a sum of convolutions that could be computed efficiently using the Fast Fourier Transform (FFT).
They observed that a function of the form in Eq.\ \ref{eq:cider_kernel}, which depends exclusively on the distance between two points and on a semilocal functional of the density at each point, can be approximately expanded as~\cite{Roman-Perez2009}
\begin{equation}
    \Phi\left(a(\mathbf{r}_2), b(\mathbf{r}_1), r_{12}\right) \approx
    \sum_{\alpha\beta} p^a_{\alpha}(a(\mathbf{r}_2)) p^b_{\beta}(b(\mathbf{r}_1)) \Phi_{\alpha\beta}(r_{12}).\label{eq:rps_expansion}
\end{equation}
In Eq.\ \ref{eq:rps_expansion}, $\alpha$ and $\beta$ are indices which each correspond to interpolation control points $q_\alpha$ and $q_\beta$, respectively, and $p^a_{\alpha}(a(\mathbf{r}))$ and $p^b_{\beta}(b(\mathbf{r}))$ are interpolating functions which take the values $p^a_{\alpha}(q_{\alpha'})=\delta_{\alpha\alpha'}$ and $p^b_{\beta}(q_{\beta'})=\delta_{\beta\beta'}$ at the interpolating points. The term $\Phi_{\alpha\beta}(r)\equiv\Phi(q_\alpha, q_\beta, r)$.

The Gaussian form of Eq.\ \ref{eq:cider_kernel} is particularly conducive to the use of an even-tempered basis for interpolation, in which $q_\alpha$ takes the form
\begin{equation}
    q_{\alpha} = q_0 \lambda^\alpha \label{eq:qlambd}
\end{equation}
for some constant $\lambda>1$. Rom\'an-P\'erez and Soler~\cite{Roman-Perez2009} used cubic splines for interpolation, and for the solid-state implementation of CIDER we follow this approach, with $p^a_{\alpha}(a(\mathbf{r}))$ being a cubic spline in the transformed coordinate $\gamma(\mathbf{r})=\frac{\ln(a(\mathbf{r})/q_0)}{\ln(\lambda)}$. However, for the molecular implementation, we instead treat the control points as a basis set of Gaussians $e^{-q_{\alpha}r_{12}^2}$ onto which the function $e^{-a(\mathbf{r_1})r_{12}^2}$ is projected:
\begin{align}
    p^a_{\alpha}(a(\mathbf{r})) &\equiv \sum_{\beta} (S_a^{-1})_{\alpha\beta} s_{\alpha}(a(\mathbf{r})) \\
    (S_a)_{\alpha\beta} &= \int \dd[3]\mathbf{r} \exp\left[-(q_\alpha+q_\beta)|\mathbf{r}|^2\right] \\
    s_{\alpha}(a(\mathbf{r})) &= \int \dd[3]\mathbf{r'} \exp\left\{-\left(a(\mathbf{r})+q_\alpha\right)|\mathbf{r}'|^2\right\}.
\end{align}
As $\lambda\rightarrow 1$, the number of control points (for the interpolation scheme) and number of basis functions (for the basis expansion scheme) both become infinite, and the kernel expansion becomes exact for both approaches.
We determined that the cubic interpolation yields better numerical stability for the plane-wave implementation, while the basis expansion form is sufficiently stable for the molecular implementation.

The benefit of expanding the kernel in this manner is that it allows the nonlocal features (Eq.\ \ref{eq:cider_feature}) to be expressed as a sum of convolutions
\begin{equation}
    G(\mathbf{r}_1) = \sum_{\beta} p^b_{\beta}(b(\mathbf{r}_1)) \sum_{\alpha} \int \dd[3]\mathbf{r}_2 \Phi_{\alpha\beta}(r_{12}) \theta_{\alpha}(\mathbf{r}_2),\label{eq:rps_cider_feat}
\end{equation}
with $\theta_{\alpha}(\mathbf{r})=p^a_{\alpha}(a(\mathbf{r}))n(\mathbf{r})$. Within plane-wave DFT, the convolutions can be performed with quasi-linear computational complexity by using FFTs~\cite{Roman-Perez2009}. In addition, because CIDER uses a squared exponential for $\Phi_{\alpha\beta}(r_{12})$ (Eq.\ \ref{eq:cider_kernel}), the integrals of Eq.\ \ref{eq:rps_cider_feat} can be evaluated analytically if $\theta_{\alpha}(\mathbf{r})$ is projected onto a Gaussian-type orbital basis. We use this approach to design an efficient implementation of nonlocal features for molecular DFT (Section \ref{sec:theory_cidermol}).

The Rom\'an-P\'erez and Soler technique effectively solves the problem of implementing CIDER for an all-electron plane-wave DFT calculation. However, such calculations are infeasible for all but the smallest atoms because a prohibitively large plane-wave basis would be required to describe the core electrons. In addition, pseudopotentials cannot be reliably used with nonlocal features because $G(\mathbf{r})$ calculated on the pseudo-density is not sufficiently similar to the exact, all-electron feature to get accurate results. To solve this problem, in Section~\ref{sec:theory_ciderpaw} a method is introduced to implement nonlocal features within the Projector-Augmented Wave (PAW) formalism, which requires careful attention to the contributions to $G(\mathbf{r})$ from the core regions around the atoms.

Before proceeding, we note that while we borrow and expand on numerical techniques developed for nonlocal van der Waals functionals, we emphasize that the functionals in this work are still fit only to the exchange energy and therefore do not account for the correlation effects that give rise to van der Waals interactions. Due to their nonlocality, the nonlocal features in this work could prove useful for learning van der Waals interactions, but demonstrating this will require further research.

\subsection{Efficient Implementation of Nonlocal CIDER Features for Molecular DFT}\label{sec:theory_cidermol}

Computing Eq.\ \ref{eq:cider_feature} using simple numerical integration is prohibitively expensive because the number of operations is proportional to the square of the number of grid points. While van der Waals functionals like VV10~\cite{Vydrov2010} are usually computed this way within Gaussian-type orbital codes, the contribution of such functionals is sufficiently small and smooth that a very sparse grid can be used. The nonlocal features require a grid of the same density as the semilocal XC grid, and the number of operations per pair of grid points is higher for evaluating Eq.\ \ref{eq:cider_feature} than, for example, the VV10 kernel. It is therefore necessary to reduce the computational complexity of evaluating the nonlocal features for molecular systems.

In this section, we present a charge-partitioning and density fitting-based approach to compute the nonlocal features within Gaussian-type orbital DFT codes for molecular systems, which we have implemented for use in the PySCF code~\cite{Sun2018,Sun2020}. Much of this approach is based on the work of Franchini et al.~\cite{Franchini2014}, who developed a similar method for computing the Coulomb energy. To start, the density is partitioned into atomic contributions using Becke partitioning~\cite{Becke1988},
\begin{align}
    n_A(\mathbf{r}) &= W_A(\mathbf{r}) n(\mathbf{r}) \\
    \sum_A W_A(\mathbf{r}) &= 1,
\end{align}
where $A$ is the atom index and $W_A(\mathbf{r})$ are the atomic Becke weights.
The first step in evaluating $G(\mathbf{r})$ is to project $\theta_{\alpha}(\mathbf{r})$ onto spherical harmonics for each atom (with $L$ being the combined $l$ and $m$ indices and $\mathbf{R}_A$ being the position of atom $A$):
\begin{align}
    \theta_{\alpha,L}^A(r) &= \int \dd\Omega_A \theta_{\alpha}(\mathbf{r}) W_A(\mathbf{r}) Y_L(\Omega_A)\\
    \mathbf{r}&=\mathbf{R}_A+(r\cos\phi\sin\theta, r\sin\phi\sin\theta, r\cos\theta),
\end{align}
where $\Omega_A=(\theta,\phi)$ is the solid angle and $Y_L(\Omega)$ are the spherical harmonics. This projection is done numerically using Lebedev grids~\cite{Lebedev1999}. Then each $L$ channel is projected onto an even-tempered Gaussian basis $\{\chi_{A,L,\mu}(\mathbf{r})\}$ as follows:
\begin{align}
    \chi_{A,L,\mu}(\mathbf{r}) &= Y_{L}(\hat{\mathbf{r}}) N_{l,\mu} r^l \exp(-\mu r^2)\label{eq:theta_proj} \\
    \tilde{\theta}_{\alpha,L,\mu}^A &= \int \dd r\,r^{2+l} N_{l,\mu} \exp(-\mu r^2) \theta_{\alpha,L}^A(r),
\end{align}
where $N_{l,\mu}$ is a normalization constant. The exponents $\mu$ are constructed using the even-tempered Gaussian approach with tunable parameter $\beta^\text{aux}$
\begin{equation}
    \mu_n = \mu_\text{min} \left(\beta^\text{aux}\right)^n\,\,\,\,\,\,\,n=\{0,1,...,N\} \label{eq:auxiliary_etb}
\end{equation}
A transformation to $\theta_{\alpha,L,\mu}^A$ is then performed by multiplying $\tilde{\theta}_{\alpha,L,\mu}^A$ by the inverse overlap matrix of the $\chi$ basis for each $l$ channel:
\begin{align}
    \theta_{\alpha,L,\mu}^A &= \sum_\nu \left[\left(\mathbf{S}_{A,L}^\chi\right)^{-1}\right]_{\mu\nu} \tilde{\theta}_{\alpha,L,\nu}^A \label{eq:theta_transform} \\
    \left(\mathbf{S}_{A,L}^\chi\right)_{\mu\nu} &= \braket{\chi_{A,L,\mu}}{\chi_{A,L,\nu}}.
\end{align}

For each atom, a second even-tempered basis $\{\xi_{A,L,\mu}(\mathbf{r})\}$ is introduced to serve as a basis for the convolutions of $\chi_{A,L,\mu}(\mathbf{r})$. The following projections are computed using analytical Gaussian integrals:
\begin{align}
    \tilde{F}_{\beta,L,\nu}^A =& \sum_{\alpha,\mu} \theta_{\alpha,L,\mu}^A \notag\\ &\times \int \dd[3]\mathbf{r}_1 \dd[3]\mathbf{r}_2\, \Phi_{\alpha\beta}(r_{12}) \chi_{A,L,\mu}(\mathbf{r}_1) \xi_{A,L,\nu}(\mathbf{r}_2). \label{eq:xi_proj}
\end{align}
The transformation to $F_{\beta,L,\nu}^A$ is then performed by multiplying $\tilde{F}_{\beta,L,\nu}^A$ by the inverse overlap matrix of the $\xi$ basis, similarly to Eq.\ \ref{eq:theta_transform}.

The second to last step is to transform $F_{\beta,L,\nu}^A$ to $F_{\beta}(\mathbf{r}_g)$, where $\mathbf{r}_g$ are the DFT integration grid points. Because evaluating every $\xi$ orbital of every atom at every $\mathbf{r}_g$ would be computationally expensive, $F_{\beta,L,\nu}^A$ is first transformed to a cubic spline $F_{\beta,L,G,p}^A$, where $G$ is a radial coordinate index and $p$ indexes the polynomial coefficients of the spline in the interval between points $G$ and $G+1$. The transformation from $F_{\beta,L,G,p}^A$ to $F_{\beta}(\mathbf{r}_g)$ is then performed as
\begin{equation}
    F_{\beta}(\mathbf{r}_g) = \sum_A \sum_L Y_L(\hat{\mathbf{r}}_{gA}) \sum_{p=0}^3 (r_{gA} - r_G)^p F^A_{\beta,L,G,p},\label{eq:spline_eval}
\end{equation}
where $\mathbf{r}_{gA}=\mathbf{r}_g - \mathbf{R}_A$. The spline index $G$ is a step function of $|\mathbf{r}_{gA}|$. The computational cost of evaluating the splines depends heavily on the efficiency of memory access patterns on the processor, so the coordinates are sorted by $G$ before evaluating the contributions to $F_{\beta}(\mathbf{r}_g)$ from each atom.


The last and simplest step to compute $G(\mathbf{r})$ is to evaluate the sum
\begin{equation}
    G(\mathbf{r}) = \sum_{\beta} p^b_\beta(\mathbf{r}_g) F_{\beta}(\mathbf{r}_g).
\end{equation}
After computing the XC energy from $G(\mathbf{r})$ and the semilocal features, the contributions to the XC potential from the $\theta_{\alpha}(\mathbf{r})$ terms can then be computed by passing backward through the matrix multiplications and spline evaluations detailed above.

The gradient of the total energy with respect to nuclear positions is required for the evaluation of forces. For this implementation of the CIDER functional with nonlocal features, the energy gradient with respect to the nuclear position of atom $A$ is given by
\begin{align}
    \nabla_A E_{\text{xc}} =& \nabla_A^{\text{sl}} E_{\text{xc}} + \sum_g \left(\sum_{\alpha} \fdv{E_{\text{xc}}}{\theta_{\alpha}(\mathbf{r}_g)} \theta_{\alpha}(\mathbf{r}_g) \right) \nabla_A w_g \notag\\
    &+ \sum_g \left(\sum_\beta \fdv{E_{\text{xc}}}{F_{\beta}(\mathbf{r}_g)} \nabla_A F_\beta(\mathbf{r}_g)\right) w_g,
\end{align}
where $w_g$ is the numerical integration quadrature weight corresponding to the coordinate $\mathbf{r}_g$. The analytical energy gradient for a semilocal functional $\nabla_A^{\text{sl}}E_{\text{xc}}$ is given by Johnson \emph{et al.}~\cite{Johnson1993} Eq.\ 11.

The computational bottleneck of this algorithm is the spline evaluation in Eq.\ \ref{eq:spline_eval}, whose cost scales quadratically with system size. Specifically, the cost scales as $\Theta(N_\text{grid}\, N_\text{atom}\, N_\beta\, l_\text{max}^2)$, where $N_\text{grid}$ is the number of integration grid points, $N_\text{atom}$ is the number of atoms, $N_\beta$ is the number of kernel interpolation points in Eq.\ \ref{eq:rps_expansion}, and $l_\text{max}$ is the maximum spherical harmonic order used in the $\xi$ basis. All other components of the algorithm are performed on individual atoms and therefore scale linearly with system size.

\subsection{Quasi-linear-scaling Implementation of Nonlocal CIDER Features for Plane-Wave DFT with PAW}\label{sec:theory_ciderpaw}

This section covers our algorithm for evaluating the nonlocal CIDER features within the Projector-Augmented Wave (PAW) method~\cite{Blochl1994}, a generalization of the pseudopotential method that can recover the full wave function and electron density. We have implemented this algorithm in the GPAW code~\cite{Mortensen2005,Enkovaara2010}. As discussed above, Eqs.\ \ref{eq:rps_expansion} and \ref{eq:rps_cider_feat} enable the evaluation of $G(\mathbf{r})$ in $O(N\log N)$ complexity if the density $n(\mathbf{r})$ is represented on an FFT grid. However, in the PAW formalism of Bl\"ochl~\cite{Blochl1994}, the all-electron density $n(\mathbf{r})$ is represented as
\begin{equation}
    n(\mathbf{r}) = \tilde{n}(\mathbf{r}) + \sum_A (n_A^1(\mathbf{r}) - \tilde{n}_A^1(\mathbf{r})),\label{eq:paw_dens}
\end{equation}
where $A$ indexes the atoms, $\tilde{n}(\mathbf{r})$ is the pseudo-density on the FFT grid, and the terms with the 1 superscript are represented on radial support grids centered around each atom and extending to some cutoff radius $r_c$. Within the cutoff radius of atom $A$, $\tilde{n}(\mathbf{r})=\tilde{n}_A^1(\mathbf{r})$ and $n_A^1(\mathbf{r})$ is the all-electron density. Outside the cutoff radius of atom $A$, $n_A^1(\mathbf{r})=\tilde{n}_A^1(\mathbf{r})$.

For a semilocal functional, Eq.\ \ref{eq:paw_dens} yields a simple expression for the XC energy $E_{\text{xc}}[n]$:
\begin{equation}
    E_{\text{xc}}[n] = E_{\text{xc}}[\tilde{n}] + \sum_A (E_{\text{xc}}[n_A^1] - E_{\text{xc}}[\tilde{n}_A^1]).
\end{equation}
However, the case is not so simple for a nonlocal functional such as CIDER. If Eq.\ \ref{eq:cider_feature} is evaluated on the pseudo-density $\tilde{n}(\mathbf{r})$, the resulting $G(\mathbf{r})$ is not equal to the all-electron feature even outside the cutoff radius because the convolutions used to compute the feature are nonlocal. For the same reason, evaluating Eq.\ \ref{eq:cider_feature} on the atomic density $n_A^1(\mathbf{r})$ does not yield the all-electron feature inside the cutoff radius. Therefore, an implementation of CIDER (and other nonlocal functionals with similar forms, such as some kinetic energy and van der Waals functionals) for PAW must construct a ``pseudo-feature'' $\tilde{G}(\mathbf{r})$ that is smooth but yields the correct feature outside the cutoff radius, as well as construct the ``all-electron feature'' $G(\mathbf{r})$ on the radial support grids inside the cutoff radius. Because of this difficulty, current plane-wave implementations of nonlocal functionals like van der Waals correlation~\cite{Klimes2011} use the pseudo-density rather than the full PAW density. When this approach is used, one may need to use PAW datasets with more valence electrons, increasing computational cost~\cite{Klimes2011}. In addition, because CIDER computes the exchange energy, which is of much larger magnitude than the van der Waals correlation energy, higher precision is required. A similar problem exists for kinetic energy density functionals. There is an implementation of semilocal kinetic energy functionals with PAW~\cite{Lehtomaki2014}, and of nonlocal kinetic energy functionals with ultrasoft pseudopotentials~\cite{Xu2022}, but no nonlocal kinetic energy functional implementation within PAW. Therefore, by designing an algorithm for computing nonlocal features for CIDER within the PAW formalism, we also enable the full integration of van der Waals functionals and orbital-free DFT~\cite{Witt2018} into the PAW method.

Starting from Eq.\ \ref{eq:paw_dens}, the all-electron value of $G(\mathbf{r})$ \emph{outside the core region} is given by
\begin{align}
    \tilde{G}(\mathbf{r}) &= \sum_\beta p^b_\beta\left(\tilde{b}(\mathbf{r})\right) \left[ \tilde{F}_\beta(\mathbf{r})
    + \sum_A \Delta F_{A,\beta}(\mathbf{r}) \right] \\
    \tilde{F}_\beta(\mathbf{r}_1) &= \sum_\alpha \int \dd[3]\mathbf{r}_2 \, \Phi_{\alpha\beta}(r_{12}) \tilde{\theta}_{\alpha}(\mathbf{r}_2)\\
    \Delta F_{A,\beta}(\mathbf{r}_1) &= \sum_\alpha \int \dd[3]\mathbf{r}_2 \, \Phi_{\alpha\beta}(r_{12}) \left(\theta^1_{A,\alpha}(\mathbf{r}_2) - \tilde{\theta}^1_{A,\alpha}(\mathbf{r}_2)\right).
\end{align}
The quantities $\tilde{b}(\mathbf{r})$ and $\tilde{\theta}_\alpha(\mathbf{r})$ are evaluated on the pseudo-density $\tilde{n}(\mathbf{r})$; $\theta^1_{A,\alpha}(\mathbf{r})$ and $\tilde{\theta}^1_{A,\alpha}(\mathbf{r})$ are evaluated on $n_A^1(\mathbf{r})$ and $\tilde{n}_A^1(\mathbf{r})$, respectively. $\tilde{F}_\beta(\mathbf{r})$ can be evaluated simply using FFTs. To compute $\Delta F_{A,\beta}(\mathbf{r})$, the term $\Delta\theta^1_{A,\alpha}(\mathbf{r})=\theta^1_{A,\alpha}(\mathbf{r}) - \tilde{\theta}^1_{A,\alpha}(\mathbf{r})$ is first projected onto spherical harmonics and Gaussians as described in the previous section. Because $\Delta\theta^1_{A,\alpha}(\mathbf{r})$ is localized within the augmentation region and mostly spherical, far fewer spherical harmonics and Gaussians are required per atom than for the Gaussian-type orbital implementation. This projection yields $\Delta\theta_{A,\alpha,L,\mu}^1$. The Gaussians are then analytically projected into reciprocal space to yield $\Delta\theta_{A,\alpha,L}^1(k)$. In reciprocal space,
\begin{equation}
    \Delta F_{A,L,\beta}(k) = \sum_{\alpha} \Phi_{\alpha\beta}(k) \Delta\theta_{A,\alpha,L}^1(k).\label{eq:dfalbk}
\end{equation}
The above equation applies because the convolution $\Phi_{\alpha\beta}(k)$ has angular momentum number $l=0$. Note that in this section, denoting a symbol as function of $k$ implies that it is the Fourier transform of the real-space function corresponding to that symbol. For example, $\Phi_{\alpha\beta}(k)$ is the Fourier transform of $\Phi_{\alpha\beta}(r)$.

Inspired by the PAW method itself, we perform the following trick to approximate $\Delta F_{A,\beta}(\mathbf{r})$ on the FFT grid. We introduce two sets of functions, $\{g_i^A(\mathbf{r})\}$ and $\{h_j^A(\mathbf{r})\}$. The $h_j^A(\mathbf{r})$ functions form a localized basis (i.e. each function vanishes for $r>r_c$) and are meant to fit the high-frequency components of $\Delta F_{A,\beta}(\mathbf{r})$ introduced by the all-electron density inside the core region. For the current implementation of these functions, we use differences between Gaussians and smooth polynomials that match their derivatives at $r_c$. The other functions $g_i^A(\mathbf{r})$ are both localized and smooth, and are meant to augment $\Delta\theta^1_{A,\alpha}(\mathbf{r})$ to yield the correct values of $\Delta F_{A,\beta}(\mathbf{r})$ outside of the augmentation region. To do this, least-squares linear regression is used to fit the following function to $\Delta F_{A,\beta}(\mathbf{r})$:
\begin{align}
    \hat{y}_{A,\beta}(\mathbf{r}) &= \hat{y}^1_{A,\beta}(\mathbf{r}) + \hat{y}^2_{A,\beta}(\mathbf{r})\\
    \hat{y}^1_{A,\beta}(\mathbf{r}) &= \sum_j D_{j\beta} h_j^A(\mathbf{r}) \label{eq:djb}\\
    \hat{y}^2_{A,\beta}(\mathbf{r}_1) &= \sum_{i,\alpha} C_{i\alpha} \int \dd[3]\mathbf{r}_2 \phi_{\alpha\beta}(r_{12}) g_i^A(\mathbf{r}_2), \label{eq:cia}
\end{align}
where the coefficients $C_{i\alpha}$ and $D_{j\beta}$ are determined by the regression. The necessary integrals are actually performed in reciprocal space since Eq.\ \ref{eq:dfalbk} yields $\Delta F_{A,L,\beta}(k)$ rather than $\Delta F_{A,\beta}(\mathbf{r})$. At first glance, this appears to be a complicated and expensive fitting problem, but by separating the basis functions into their individual angular momentum channels and doing a separate regression for each channel, it becomes tractable. The details for this procedure are described in Appendix \ref{app:feat_smooth_const}.

Because the functions $h_j^A(\mathbf{r})$ are localized, in the complete basis set limit $\Delta F_{A,\beta}(\mathbf{r}) = \hat{y}^2_{A,\beta}(\mathbf{r})$ outside the augmentation region. Therefore, by defining
\begin{equation}
    \tilde{\theta}^{\text{aug}}_{\alpha}(\mathbf{r}) \equiv \tilde{\theta}_\alpha(\mathbf{r}) + \sum_A \sum_{i} C_{i\alpha}^A g_i^A(\mathbf{r}),\label{eq:theta_aug}
\end{equation}
one can compute
\begin{align}
    \tilde{G}(\mathbf{r}) &= \sum_\beta p^b_\beta\left(\tilde{b}(\mathbf{r})\right) \tilde{F}^{\text{aug}}_\beta(\mathbf{r}) \\
    \tilde{F}^{\text{aug}}_\beta(\mathbf{r}) &= \sum_\alpha \int \dd[3]\mathbf{r}_2 \Phi_{\alpha\beta}(r_{12}) \tilde{\theta}^{\text{aug}}_{\alpha}(\mathbf{r}),\label{eq:feat_convolutions}
\end{align}
which satisfies the condition that the pseudo-feature $\tilde{G}(\mathbf{r})$ is equal to the all-electron feature $G(\mathbf{r})$ outside the augmentation region (within the limit of complete basis sets).

Within the core regions, $\tilde{F}^{\text{aug}}_\beta(\mathbf{r})$ is projected onto the radial support grids using a set of projectors in a very similar formalism to PAW itself. This quantity will be called $\tilde{F}^{1,\text{aug}}_{A,\beta}(\mathbf{r})$, in keeping with the PAW notation, and it takes the form
\begin{align}
    \tilde{F}^{1,\text{aug}}_{A,\beta}(\mathbf{r}) =& \sum_j B_{j\beta}^A \tilde{f}^A_j(\mathbf{r}) + \tilde{F}_{A,\beta}^{1,0}(\mathbf{r}) \notag\\
    &- \sum_j \tilde{f}_j^A(\mathbf{r}) \left[\int_1 \dd[3]\mathbf{r}' \tilde{p}^A_j(\mathbf{r}') \tilde{F}_{A,\beta}^{1,0}(\mathbf{r}')\right] \label{eq:feat_paw_aug} \\
    B_{j\beta}^A =& \dd v \sum_{g\in A} \tilde{F}^{\text{aug}}_\beta(\mathbf{r}_g) \tilde{p}_j^A(\mathbf{r}_g). \label{eq:pa_numint}
\end{align}
In the above equations, $\tilde{f}_i^A(\mathbf{r})$ and $\tilde{p}_j^A(\mathbf{r})$ are sets of atom-centered functions satisfying $\braket{\tilde{f}_i^A}{\tilde{p}_j^A}=\delta_{ij}$, and $\tilde{p}_j^A(\mathbf{r})$ vanish for $r>r_c$ so that the numerical integral of Eq.\ \ref{eq:pa_numint} can be performed efficiently on the FFT grid points inside the cutoff sphere of atom $A$. The symbol $\dd v$ represents the FFT grid volume element. This is directly analogous to the projector-partial wave pairs in PAW~\cite{Blochl1994}. The construction of $\tilde{f}_i^A(\mathbf{r})$ and $\tilde{p}_j^A(\mathbf{r})$ is described in Appendix \ref{app:feat_aug_const}. $\tilde{F}_{A,\beta}^{1,0}(\mathbf{r})$ is an approximate convolution of $\tilde{\theta}^1_{A,\alpha}(\mathbf{r})$, and while arbitrary, it helps make the $\tilde{f}_j^A(\mathbf{r})$ expansion more efficient. Using this construction, the all-electron and pseudo-features on the radial support grids are, respectively,
\begin{align}
    G_A^1(\mathbf{r}) &= \sum_\beta p^b_\beta\left(b_A^1(\mathbf{r})\right) \left(\tilde{F}^{1,\text{aug}}_{A,\beta}(\mathbf{r}) + \hat{y}^1_{A,\beta}(\mathbf{r})\right)\\
    \tilde{G}_A^1(\mathbf{r}) &= \sum_\beta p^b_\beta\left(\tilde{b}_A^1(\mathbf{r})\right) \tilde{F}^{1,\text{aug}}_{A,\beta}(\mathbf{r}).
\end{align}
The exchange correlation energy is then
\begin{equation}
    E_\text{xc}[n] = E_\text{xc}[\tilde{n},\tilde{G}] + \sum_A \left(E_\text{xc}[n^1_A, G^1_A] - E_\text{xc}[\tilde{n}^1_A, \tilde{G}^1_A] \right),
\end{equation}
where all three terms are semilocal functionals of the density and of the nonlocal features. The first term is evaluated on the FFT grid and the second two on the radial support grids on the atoms.

The use of nonlocal features introduces a few additional terms to the forces and stresses not present for semilocal DFT. The force terms are
\begin{align}
    \mathbf{F}_A &= \mathbf{F}^\text{sl}_A + \mathbf{F}^{\text{PASDW}}_A \\
    \mathbf{F}^{\text{PASDW}}_A &= \sum_{\alpha} \dd v \sum_{ig}  C^A_{i\alpha} \nabla g_i^A(\mathbf{r}_g) \pdv{E_{\text{xc}}}{\theta_\alpha(\mathbf{r}_g)}\notag\\
    &+ \sum_\beta \dd v \sum_{jg} \pdv{E_{\text{xc}}}{B^A_{j\beta}} \nabla \tilde{p}_j^A(\mathbf{r}_g) \tilde{F}_\beta(\mathbf{r}_g).
\end{align}
The stress terms are
\begin{widetext}
\begin{align}
    \sigma_{\mu\nu} &= \sigma_{\mu\nu}^{\text{sl}} + \sigma_{\mu\nu}^{\text{PASDW}} \\
    \sigma_{\mu\nu}^{\text{PASDW}} &= \sum_\alpha \sum_A \dd v \sum_g \left\{ \left[ \left(\mathbf{r}_g-\mathbf{R}_A\right)_\mu \pdv{E_{\text{xc}}}{\theta_\alpha(\mathbf{r}_g)} \right]  \left[ \sum_i C_{i\alpha}^A \nabla_\nu g_i^A(\mathbf{r}_g) \right] \right\} 
    + \delta_{\mu\nu} \sum_\beta \sum_A \sum_j \left(\pdv{E_{\text{xc}}}{B_{j\beta}^A} B_{j\beta}^A\right) \notag\\
    &+ \sum_\beta \sum_A \dd v \sum_g \left\{ \left[ (\mathbf{r}_g-\mathbf{R}_A)_\mu F_\beta(\mathbf{r}_g) \right] \left[ \sum_j \pdv{E_{\text{xc}}}{B_{j\beta}^A} \nabla_\nu \tilde{p}_j^A(\mathbf{r}_g) \right] \right\} \notag\\
    &+ \sum_k \frac{G_{k\mu}G_{k\nu}}{|\mathbf{G}_k|} \sum_{\beta} \left(\pdv{E_{\text{xc}}}{\tilde{F}_\beta(\mathbf{G}_k)}\right)^* \sum_\alpha \Phi_{\alpha\beta}'(|\mathbf{G}_k|) \theta_{\alpha}(\mathbf{G}_k).
\end{align}
\end{widetext}
In the above equations, $\mathbf{F}^\text{sl}_A$ and $\sigma_{\mu\nu}^{\text{sl}}$ refer to the force and stress equations for semilocal functionals within the PAW formalism (see Section IIID of Ref.~\cite{Kresse1999a}).

There are several significant contributions to the computational cost of the above approach for evaluating $G(\mathbf{r})$, which all scale linearly or quasi-linearly with system size. The first contribution comes from computing the Fourier transforms of $\tilde{\theta}_\alpha^\text{aug}(\mathbf{r})$ and $\tilde{F}^{\text{aug}}_\beta(\mathbf{r})$ (as well as the functional derivatives with respect to these quantities), which is necessary to compute the convolutions like those in Eq.\ \ref{eq:feat_convolutions} in reciprocal space. The cost of these operations scales as $\Theta(N_\alpha \,N_\text{FFT}\,\log(N_\text{FFT}))$, where $N_\alpha$ is the number of kernel interpolation points in Eq.\ \ref{eq:rps_expansion} and $N_\text{FFT}$ is the size of the FFT grid used for integrating the XC energy. The second contribution arises from computing the convolutions in Eq.\ \ref{eq:feat_convolutions} in reciprocal space, and scales as $\Theta(N_\alpha^2\,N_\text{FFT})$. The third contribution comes from computing the PAW correction to $\tilde{\theta}_\alpha^\text{aug}(\mathbf{r})$ (the sum in Eq.\ \ref{eq:theta_aug}). This and related terms scale as $\Theta(N_\text{FFT,atom}\,N_\text{atom}\,N_\alpha)$, where $N_\text{FFT,atom}$ is the number of FFT grid points inside the core region of an atom and $N_\text{atom}$ is the number of atoms. The last contribution consists of all the other PAW correction-related routines (e.g. computing $C_{i\alpha}^A$), which depend only on the density on the atomic radial grids and therefore scale linearly with $N_\text{atom}$. The overall scaling with system size $N$ is then $\Theta(N \log(N))$. Notably, none of the components of the algorithm scale with the number of k-points.

\section{Results and Discussion}\label{sec:results}

\begin{figure*}[btp]
    \includegraphics[width=1.5\columnwidth]{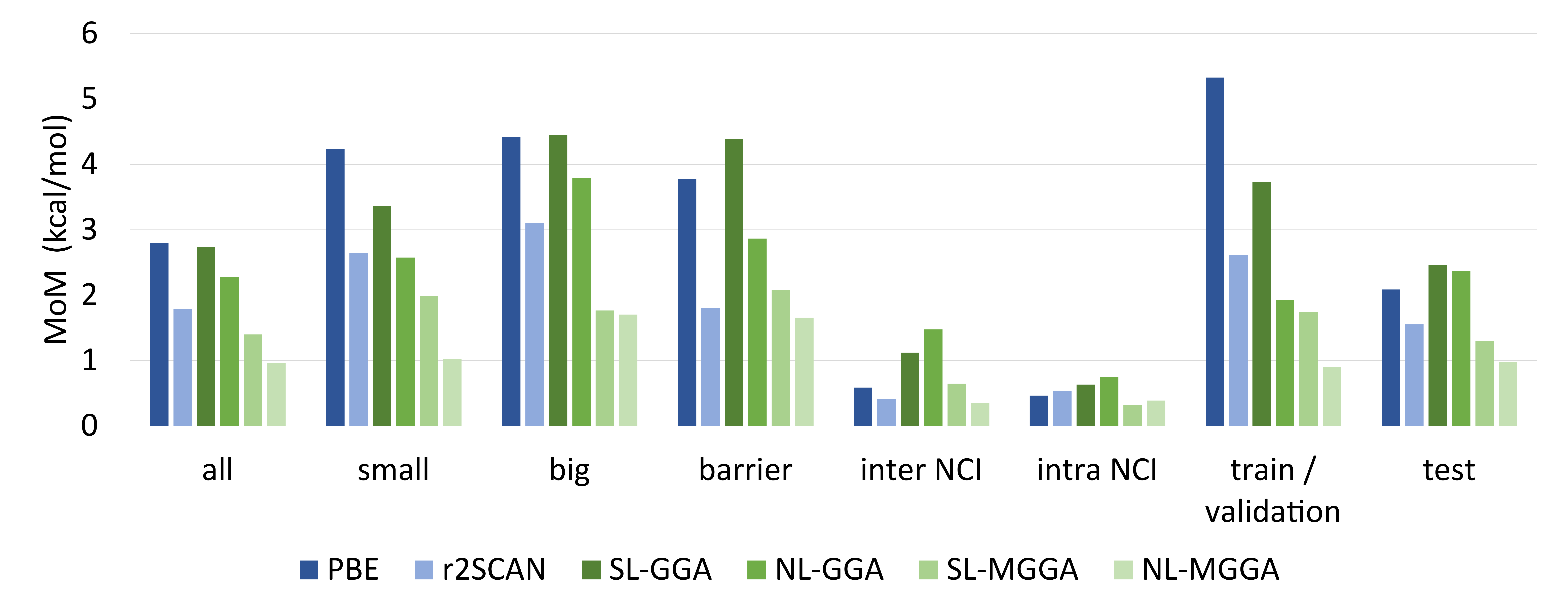}
    \caption{Accuracy of PBE0/CIDER (Eq.~\ref{eq:pbe0cider}) compared to PBE0 for each functional type on the GMTKN55 database of molecular properties, using the MoM error metric. The database subsets are~\cite{Goerigk2017}: ``all'': Full database; ``small'': Basic properties and reaction energies for small systems; ``big'': Reaction energies for large systems and isomerization reactions; ``barrier'': Barrier heights; ``inter NCI'': Intermolecular noncovalent interactions; ``intra NCI'': Intramolecular noncovalent interactions.}
    \label{fig:pbe0_ref_1}
\end{figure*}

Using Gaussian process regression, we fit each of the model types in Section~\ref{sec:gp} (SL-GGA, NL-GGA, SL-MGGA, and NL-MGGA) to match the exact exchange energy on subsets of the GMTKN55~\cite{Goerigk2017} and SOL62~\cite{Zhang2018,Trepte2022} databases, as well as a few dozen other training points. We provide a summary of our methods in this section and explain them in more technical detail in Appendix~\ref{sec:methods}.

To train the ML functionals, ground-state DFT calculations were first performed using the PBE functional~\cite{Perdew1996}. Then, for each chemical system in the training set, the density and orbitals obtained from the PBE calculation were used to compute the features described in Section~\ref{sec:feature_defs}, as well the exact exchange energy of Eq.\ \ref{eq:exact_exchange_energy}. The computed features were used as the inputs to the Gaussian process to predict the exact exchange energy. See Appendices~\ref{sec:molcalc_settings} and~\ref{sec:ssrefgpaw} for details.

The training and validation systems were chosen as described in Appendix~\ref{sec:trval_selection}. The SOL62 database~\cite{Zhang2018} contains cohesive energies for a mix of metals and non-metals, for which 9 were used for training, 22 for validation, and 31 for testing. The GMTKN55 database contains 55 sub-databases blocked into five categories: small-molecule properties (``small''), larger molecule reactions and isomerizations (``big''), barrier heights (``barrier''), intermolecular noncovalent interactions (``inter NCI''), and intramolecular noncovalent interactions (``intra NCI''). The training and validation data consist of twelve sub-databases of GMTKN55 and contain some representative systems from each category, but disproportionately taken from the ``small'' category. All data not used for training or validation is considered the GMTKN55 test set. This GMTKN55 partitioning results in 280 train data, 163 validation data, and 1062 test data. Further technical details for training are provided in Appendices~\ref{sec:control_points} and~\ref{sec:noise_hparam}.

The SL-GGA, NL-GGA, SL-MGGA, and NL-MGGA functionals used for benchmarking below are the functionals of each type with the best validation set performance, as described in Appendix~\ref{sec:hparam_sweep}. One important technical note is that each functional was trained via delta-learning on top of some baseline exchange functional. We tried PBE and the Chachiyo GGA~\cite{Chachiyo2020} as baselines, and Chachiyo proved more accurate for each functional class. However, as described in Section~\ref{sec:results_finetune} below, slight numerical stability improvements were found when using the NL-MGGA with a PBE baseline (dubbed NL-MGGA-PBE). As described in more detail in Section~\ref{sec:results_finetune}, we also found that retraining NL-MGGA-PBE with a more diverse training set from GMTKN55 (which is described in Appendix~\ref{sec:dtr_method}) results in improved numerical stability. Therefore, the resulting functional, dubbed NL-MGGA-DTR, was used for the applications in Sections~\ref{sec:results_hybrid}-\ref{sec:results_defects}.

When testing and benchmarking the functionals, all calculations (both conventional and CIDER functionals) were performed self-consistently, with the exception of the SOL62 cohesive energies. These were obtained non-self-consistently from PBE orbitals, due to the difficulty of computing the isolated atom energies in plane-wave DFT. This is a small technical barrier that will be addressed in future work. All calculations in the main text used static geometries, with the exception of the defect calculations in Section~\ref{sec:results_defects}; in this section, the silicon lattice parameters and defect structures were optimized with both PBE and CIDER.

In Section \ref{sec:results_mtype}, we demonstrate that the NL-MGGA is capable of learning the exchange functional more accurately than the other feature sets, and in Section \ref{sec:results_1e2e} we use 1 and 2-electron systems to provide intuition for why this is. Next, we fine-tune the accuracy and numerical stability of the NL-MGGA (Section \ref{sec:results_finetune}) and demonstrate that the resulting exchange functional can be used to match the accuracy of hybrid DFT (Section \ref{sec:results_hybrid}).

Having designed an accurate exchange functional, we demonstrate that it can be used to improve band gap predictions over standard meta-GGAs in Section \ref{sec:results_bandgap}. In Section \ref{sec:results_speed}, we show that for large systems in plane-wave DFT, our implementation of the NL-MGGA form is similarly efficient to semilocal meta-GGAs and much more efficient than hybrid DFT. Lastly, in Section \ref{sec:results_defects}, we take advantage of our new functional's efficiency and accurate band gap predictions to compute charge transition levels of point defects in silicon, which is a notoriously challenging task for DFT due to the underestimation of band gaps and the large supercell sizes required to simulate isolated point defects~\cite{Freysoldt2014}.

Taken together, the benchmarks provided below test the capacity of our ML models as well as their transferability across different chemical compositions and structures, including more complicated geometries like defect sites that are not present in the training data. However, our benchmarks mostly lack some system types and properties that play important roles in chemistry and materials science, including out-of-equilibrium geometries (with the exception of the barrier heights in GMTKN55), more complex materials like ternary and quaternary compounds and disordered phases, elastic and vibrational properties, etc. Transferability to these types of systems should not be assumed until future work provides more insight and further improvements to the ML models. For now, our benchmarks of CIDER functionals for predicting  bond lengths and lattice constants in Supplemental Material Section S6 provide some initial insight into the ability of CIDER functionals to accurately predict structural properties in addition to energetic properties.

\subsection{Accuracy of Different Model Types for Molecular and Solid-state Benchmarks} \label{sec:results_mtype}

We assess the accuracy of our exchange functionals by substituting them for exact exchange in hybrid functionals to create a ``surrogate hybrid'' functional. For example, one of the simplest and most popular hybrid functionals is PBE0~\cite{Adamo1999}, which mixes a 1/4 fraction of exact (Hartree-Fock, HF) exchange into the PBE functional~\cite{Perdew1996}:
\begin{equation}
    E_{\text{xc}}[n] = \frac{3}{4} E_\text{x}^\text{PBE}[n] + \frac{1}{4} E_\text{x}^\text{HF}[n] + E_\text{c}^\text{PBE}[n]. \label{eq:pbe0}
\end{equation}
In this section we perform so-called PBE0/CIDER calculations, where the exchange-correlation energy is
\begin{equation}
    E_{\text{xc}}[n] = \frac{3}{4} E_\text{x}^\text{PBE}[n] + \frac{1}{4} E_\text{x}^\text{CIDER}[n] + E_\text{c}^\text{PBE}[n]. \label{eq:pbe0cider}
\end{equation}
One can also generalize PBE0/CIDER with different fractions of CIDER exchange, giving the PBE0($\alpha$)/CIDER form, which is used in later sections:
\begin{equation}
    E_{\text{xc}}[n] = (1-\alpha) E_\text{x}^\text{PBE}[n] + \alpha E_\text{x}^\text{CIDER}[n] + E_\text{c}^\text{PBE}[n]. \label{eq:pbe_alpha_cider}
\end{equation}
Throughout the results, we will also use abbreviations like PBE0/SL-GGA and PBE0/NL-MGGA, which indicate that the specific CIDER functionals used in PBE0/CIDER are the SL-GGA and NL-MGGA, respectively.

All ML functional results in this section refer to the PBE0/CIDER surrogate hybrid form. Because our exchange functionals are trained on exact exchange energies, they are ``perfect'' if they exactly reproduce the predicted reaction energies of PBE0 on the test set. To measure the error of the exchange functionals, we calculate the average error between PBE0/CIDER and PBE0 for each functional type on GMTKN55 and SOL62. All calculations on GMTKN55 are SCF calculations, but the SOL62 cohesive energy CIDER and PBE0 calculations are performed non-self-consistently with PBE orbitals. See Appendices~\ref{sec:molcalc_settings} and \ref{sec:ssrefgpaw} for details.

The average deviations of the machine-learned functionals from PBE0, using the mean of means (MoM) metric (i.e. the mean of the mean absolute deviations, or MADs, for the sub-databases of GMTKN55), are presented in Fig.\ \ref{fig:pbe0_ref_1}. Results for all functionals include D4 dispersion corrections~\cite{Caldeweyher2019}, with the PBE0/CIDER functionals using the PBE0 dispersion parameters. The nonempirical PBE and r$^2$SCAN~\cite{Furness2020} functionals, while not designed to reproduce PBE0, are included in Fig.\ \ref{fig:pbe0_ref_1} to show typical deviations between GGAs, meta-GGAs, and hybrids.

\begin{figure}[btp]
    \includegraphics[width=\columnwidth]{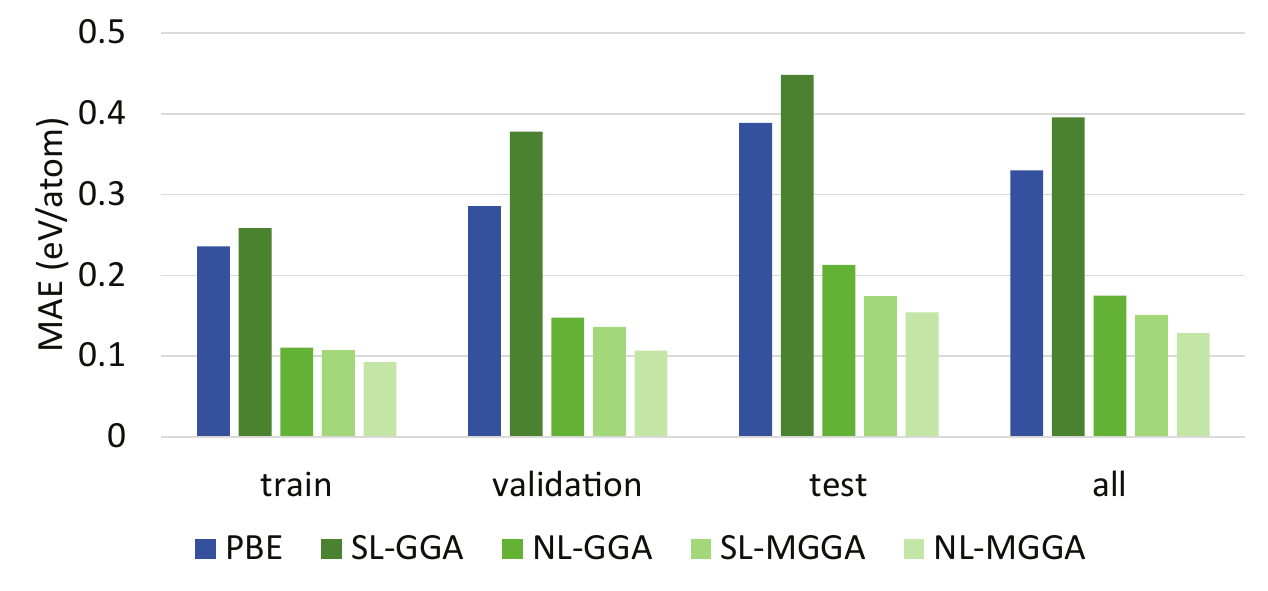}
    \caption{Average deviation of PBE and the PBE0/CIDER ML functionals from PBE0 on the SOL62 cohesive energies database.}
    \label{fig:sol62_nscf_tvt}
\end{figure}

While the trends vary slightly in the different subsets of GMTKN55, the key trend is that the quality of the fit to exact exchange improves going from SL-GGA to NL-GGA to SL-MGGA to NL-MGGA. Notably, the SL-GGA and NL-GGA suffer from out-of-distribution transferability issues, resulting in larger deviations from PBE0 even compared to the PBE functional on the test set. On the other hand, the ML meta-GGAs are more transferable outside the training data than the ML GGAs, so they match PBE0 more closely than PBE and r$^2$SCAN on the test set. The NL-MGGA is more accurate than the SL-MGGA, achieving an MoM of less than 1 kcal/mol on the test set. The fact that the NL-MGGA outperforms both the NL-GGA and SL-MGGA suggests that the kinetic energy and nonlocal features provide non-redundant information, meaning that nonlocal density features cannot replace the kinetic energy density nor vice versa. As shown in Fig.\ \ref{fig:sol62_nscf_tvt}, the same general trends apply for SOL62; the NL-MGGA is the most accurate and transferable ML functional for reproducing PBE0, followed by the SL-MGGA and then the NL-GGA.

Because GMTKN55 contains a variety of properties and systems, the performance of the functionals is somewhat dependent on the choice of error metric. However, the general trends are the same as for the MoM metric. See Supplemental Material Section S2A for a discussion of the impact of error metric on the results.

\subsection{Impact of Nonlocality and Derivative Discontinuity on Model Accuracy} \label{sec:results_1e2e}

\begin{figure}[btp]
    \centering
    \includegraphics[width=\columnwidth]{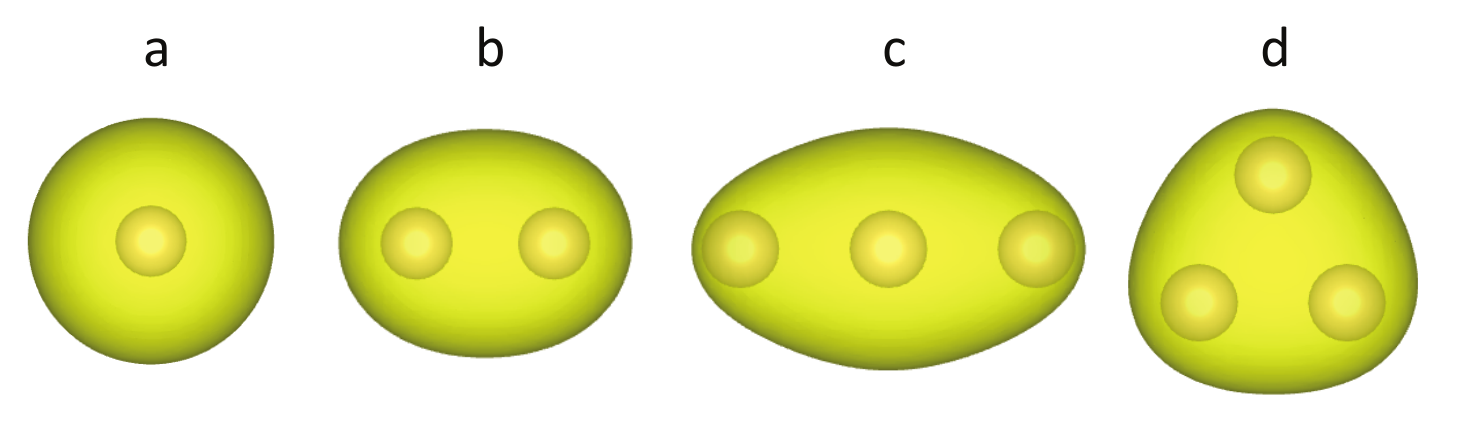}
    \caption{Geometries and density distributions of the 1-electron systems dataset. a) Hydrogen atom. b) \cf{H2+}. c) \cf{H3^{2+}}-linear. d) \cf{H3^{2+}}-triangular.}
    \label{fig:1e_densities}
\end{figure}

Overall, the SL-MGGA model is more accurate than the NL-GGA model across the databases tested here. However, the fact that NL-GGA improves over SL-GGA and that NL-MGGA improves over SL-MGGA suggests that the nonlocal features contribute useful, non-redundant information about the density for exchange energy prediction. This raises the question: What types of systems require the nonlocal features, and which require the semilocal orbital dependence of the kinetic energy density? We hypothesize that the derivative discontinuity of the exchange functional~\cite{Mori-Sanchez2014} plays a key role. Because only the kinetic energy density---and not the nonlocal features of the NL-GGA---includes a derivative discontinuity passing through integer electron number, the NL-GGA (as well as any smooth, pure density functional) cannot be expected to accurately capture energy differences involving a change in electron number, such as ionization energies and electron affinities. On the other hand, existing semilocal meta-GGAs are fundamentally limited by their locality and cannot be expected to improve over GGAs for energy differences between systems of like electron number. This is especially true of one-electron systems, for which the kinetic energy is simply determined by the density gradient.

To explore this hypothesis, we set up two simple sets of systems. The first set is one-electron systems for different arrangements of protons: the hydrogen atom, the \cf{H2+} molecular ion, the linear \cf{H3^2+} molecular ion, and the triangular \cf{H3^2+} molecular ion, all with bond lengths of 0.7 \AA, as illustrated in Fig.\ \ref{fig:1e_densities}. Because each system contains only one electron with different degrees of delocalization, we expect that only the functionals with nonlocal features can accurately predict energy differences between these systems. Table \ref{tab:1e2e_systems} shows the MAD between the ML exchange functionals and Hartree-Fock for the energy differences between the hydrogen atom and the molecular ions. As expected, the SL-GGA and SL-MGGA are both less accurate than NL-GGA and NL-MGGA. This confirms that the nonlocal features successfully characterize single-electron delocalization in a way that is not feasible with semilocal functionals.

\begin{table}[btp]
    \caption{Mean absolute deviation (kcal/mol) of the four types of ML functional for simple 1 and 2-electron systems compared to Hartree-Fock.}
    \begin{ruledtabular}
    \begin{tabular}{cdd}
        Model&
        \multicolumn{1}{c}{1e Systems\footnote{The 1-electron system dataset consists of the energies of \cf{H2+}, \cf{H3^{2+}}-linear, and \cf{H3^{2+}}-triangular referenced to the energy of the isolated hydrogen atom. All nearest-neighbor bond lengths are 0.7 \AA.}}&
        \multicolumn{1}{c}{2e Systems\footnote{The 2-electron system dataset consists of three energy differences between \cf{He+}, \cf{He}-singlet, and \cf{He}-triplet, where \cf{He}-triplet is the neutral helium atom excited state with both electrons having the same spin.}} \\
        \colrule
        SL-GGA  & 2.9 & 8.7 \\
        NL-GGA  & 2.5 & 5.5 \\
        SL-MGGA & 4.8 & 3.9 \\
        NL-MGGA & 1.1 & 2.7 \\
    \end{tabular}
    \end{ruledtabular}
    \label{tab:1e2e_systems}
\end{table}

The second set of systems consists of the \cf{He+} ion, the \cf{He} singlet (closed-shell ground state), and the \cf{He} triplet (open-shell excited spin state with two same-spin electrons). Because the exchange energy has a discontinuous derivative as the number of electrons of a given spin passes through an integer~\cite{Cohen2008,Mori-Sanchez2014}, the energy differences between \cf{He+} and \cf{He} triplet and between \cf{He} singlet and \cf{He} triplet can only be described accurately by functionals that capture this derivative discontinuity. As shown in Table \ref{tab:1e2e_systems}, the absolute deviations from Hartree Fock for the energy differences between these three systems are larger for SL-GGA and NL-GGA than for SL-MGGA and NL-MGGA. This indicates that the derivative discontinuity included in the kinetic energy density is essential for capturing energy differences related to electron addition/removal. Notably, the NL-MGGA is the most accurate of the four functional types for both datasets.

These observations regarding the role of the derivative discontinuity explain why a previous attempt at machine learning a GGA resulted in more accurate atomization energies but less accurate ionization potentials than PBE~\cite{Kasim2021}. It also provides a possible reason for the poor generalization of the SL-GGA and NL-GGA in Figure~\ref{fig:pbe0_ref_1}; training on properties like ionization potentials and electron affinities that are difficult to describe with orbital-free functionals could lead to over-fitting and decrease the accuracy of other properties. Likewise, the role of nonlocality explains why Kov\'acs \emph{et al.}~\cite{Kovacs2022} found that existing meta-GGAs, as well as 25 meta-GGAs they empirically trained, suffer from a trade-off between cohesive energy accuracy and band gap accuracy. Meta-GGAs are limited in their ability to describe both these properties at once due their semilocality. In addition, while we do not exhaustively try many possible NLDFs to conclusively show that they can only yield accurate models when paired with orbital-dependent features, the NLDFs we have implemented do need to be paired with a meta-GGA formalism to improve their accuracy. These findings provide evidence that to accurately and transferably fit a nonlocal, orbital-dependent quantity like the exchange energy, an ML functional should contain both orbital dependence and nonlocality in its feature set.

\subsection{Fine-tuning for Numerical Stability} \label{sec:results_finetune}

\begin{table}[tbp]
    \caption{The number of systems in the GMTKN55 database for which a given functional did not achieve SCF convergence with thresholds of $10^{-8}$ Eh for total energy and $10^{-4}$ Eh for the orbital gradient of the energy. All functionals use the PBE0/CIDER form (Eq.\ \ref{eq:pbe0cider}).}
    \begin{ruledtabular}
    \begin{tabular}{cc}
        Functional & No. Unconverged\\
        \colrule
        SL-GGA & 10\\
        NL-GGA & 16\\
        SL-MGGA & 5\\
        NL-MGGA & 3\\
        NL-MGGA-PBE & 1\\
        NL-MGGA-DTR & 0\\
    \end{tabular}
    \end{ruledtabular}
    \label{tab:conv_problems}
\end{table}

Because XC functionals are used within SCF calculations, they must be numerically stable and consistently achieve electronic convergence on real systems to be useful for practical applications. For the GMTKN55 dataset, we found that some systems do not quite converge to our default convergence thresholds, which were $10^{-8}$ Hartree atomic unit (Eh) for the total energy and $10^{-4}$ Eh for the orbital gradient of the energy. (See Appendix~\ref{sec:convergence_issues} for more details on how we remedied convergence problems.) We found that these calculations could only be converged with a $10^{-7}$ Eh energy threshold and $2\times 10^{-3}$ Eh gradient threshold. Interestingly, this problem is least severe for the NL-MGGA, as illustrated in Table \ref{tab:conv_problems}, which shows the number of systems for each functional type which did not converge to the $10^{-8}$ Eh energy and $10^{-4}$ Eh gradient thresholds. We believe this is because the increased model dimensionality of the NL-MGGA allows it to fit the exchange energy with a smoother function of the input features. 

\begin{figure}[tbp]
    \includegraphics[width=\columnwidth]{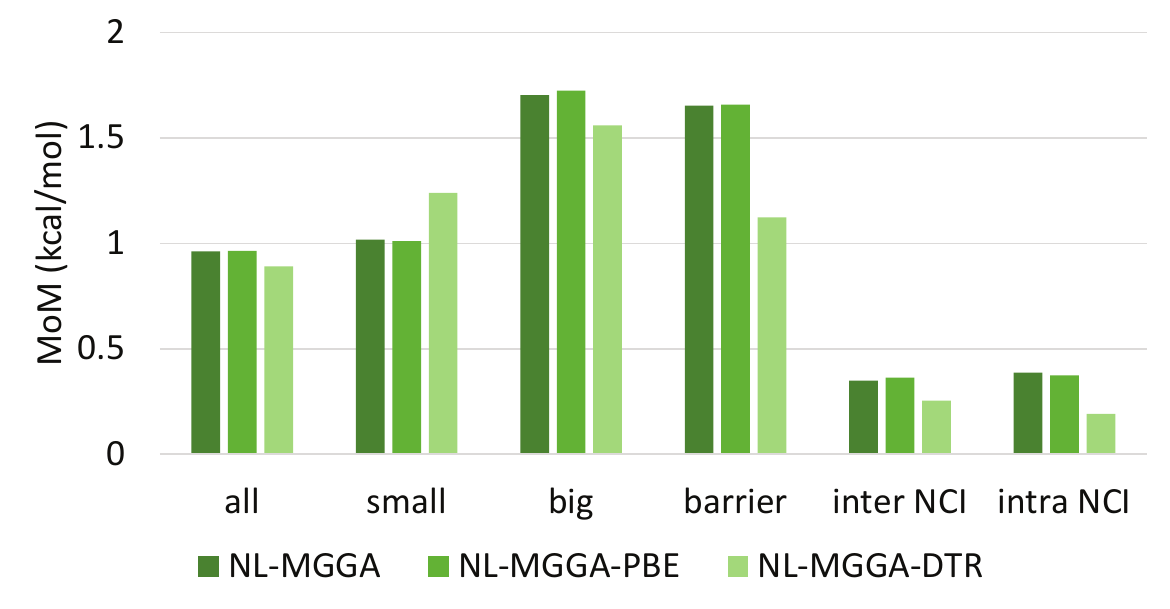}
    \caption{Accuracy of GMTKN55 predictions for the NL-MGGA and the NL-MGGA-PBE and NL-MGGA-DTR variants described in the text (all using the PBE0/CIDER form), compared to PBE0. Train/test partition results are not shown because the train and test data are different for the NL-MGGA-DTR functional.}
    \label{fig:finetune_mol}
\end{figure}

Because the NL-MGGA still sees some convergence problems, we decided to explore two modifications of the NL-MGGA for improved numerical stability. The first is the NL-MGGA-PBE, which has the same hyperparameters and feature shape as the NL-MGGA but uses the PBE functional as a baseline rather than the Chachiyo functional~\cite{Chachiyo2020}. NL-MGGA-PBE was screened during the validation procedure (Appendix~\ref{sec:hparam_sweep}) but was found to be slightly less accurate than the NL-MGGA with the Chachiyo baseline. However, only 1 system sees a convergence issue in Table \ref{tab:conv_problems} with NL-MGGA-PBE, compared to 3 with NL-MGGA. We also tried training an NL-MGGA model with the PBE baseline and a more chemically diverse subset of GMTKN55, as described in Appendix~\ref{sec:dtr_method}. We refer to this new train/validation/test partition as DTR (for ``diverse training''), and it yields the NL-MGGA-DTR, for which all GMTKN55 systems achieve an orbital gradient convergence of $10^{-4}$ Eh or lower. In fact, all but 10 of the 2462 GMTKN55 systems converge to within $10^{-9}$ Eh for energy and $10^{-4.5}$ Eh for orbital gradient with NL-MGGA-DTR, and the rest to within $10^{-8}$ and $10^{-4}$ Eh, respectively.

As shown in Fig.\ \ref{fig:finetune_mol}, which displays MoM errors of PBE0/CIDER versus PBE0, the accuracy of these three NL-MGGA variants is similar across the GMTKN55 dataset. The three functional variants also have nearly identical performance for the SOL62 cohesive energies, with MADs compared to PBE0 of 0.13 eV/atom for all three. Therefore, for the remainder of this work, we adopt the NL-MGGA-DTR model as our primary functional of choice for applications, and we name it CIDER23X-NL-MGGA-DTR. This functional is provided (along with several others used in this work) with an early release version of the CiderPress package as described in Section~\ref{sec:data_availability}, and it is the recommended functional for any researchers interested in running DFT calculations with CIDER.

In addition to the stability of a functional for the purpose of SCF convergence, one might also be interested in other metrics of numerical stability, such as the accuracy of the approximate evaluation of nonlocal density features described in Sections~\ref{sec:theory_cidermol} and~\ref{sec:theory_ciderpaw} and the convergence of the CIDER functional with respect to the size of the grid used to integrate the XC energy. These characteristics are discussed in the Supplemental Material Section S8.

\subsection{Obtaining Hybrid DFT Accuracy} \label{sec:results_hybrid}

\begin{figure}[tpb]
    \includegraphics[width=\columnwidth]{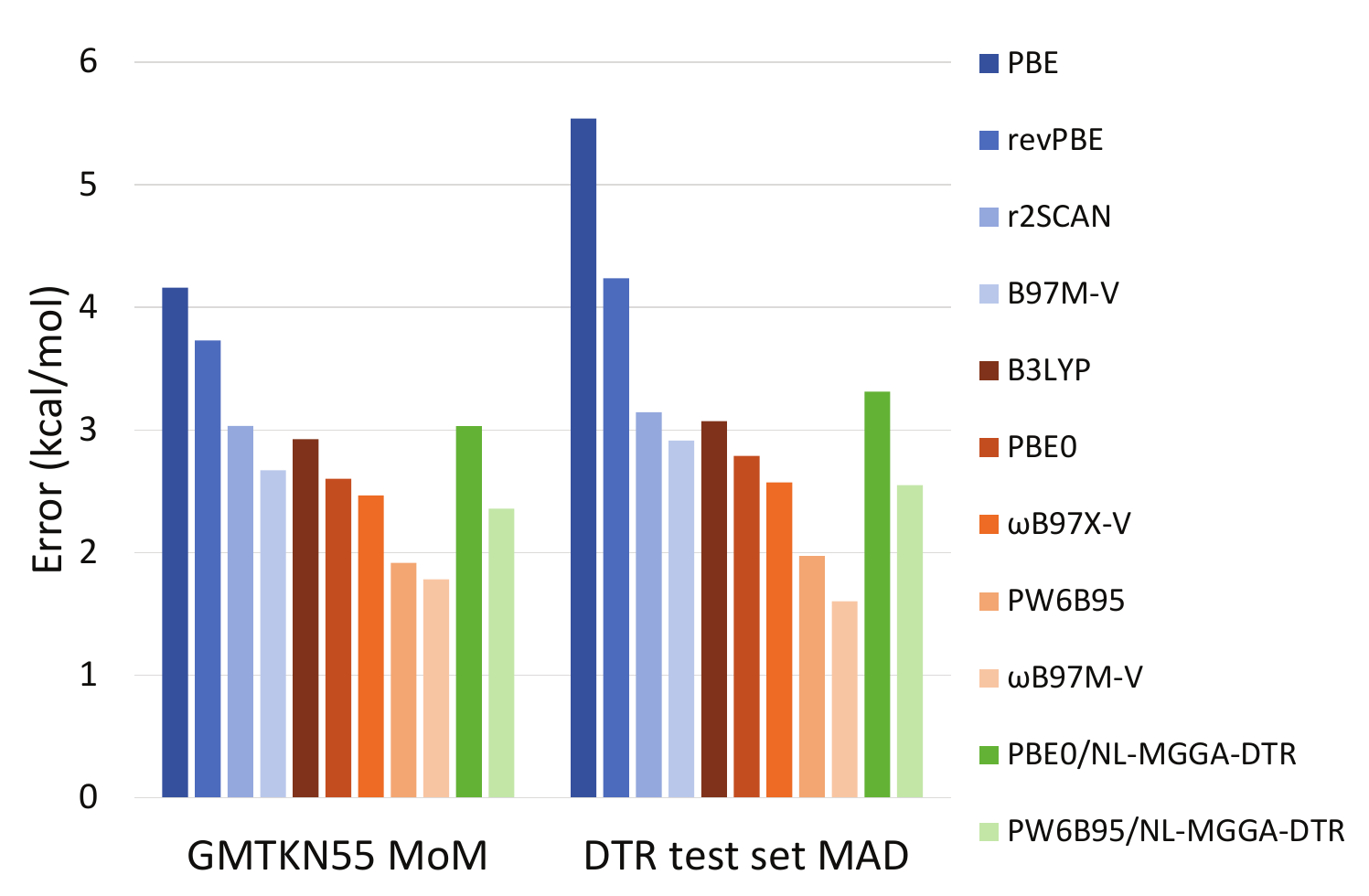}
    \caption{MoM error for all of GMTKN55 (left) and MAD on the DTR test set (right) relative to GMTKN55 reference values, for semilocal, hybrid, and CIDER functionals.}
    \label{fig:hybrid_acc}
\end{figure}
In this section, we show that by substituting NL-MGGA-DTR for exact exchange in hybrid functionals, accurate predictions of the XC energy can be obtained for molecular systems without additional parameter tuning. In particular, we look at PBE0/CIDER (Eq.\ \ref{eq:pbe0cider}) and another type of functional called PW6B95/CIDER. The original PW6B95 functional~\cite{Zhao2005} is
\begin{equation}
    E_{\text{xc}}^\text{PW6B95}[n] = 0.28 E_\text{x}^\text{HF}[n] + E_\text{xc,sl}^\text{PW6B95}[n],
\end{equation}
where $E_\text{xc,sl}^\text{PW6B95}$ is a meta-GGA term. Just like with PBE0/CIDER, we substitute CIDER in for Hartree-Fock exchange to obtain PW6B95/CIDER:
\begin{equation}
    E_{\text{xc}}^\text{PW6B95}[n] = 0.28 E_\text{x}^\text{CIDER}[n] + E_\text{xc,sl}^\text{PW6B95}[n].
\end{equation}
We chose to construct PW6B95/CIDER functionals because the original PW6B95~\cite{Zhao2005} is among the most accurate functionals on the GMTKN55 database~\cite{Goerigk2017,Kirkpatrick2021}.

For PBE0/NL-MGGA-DTR, PW6B95/NL-MGGA-DTR, and several semilocal~\cite{Perdew1996,Zhang1998,Furness2020,Mardirossian2015} and hybrid~\cite{Stephens1994,Adamo1999,Zhao2005,Mardirossian2014,Mardirossian2016} functionals, Fig.\ \ref{fig:hybrid_acc} shows the MoM error on all of GMTKN55 as well as the mean absolute error (MAE) on the test partition used for NL-MGGA-DTR. The errors are relative to the high-accuracy benchmark data collected in the GMTKN55 database~\cite{Goerigk2017}. All functionals include D4 dispersion corrections~\cite{Caldeweyher2019} except for B97M-V~\cite{Mardirossian2015}, $\omega$B97X-V~\cite{Mardirossian2014}, and $\omega$B97M-V~\cite{Mardirossian2016}, which all use VV10 dispersion~\cite{Vydrov2010}. PBE0/CIDER calculations use PBE0 dispersion parameters, and PW6B95/CIDER calculations use PW6B95 dispersion parameters. We show conventional semilocal functionals in shades of blue, conventional hybrid functionals in orange, and CIDER functionals in green.

As Fig.\ \ref{fig:hybrid_acc} shows, the PBE0/NL-MGGA-DTR functional is comparable in accuracy to r$^2$SCAN, but the combination of systematic errors in the NL-MGGA-DTR functional and in PBE0 itself prevent it from outperforming the most accurate meta-GGA (B97M-V) or any of the hybrid functionals. On the other hand, because PW6B95 is much more accurate than PBE0 on the GMTKN55 dataset, PW6B95/NL-MGGA-DTR is more accurate than all semilocal functionals explored here. It is also more accurate than the PBE0 and B3LYP~\cite{Stephens1994} hybrid functionals and equivalent in accuracy to the range-separated hybrid $\omega$B97X-V~\cite{Mardirossian2014}. This shows that the NL-MGGA-DTR exchange functional can be used to calculate molecular reaction energies with an accuracy that was previously not feasible without exact exchange mixing.

As with Section \ref{sec:results_mtype}, the analysis of the errors on GMTKN55 depends somewhat on the error metric used. These nuances are discussed in further detail in Supplemental Material Section S2B, which also provides a comparison between the nonlocal and semilocal meta-GGA models.

\subsection{Using CIDER for Improved Band Gap Prediction} \label{sec:results_bandgap}

\begin{table*}[tbp]
\caption{Error statistics for 255 band gaps from the database published by Borlido \emph{et al.}~\cite{Borlido2019}, in eV. The ML functionals use the PBE0($\alpha$)/CIDER formalism of Eq.\ \ref{eq:pbe_alpha_cider}.}
\begin{ruledtabular}
\begin{tabular}{cddddddd}
{}&
\multicolumn{1}{c}{PBE\footnote{This work.}}&
\multicolumn{1}{c}{r$^2$SCAN\footnotemark[1]}&
\multicolumn{1}{c}{SL-MGGA\footnotemark[1]}&
\multicolumn{2}{c}{NL-MGGA-DTR\footnotemark[1]}&
\multicolumn{1}{c}{HSE06\footnote{Borlido \emph{et al.}~\cite{Borlido2019}}}&
\multicolumn{1}{c}{PBE0\footnotemark[2]} \\
{} &  \centering {} &  {} &  \multicolumn{1}{c}{\textrm{$\alpha=0.25$}} &  \multicolumn{1}{c}{\textrm{$\alpha=0.25$}} &  \multicolumn{1}{c}{\textrm{$\alpha=0.35$}} &  {} &  {} \\
\colrule
ME\footnote{Mean error} (eV)   & -1.14 &   -0.81 &    -0.74 &        -0.61 &         -0.38 &  -0.17 &  0.40 \\
MAE\footnote{Mean absolute error} (eV)  &  1.16 &    0.87 &     0.82 &         0.71 &          0.58 &   0.53 &  0.73 \\
RMSE\footnote{Root mean square error} (eV) &  1.50 &    1.18 &     1.10 &         1.02 &          0.86 &   0.80 &  0.90 \\
MARE\footnote{Mean absolute relative error}      &  0.53 &    0.41 &     0.41 &         0.35 &          0.33 &   0.33 &  0.67 \\
\end{tabular}
\end{ruledtabular}
\label{tab:band_gaps}
\end{table*}

The band gap problem~\cite{Sham1983,Perdew1986}, in which semilocal DFT tends to drastically underestimate the band gaps of solids, stands in the way of a variety of important applications like high-throughput screening for optical and electronic properties and semiconductor point defect physics~\cite{Lany2008,Freysoldt2014,Schultz2006}. By introducing a derivative discontinuity into the XC functional, meta-GGAs partly mitigate this problem~\cite{Perdew2017}, but they still tend to significantly underestimate band gaps~\cite{Borlido2020}. Exceptions include mBJ~\cite{Tran2009} and TASK~\cite{Aschebrock2019}, which each have important drawbacks. The mBJ functional provides the exchange potential only, not the energy~\cite{Tran2009}, and TASK does not predict atomization energies and lattice constants as well as SCAN~\cite{Aschebrock2019,Kovacs2022,Lebeda2022}. This reinforces the discussion in Section~\ref{sec:results_1e2e} regarding the accuracy trade-offs meta-GGAs must make due to their semilocality.

By introducing nonlocality into the exchange energy, PBE0/NL-MGGA-DTR (Eq.\ \ref{eq:pbe0cider} or Eq.\ \ref{eq:pbe_alpha_cider} with $\alpha=0.25$) improves band gap predictions over PBE and r$^2$SCAN, as shown in Table \ref{tab:band_gaps}. This improvement occurs even though the CIDER functionals are not fit to solid-state band gaps. The PBE0/SL-MGGA also improves band gap predictions over r$^2$SCAN, but significantly less so than PBE0/NL-MGGA-DTR. The database used for testing covers 255 of the 472 band gaps in the database published by Borlido \emph{et al.}~\cite{Borlido2019}, with La, Yb, and Th-containing systems being excluded as well as systems with a greater than $0.05$ eV difference between the spin-orbit-corrected and non-spin-orbit-corrected band gaps (as computed with PBE by Borlido \emph{et al.}~\cite{Borlido2019}). See Appendix~\ref{sec:band_gap_methods} for details, and see Supplemental Material Section S3 for error metrics for all 453 solids in the database that do not contain La, Yb, or Th.

\begin{figure*}[tbp]
    \subfloat[\label{fig:pbe_si_bs}]{\includegraphics[width=0.32\textwidth]{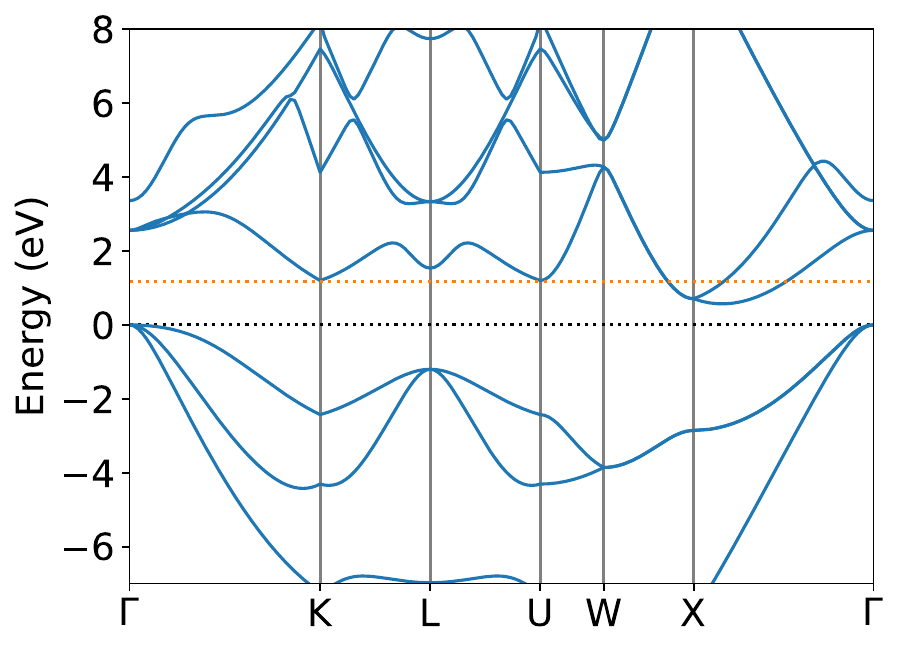}}
    \subfloat[\label{fig:r2scan_si_bs}]{\includegraphics[width=0.32\textwidth]{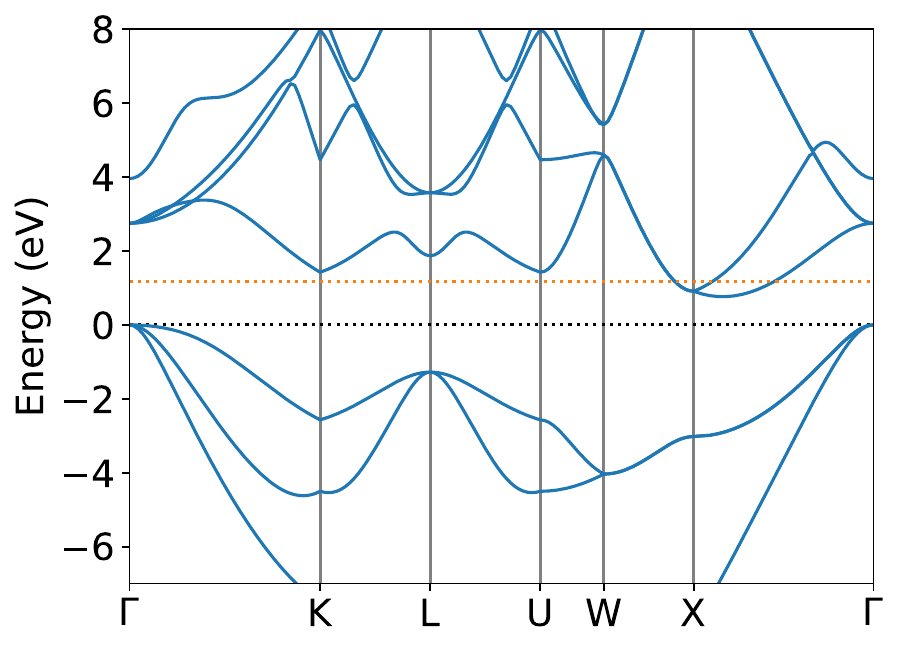}}
    \subfloat[\label{fig:cider_si_bs}]{\includegraphics[width=0.32\textwidth]{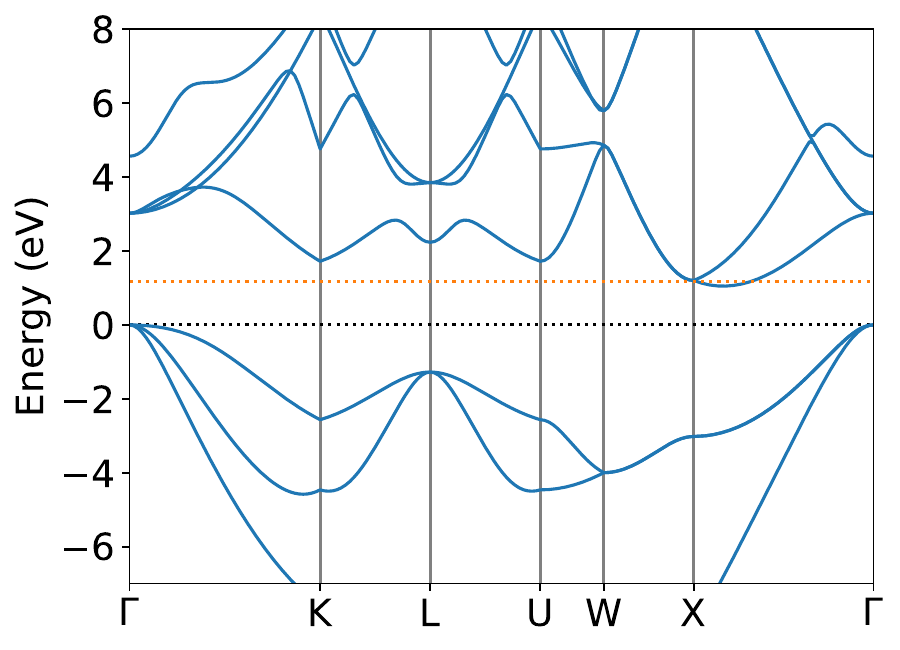}}
    \caption{Band structures of silicon with three different functionals. The black dotted line is positioned at the valence band maximum, and the orange dotted line is located at the experimental band gap of 1.17 eV. (a) PBE. (b) r$^2$SCAN. (c) PBE0/NL-MGGA-DTR.}
    \label{fig:band_structures1}
\end{figure*}

\begin{figure*}[tbp]
    \subfloat[\label{fig:pbe_bp_bs}]{\includegraphics[width=0.32\textwidth]{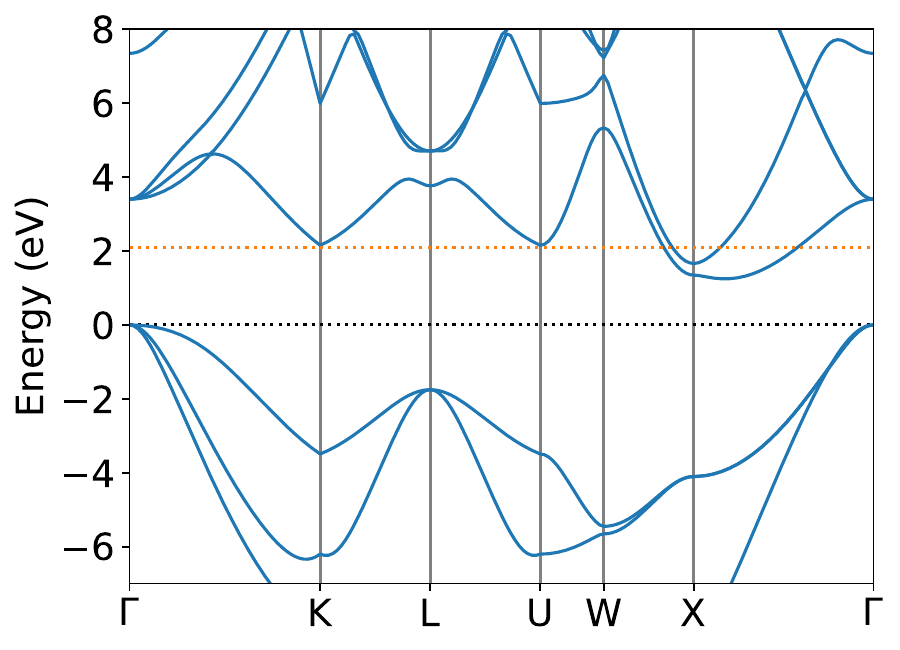}}
    \subfloat[\label{fig:r2scan_bp_bs}]{\includegraphics[width=0.32\textwidth]{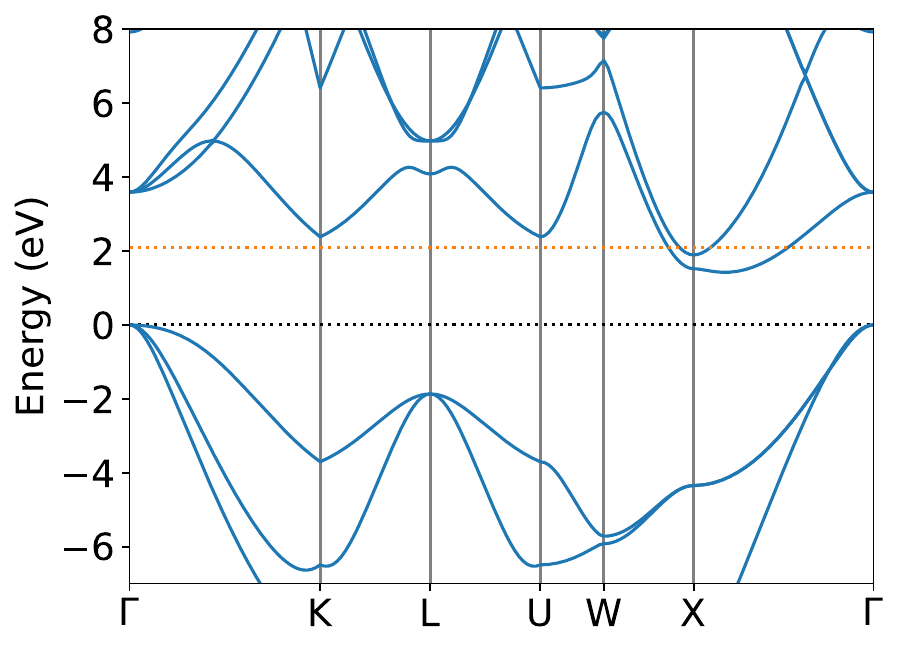}}
    \subfloat[\label{fig:cider_bp_bs}]{\includegraphics[width=0.32\textwidth]{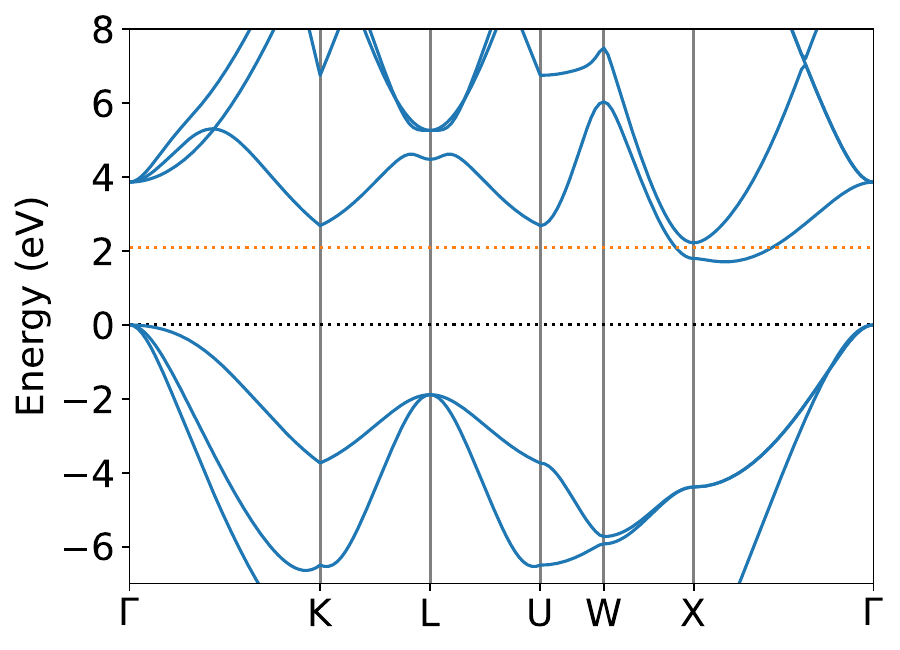}}
    \caption{Band structures of boron phosphide with three different functionals. The black dotted line is positioned at the valence band maximum, and the orange dotted line is located at the experimental band gap of 2.10 eV. (a) PBE. (b) r$^2$SCAN. (c) PBE0/NL-MGGA-DTR.}
    \label{fig:band_structures2}
\end{figure*}

To provide a clearer picture of the improved band gap prediction provided by the NL-MGGA-DTR functional, Table \ref{tab:bgsubset} shows the band gap predictions for a selected subset of systems that are commonly studied~\cite{Yang2016}. In addition, Figs.\ \ref{fig:band_structures1} and \ref{fig:band_structures2} contain band structures for silicon (Si) and boron phosphide (BP), respectively, which show that PBE0/NL-MGGA-DTR provides similar band structures to PBE and r$^2$SCAN but with a wider band gap.

\begin{table}[tbp]
\caption{Band gaps for a small subset of solids (in eV). CIDER refers to PBE0/NL-MGGA-DTR.}
\begin{ruledtabular}
\begin{tabular}{cddddd}
{} &   \multicolumn{1}{c}{PBE} &  \multicolumn{1}{c}{r$^2$SCAN} &  \multicolumn{1}{c}{CIDER} &  \multicolumn{1}{c}{HSE06} &  \multicolumn{1}{c}{Expt.~\cite{Borlido2019}} \\
\colrule
Si        &  0.57 &    0.76 &         1.05 &   1.15 &          1.17 \\
Ge        &  0.12 &    0.48 &         0.71 &   0.81 &          0.74 \\
InP       &  0.69 &    1.12 &         1.47 &   1.52 &          1.42 \\
GaAs      &  0.59 &    1.08 &         1.36 &   1.43 &          1.52 \\
CdSe      &  0.71 &    1.19 &         1.48 &   1.69 &          1.74 \\
BP        &  1.25 &    1.42 &         1.71 &   1.98 &          2.10 \\
GaP       &  1.61 &    1.88 &         2.09 &   2.29 &          2.35 \\
CdS       &  1.20 &    1.65 &         2.02 &   2.27 &          2.48 \\
GaN       &  1.88 &    2.33 &         2.65 &   3.18 &          3.50 \\
ZnS       &  2.13 &    2.69 &         3.04 &   3.37 &          3.72 \\
C         &  4.13 &    4.34 &         4.70 &   5.33 &          5.50 \\
BN        &  4.21 &    4.79 &         5.08 &   5.63 &          5.96 \\
CaO       &  3.68 &    4.23 &         4.53 &   5.32 &          6.88 \\
MgO       &  4.73 &    5.63 &         6.11 &   6.42 &          7.67 \\
NaCl      &  5.11 &    5.90 &         6.32 &   6.45 &          8.75 \\
LiF       &  9.07 &   10.05 &        10.40 &  11.39 &         13.60 \\
Ar        &  8.71 &    9.61 &         9.71 &  10.36 &         14.15 \\
\colrule
ME   & -1.93 &   -1.42 &        -1.11 &  -0.74 &          \multicolumn{1}{c}{-} \\
MAE  &  1.93 &    1.42 &         1.11 &   0.76 &          \multicolumn{1}{c}{-} \\
RMSE &  2.41 &    1.88 &         1.66 &   1.31 &          \multicolumn{1}{c}{-} \\
MARE      &  0.45 &    0.29 &         0.18 &   0.11 &          \multicolumn{1}{c}{-} \\
\end{tabular}
\end{ruledtabular}
\label{tab:bgsubset}
\end{table}


There is room for improvement in these band gaps, since PBE0/NL-MGGA-DTR still systematically underestimates band gaps compared to the state-of-the-art HSE06~\cite{Heyd2003,Krukau2006} range-separated hybrid functional, as well as the PBE0 functional to which PBE0/CIDER is an approximation. One possible solution is to tune the fraction of CIDER exchange, which is already done with exact exchange within hybrid DFT to optimize functional accuracy for different applications~\cite{Moussa2012}. As shown in Table \ref{tab:band_gaps}, PBE0($\alpha$)/NL-MGGA-DTR with $\alpha=0.35$ has increased band gaps over $\alpha=0.25$ and similar error statistics to HSE06. However, one drawback of this approach is that because NL-MGGA-DTR increases band gaps less than exact exchange for a given mixing fraction, one might need a large fraction of NL-MGGA-DTR to fit band gaps. This large fraction might not provide accurate predictions of other material properties. Alternative approaches to improve CIDER band gaps include using physical intuition and exact constraints to tune the derivative discontinuity contribution from the kinetic energy density (as is done in the TASK functional~\cite{Aschebrock2019}), introducing more nonlocal features to improve the accuracy of the functional in general, and training on fundamental band gaps (i.e. differences between ionization potentials and electron affinities). However, considering the drastic increase in computational cost associated with using HSE06 or PBE0 compared to semilocal DFT, the ability to mitigate the underestimation of band gaps without explicitly fitting them---and while accurately predicting other properties like bond energies---already serves as a major step toward solving the band gap problem.

\subsection{Benchmarking Performance on Large Condensed Matter Systems} \label{sec:results_speed}

\begin{figure*}[tbp]
    \subfloat[\label{fig:perf_bulk}]{\includegraphics[width=0.98\columnwidth]{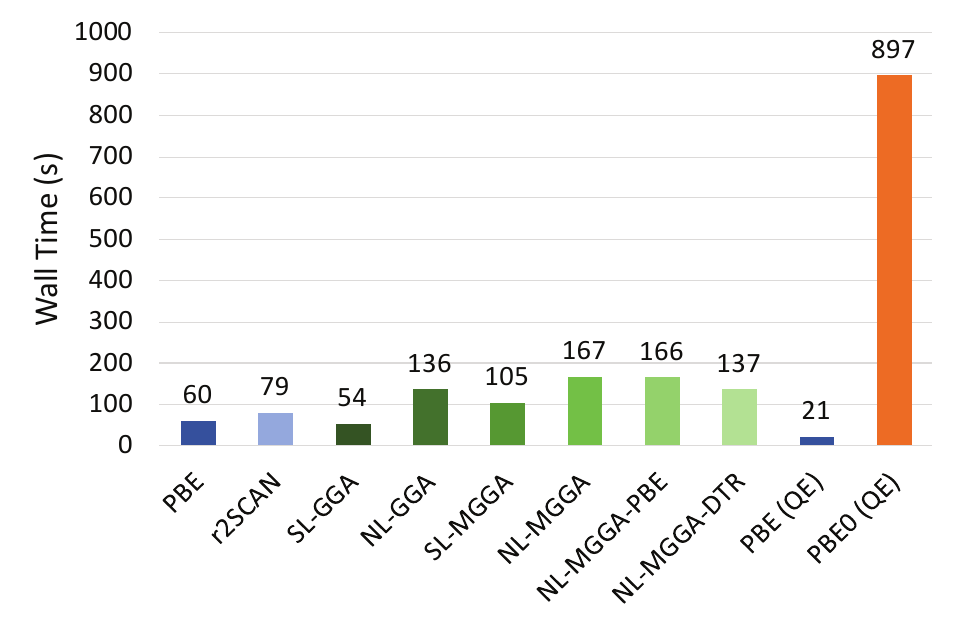}}\qquad
    \subfloat[\label{fig:perf_vac}]{\includegraphics[width=0.98\columnwidth]{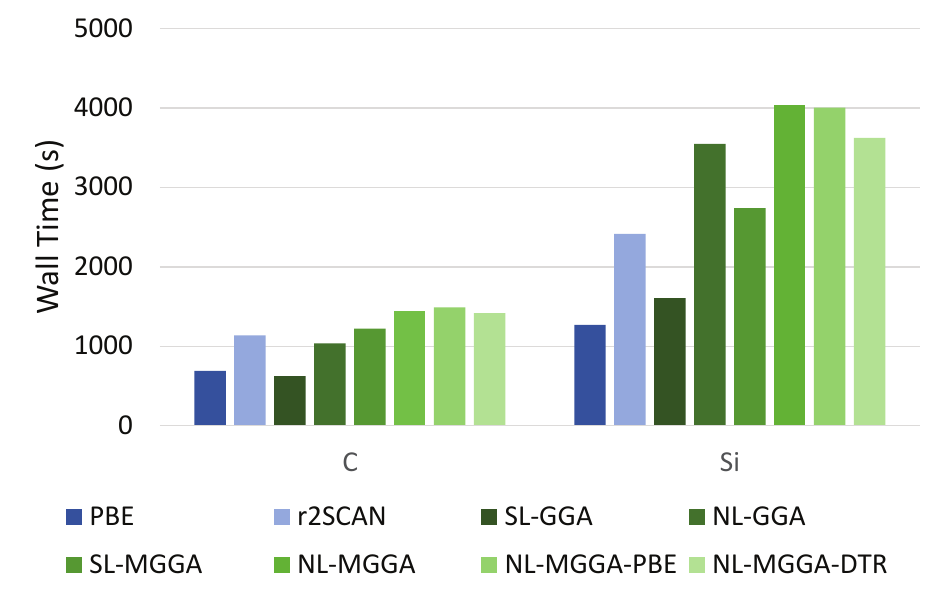}}
    \caption{Computational cost for plane-wave DFT SCF calculations of large systems with different functionals. The semilocal (blue) and CIDER (green) calculations were performed with GPAW on one 64-core Intel Ice Lake node for diamond and two 64-core nodes for silicon. The PBE0 calculation (orange) was performed on one 64-core node with the Quantum ESPRESSO (QE) code~\cite{QE-2009,QE-2017}. An additional PBE calculation was also performed with QE to compare the computational cost of GPAW and QE. (a) Total SCF calculation wall time for diamond 216-atom bulk supercell with $\Gamma$-point sampling only. (b) Total SCF calculation wall times for diamond and silicon neutral vacancy formation energies in a 216-atom supercell with $2\times 2\times 2$ Monkhorst-Pack k-point mesh~\cite{Monkhorst1976}.}
    \label{fig:performance216}
\end{figure*}

\begin{table}[tbp]
    \caption{The formation energies of the neutral vacancies in diamond and silicon (in eV), computed with 216-atom supercells and a $2 \times 2 \times 2$ Monkhorst-Pack k-point mesh~\cite{Monkhorst1976}. All CIDER functionals use the PBE0/CIDER surrogate hybrid form.}
    \begin{ruledtabular}
    \begin{tabular}{cdd}
        \textrm{Functional}&
        \multicolumn{1}{c}{C}&
        \multicolumn{1}{c}{Si} \\
        \colrule
        PBE & 6.47 & 3.61 \\
        r$^2$SCAN & 6.36 & 4.19 \\
        \colrule
        SL-GGA & 6.11 & 3.48 \\
        NL-GGA & 6.73 & 3.50 \\
        SL-MGGA & 6.66 & 4.16 \\
        NL-MGGA & 6.77 & 4.58 \\
        NL-MGGA-PBE & 6.77 & 4.58 \\
        NL-MGGA-DTR & 6.79 & 4.42 \\
        \colrule
        HSE06 & 6.96 & 4.54 \\
        PBE0 & 6.97 & 4.65 \\
    \end{tabular}
    \end{ruledtabular}
    \label{tab:neutral_vac_formen}
\end{table}

The key benefit of the nonlocal features used to learn the exchange functional is that the quasi-linear-scaling algorithm used to compute them within plane-wave DFT (Sections \ref{sec:rps} and \ref{sec:theory_ciderpaw}) is much faster than the evaluation of exact exchange for large systems and does not depend on the number of k-points. To illustrate this efficiency, Fig.\ \ref{fig:perf_bulk} shows the wall time to perform a ground-state SCF calculation for a 216-atom diamond supercell with semilocal and PBE0/CIDER functionals in GPAW~\cite{Mortensen2005,Enkovaara2010}. These wall times are compared to those of the hybrid DFT implementation in Quantum ESPRESSO~\cite{QE-2009,QE-2017}, which uses the adaptively compressed exchange (ACE) algorithm~\cite{Lin2016} to compute the exact exchange energy as efficiently as possible. A $\Gamma$-point only k-point mesh was used for these calculations. The CIDER functionals, including those with nonlocal features, all take less than 170 seconds (compared to 60 seconds for PBE and 79 seconds for r$^2$SCAN). The total wall time for PBE0 is roughly 6.5 times slower than the NL-MGGA-DTR functional, in spite of the use of the ACE algorithm to accelerate the exact exchange calculation. Also, Quantum ESPRESSO has a general performance advantage over the plane-wave version of GPAW, with the 216-atom PBE calculation in Fig.~\ref{fig:perf_bulk} requiring 60 seconds for GPAW and only 21 seconds for Quantum ESPRESSO. Therefore, more performance optimization in both the GPAW code and the CIDER feature evaluation might further increase the computational advantage of CIDER over hybrid DFT.

The efficiency of CIDER functionals for large systems makes it practical to perform more complicated calculations like defect formation energies. Figure \ref{fig:perf_vac} shows the wall time for computing the neutral vacancy formation energies in diamond and silicon using 216-atom supercells for semilocal and PBE0/CIDER functionals. These calculations used a $2\times 2\times 2$ Monkhorst-Pack k-point mesh~\cite{Monkhorst1976}. All of the functionals are of comparable computational cost, with the NL-MGGA-DTR model being only about 25\% slower than r$^2$SCAN for diamond and 50\% slower for silicon. This small decrease in efficiency comes with a significant benefit: As shown in Table \ref{tab:neutral_vac_formen}, of the functionals tested, the NL-MGGA model and its variants give the closest match to PBE0 for predicting these vacancy formation energies, with PBE0/NL-MGGA-DTR giving deviations of 0.18 eV and 0.23 eV for diamond and silicon, respectively. The HSE06 and PBE0 formation energies were computed using VASP~\cite{Kresse1993,Kresse1996,Kresse1996a}. See Appendix~\ref{sec:ss_performance} for further calculation details.

\subsection{Charged Defect Transition Levels in Silicon} \label{sec:results_defects}

\begin{figure*}[tbp]
    \subfloat[\label{fig:si_trl_fr}]{\includegraphics[width=1.7\columnwidth]{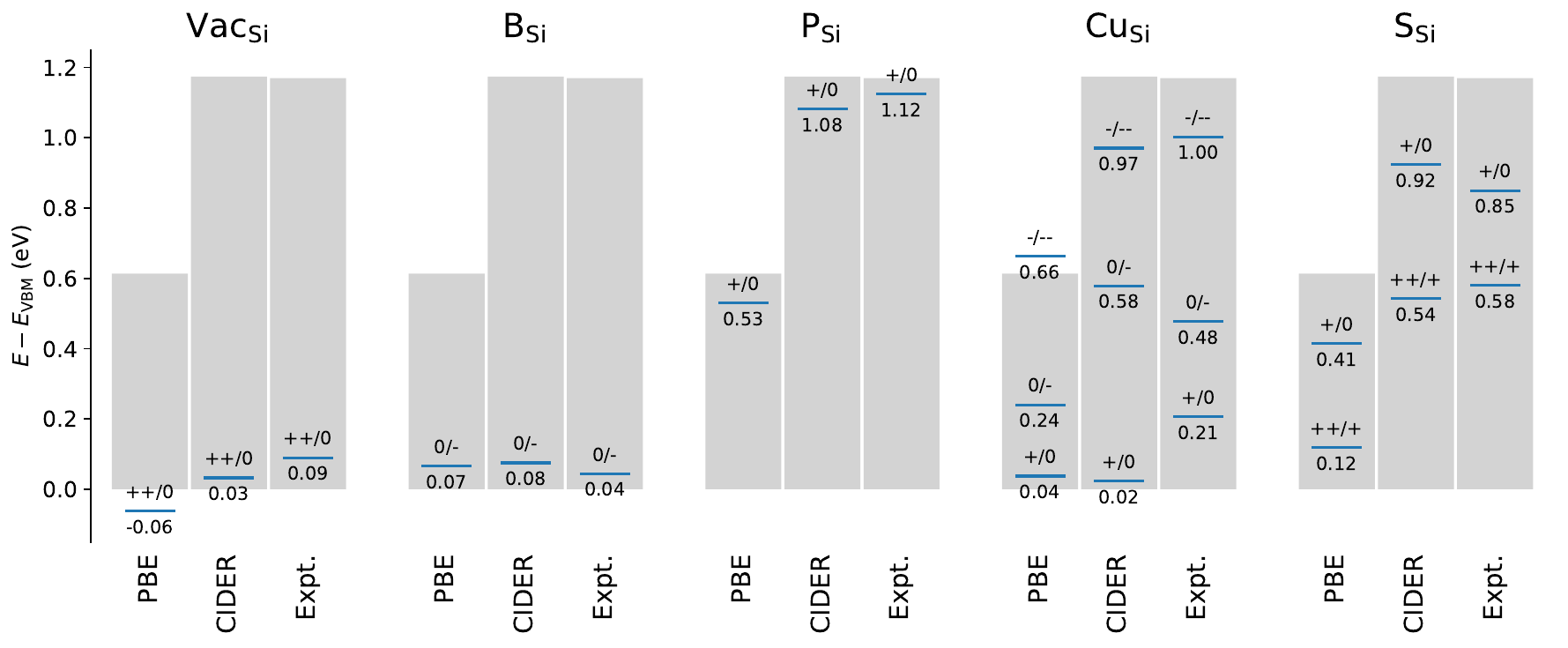}}\\
    \subfloat[\label{fig:si_trl_ls}]{\includegraphics[width=1.7\columnwidth]{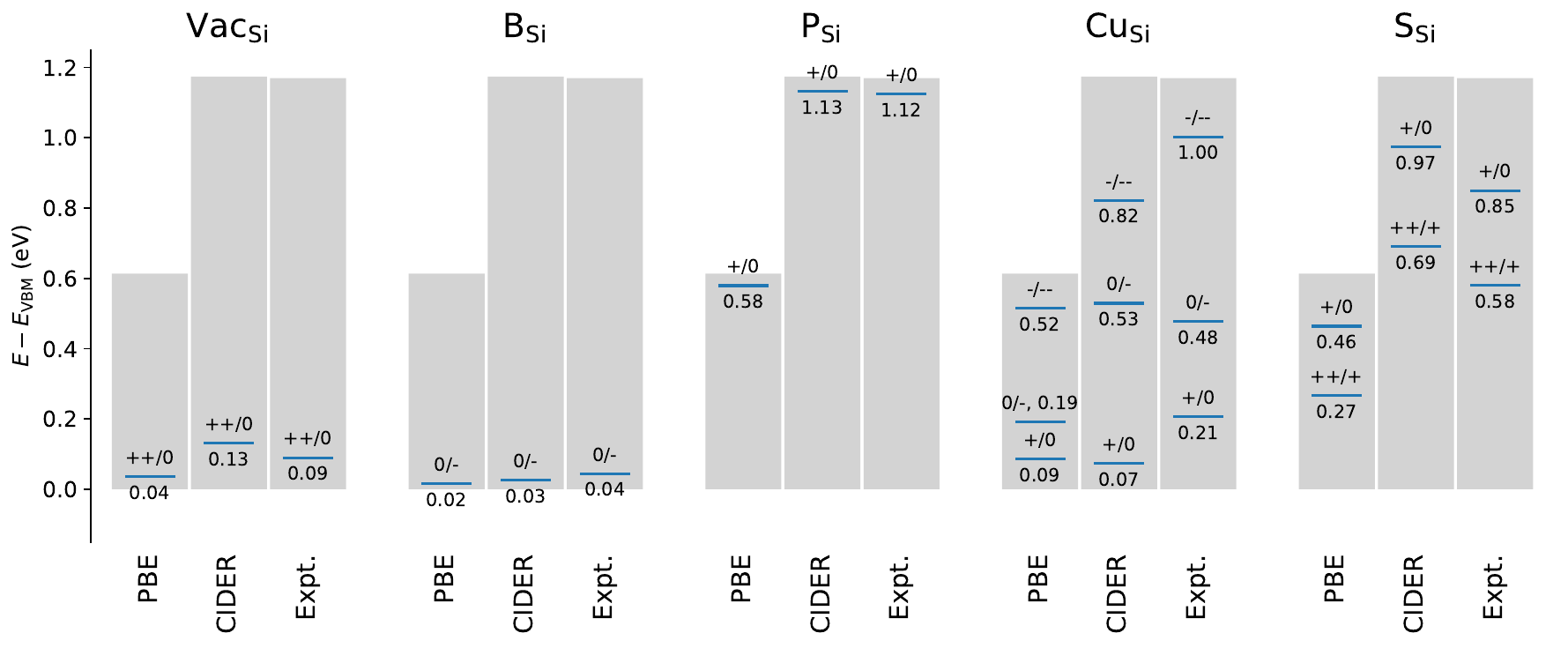}}
    \caption{Electronic transitions levels for the single vacancy and several substitutional defects in silicon computed using different functionals and finite-size correction schemes. The grey bars represent the band gap for the given functional (or experiment). The experimental references are from the following sources: vacancy~\cite{Watkins1980}, boron~\cite{Fischer1983}, phosphorus~\cite{Jagannath1981}, copper~\cite{PhysRevB.65.165203}, and sulfur~\cite{Schulz2002,Deak2010}. CIDER refers to PBE0(0.3)/NL-MGGA-DTR. (a) Transition levels computed using Freysoldt finite-size corrections~\cite{Freysoldt2009}. (b) Transition levels computed using potential alignment correction only.}
    \label{fig:defect_trl_plots}
\end{figure*}

\begin{figure*}[tbp]
    \includegraphics[width=2\columnwidth]{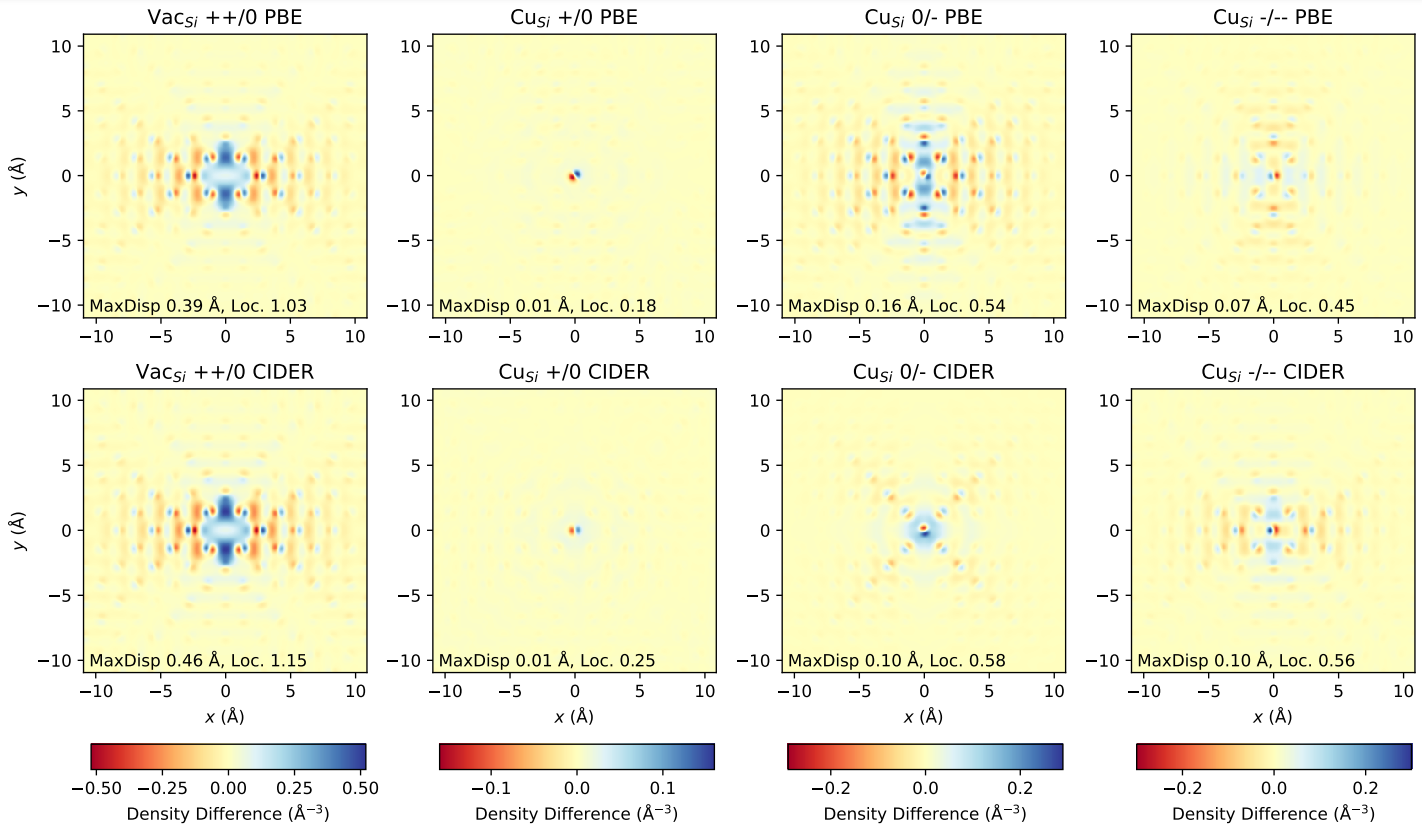}
    \caption{Change in charge distribution in the $(001)$ plane (integrated over the $z$ axis) for several charge transition levels in point defects of silicon. Blue shades represent an increase in electron density, while red shades indicate a decrease in electron density. ``MaxDisp'' is the maximum ionic displacement between the charge states, and ``Loc.'' is the fraction of the charge difference contained within a 5 \AA\ sphere around the defect site. Loc. is greater than 1 for the vacancy ++/0 transition because more than 2 electrons are drawn in to the defect region upon addition of an electron pair, depleting electron density in the surrounding region.}
    \label{fig:chg_dist}
\end{figure*}

One of the consequences of the band gap problem is that transition levels in charged defects are not accurately described by semilocal DFT, so hybrid DFT is typically used when accurate transition level predictions are needed. In this section, we show that by improving band gap prediction, CIDER is able to accurately predict defect transition levels in silicon. The transition levels associated with the single vacancy~\cite{Watkins1980}, as well as the boron~\cite{Fischer1983}, phosphorus~\cite{Jagannath1981}, copper~\cite{PhysRevB.65.165203}, and sulfur~\cite{Schulz2002,Deak2010} substitutionals, were computed as described in Appendix~\ref{sec:charged_defects}, using a 512-atom supercell, $2\times 2\times 2$ $\Gamma$-centered k-point mesh, and full structural relaxation for each defect and functional.

Figure \ref{fig:defect_trl_plots} shows the computed transition levels compared to experiment with two different finite-size correction schemes: the Freysoldt finite-size corrections~\cite{Freysoldt2009} (Fig.\ \ref{fig:si_trl_fr}) and the potential alignment correction (Fig.\ \ref{fig:si_trl_ls}). We computed the transition levels with PBE and PBE0(0.3)/NL-MGGA-DTR (Eq.\ \ref{eq:pbe_alpha_cider} with $\alpha=0.3$). The latter functional was chosen because its band gap matches the 1.17 eV experimental gap; however, PBE0/NL-MGGA-DTR yields similar results (see Supplemental Material Section S7). Supplemental Table S13 provides a more detailed list of all computed transition levels. In the rest of this section, PBE0(0.3)/NL-MGGA-DTR is referred to simply as CIDER for brevity.

PBE and CIDER both give reasonable descriptions of the ++/0 transition level in the silicon vacancy and the boron acceptor level. However, the phosphorus donor level is not well-described by PBE due to its severe underestimation of the band gap; CIDER remedies the band gap underestimation and therefore predicts a transition level in good agreement with experiment. To correct for band gap underestimation, one could predict defect levels from PBE by scaling the gap (and with it, the transition levels) to the experimental gap, but this makes it difficult to distinguish true shallow levels from deep levels that happen to sit near the (too-low) PBE conduction band edge~\cite{Lany2008}. For example, PBE predicts a transition level of 0.66 eV for the -/{-}{-} transition of $\text{Cu}_\text{Si}$ (Fig.\ \ref{fig:si_trl_fr}), 0.05 eV above the conduction band minimum. Without further information, it is unclear whether this state is resonant in the conduction band or localized in the gap. CIDER, on the other hand, places the -/{-}{-} level inside the gap at 0.97 eV, in good agreement with the 1.00 eV level from experiment.

More generally, Fig.\ \ref{fig:si_trl_fr} shows that paired with the Freysoldt correction, CIDER matches each level in the figure with an error of 0.10 eV or less compared to experiment, with the exception of the +/0 level in $\text{Cu}_\text{Si}$.
For this transition, the experimental level is 0.20 eV~\cite{PhysRevB.65.165203}, while the CIDER level is 0.02 eV with the Freysoldt correction and 0.07 eV with the potential alignment correction only. The HSE06 level was previously reported to be 0.21 eV by Sharan \emph{et al.}~\cite{Sharan2017}, in good agreement with experiment. However, Sharan \emph{et al.} used a 64-atom supercell with a $2 \times 2 \times 2$ Monkhorst-Pack k-point mesh~\cite{Monkhorst1976}. This supercell size is relatively small, and the k-point mesh is both small and fails to sample the band edges, which can interfere with the description of relatively delocalized defect states. When we recalculated the CIDER transition level using Sharan \emph{et al.}'s supercell size and k-point mesh, we obtained 0.18 eV with Freysoldt corrections, in good agreement with their result and with experiment. In the Supplemental Material (Section S7A), we provide further evidence that the HSE06 level itself is not converged with respect to k-point mesh. This finding illustrates the importance of using well-converged supercell sizes and k-point meshes in point defect calculations, a requirement that is made much easier by the efficiency of the CIDER functionals compared to hybrid DFT. Why this particular transition level is underestimated by both hybrid DFT and CIDER is an interesting question warranting future investigation.

In addition to yielding more accurate transition levels, CIDER predicts somewhat different charge distributions than PBE as well. Figure \ref{fig:chg_dist} shows the charge density differences between charge states for several of the transition levels in Fig.\ \ref{fig:defect_trl_plots}. There are a few notable takeaways from these plots. First, as expected, the charge density differences are larger for transitions that result in large ionic displacements (as indicated by the MaxDisp in the bottom of each plot). Second, these ionic displacements and resulting charge distortions are noticeable more than 5 \AA\ away from the defect site for some transitions, indicating the importance of large supercells. Lastly, a slightly larger fraction of the charge associated with the transition is contained within a 5 \AA\ sphere around the defect site for CIDER than for PBE (as indicated by the Loc. metric in the bottom of each plot), suggesting that CIDER localizes the defect charge more effectively than PBE. This is reassuring because unphysical charge delocalization is one of the key driving factors of erroneous charge transition level predictions in semiconductor point defects~\cite{Lany2008}.

In summary, our results indicate that the CIDER approach could be a route toward high-accuracy defect calculations at similar cost to semilocal DFT, which would significantly broaden the potential scope of point defects research. In particular, it could be used to improve the accuracy of high-throughput point defects studies~\cite{Broberg2023}, which are currently only computationally feasible with semilocal functionals like PBE.

\section{Conclusion}\label{sec:conclusion}

We have demonstrated that by combining the kinetic energy density (a semilocal, orbital-dependent quantity) with nonlocal, density-dependent features, it is possible to design functionals with comparable accuracy to hybrid DFT at a drastically reduced computational cost for plane-wave DFT calculations. We can do this by learning exact exchange energies, without the use of any high-accuracy reference data from experiments or wave function theory. We have shown that our nonlocal meta-GGA model has higher accuracy than semilocal functionals on both molecular benchmarks and solid-state band gap prediction, and that these improvements can assist with applications such as semiconductor point defect calculations.

The developments presented here serve a variety of purposes. As is, the exchange functional we designed can be used as an exact exchange surrogate in scenarios where hybrid DFT is impractical, such as large extended systems. This expands the scope of calculations that can be performed at hybrid DFT accuracy. In addition, the techniques we have introduced to learn smooth, physically constrained functionals and implement efficient nonlocal features could be used in other contexts, such as orbital-free DFT~\cite{Witt2018} and van der Waals functional design~\cite{Roman-Perez2009,Klimes2011,Berland2015}.

Importantly, the ability to systematically vary the descriptor complexity in our machine learning approach allows us to extract physically nontrivial information about the nature of the exchange interactions. Specifically, we find that both nonlocality and derivative discontinuity are independently important for the structure of the exchange energy density functional. This clarifies a key limitation of existing semilocal functionals.

Lastly, our approach can be generalized to learn the full XC functional, providing a path to highly transferable and universal functionals capable of describing challenging and heterogeneous chemistry. We anticipate that these developments will contribute to significantly improving the cost-accuracy trade-off of DFT, accelerating the pace and efficacy of materials and chemical research by providing a stronger theoretical backing in various sub-fields.

\section{Data Availability} \label{sec:data_availability}

The code to run CIDER calculations in both the PySCF and GPAW codes is available at the following Github repository: \url{https://github.com/mir-group/CiderPressLite}. The SL-GGA, NL-GGA, SL-MGGA, NL-MGGA, NL-MGGA-PBE, and NL-MGGA-DTR functionals are available for download using the instructions in the Github repository.

\begin{acknowledgments}
The authors thank Jennifer Coulter and Stefano Falletta for help with the hybrid DFT benchmark calculation. The authors also thank Stefan Riemelmoser and Georg Kresse for useful discussions. This work was supported by the National Defense Science and Engineering Graduate (NDSEG) Fellowship Program under contract FA9550-21-F-0003, the Camille and Henry Dreyfus Foundation Grant No. ML-22-075, and the Department of Navy award N00014-20-1-2418 issued by the Office of Naval Research. Computational resources were provided by the Harvard University FAS Division of Science Research Computing Group. 
\end{acknowledgments}

\begin{appendix}

\section{Computational Details}\label{sec:methods}

\subsection{Overview} \label{sec:methods_overview}

To explore the effectiveness of different features for learning the exchange energy, four types of functional were trained, with the model and feature vector details provided in Section \ref{sec:gp}:
\begin{itemize}
    \item SL-GGA: A semilocal GGA model based on the reduced squared gradient $s^2$ (Eq. \ref{eq:reduced_grad_s}).
    \item NL-GGA: A nonlocal model based on $s^2$ and three nonlocal features with Eqs.\ \ref{eq:anl_new} and \ref{eq:abnl_new} for the exponents.
    \item SL-MGGA: A semilocal meta-GGA model based on $s^2$ and the iso-orbital indicator $\alpha$ (Eq. \ref{eq:iso_orbital_alpha}).
    \item NL-MGGA: A nonlocal model based on $s^2$, $\alpha$, and three nonlocal features with Eqs.\ \ref{eq:anl_mgga_new} and \ref{eq:abnl_mgga_new} for the exponents.
\end{itemize}
The purpose of training semilocal GGA and meta-GGA models is to test the expressive power of the nonlocal descriptors. In our previous work (see Table S1 of Ref.~\cite{Bystrom2022}), we found that a standard, semilocal meta-GGA could not effectively learn the exchange energy density, but in this work, we determined that a semilocal meta-GGA can learn the \emph{total} exchange energy much more effectively than it can learn the exchange energy density.

The two primary datasets used in this work were the GMTKN55~\cite{Goerigk2017} and SOL62~\cite{Zhang2018,Trepte2022} databases. The GMTKN55 database contains 1505 molecular reaction energies based on 2462 single-point calculations, covering five categories: small-molecule properties, larger molecule reactions and isomerization, barrier heights, intermolecular noncovalent interactions, and intramolecular noncovalent interactions. The SOL62 dataset consists of 42 non-metals and 20 metals as categorized by Zhang \emph{et al.}~\cite{Zhang2018}. We follow Trepte and Voss~\cite{Trepte2022} in excluding Pb and Th from Zhang \emph{et al.}'s initial set of 64 solids. Twelve subsets of GMTKN55 and half of SOL62, along with a few other systems, were used for training and validation, as detailed in Section~\ref{sec:trval_selection}.

Throughout this work, we used pymatgen~\cite{Ong2013} and ase~\cite{HjorthLarsen2017} for structure manipulation and calculation setup, fireworks~\cite{Jain2015} for automated workflow management, and Scikit-learn~\cite{Pedregosa2011} for the Gaussian process models. The computational details for the PySCF and GPAW calculations are provided in Appendices \ref{sec:molcalc_settings} and \ref{sec:ssrefgpaw}, respectively. Appendices \ref{sec:trval_selection} and \ref{sec:dtr_method} describe the training and validation set selection, and Appendix \ref{sec:control_points} covers the selection of control points for the sparse Gaussian processes. Appendix \ref{sec:noise_hparam} describes the selection of noise hyperparameters and training/validation loss functions, and Appendix \ref{sec:hparam_sweep} describes the list of hyperparameters tested for each model and the model selection procedure. Appendix \ref{sec:convergence_issues} describes how we handled convergence issues for CIDER models on GMTKN55, and Appendix \ref{sec:cider_settings} describes the PySCF and GPAW settings that are specific to the CIDER functionals. Appendix \ref{sec:band_gap_methods} describes the band gap benchmark methods. Appendix \ref{sec:ss_performance} describes the computational cost benchmarking of CIDER for plane-wave DFT calculations. Finally, Appendix \ref{sec:charged_defects} describes our methodology for computing charged defect transition levels.

\subsection{Molecular Calculations and Reference Data Generation}\label{sec:molcalc_settings}

Molecular systems, including the nanoclusters discussed in Appendix \ref{sec:trval_selection}, were computed using the PySCF code~\cite{Sun2018,Sun2020}. All calculations were performed using the def2-QZVPPD basis set (with effective core potentials for atoms larger than Kr)~\cite{Weigend2003,Rappoport2010} for the atomic orbitals and the def2-universal-jkfit auxiliary basis set~\cite{Weigend2008} for the classical Coulomb energy. The seminumerical exchange (SGX) module in PySCF, which is similar to the computationally efficient chain-of-spheres exchange algorithm~\cite{Neese2009}, was used to compute the exchange energy and potential for hybrid functionals. For SGX, level 1 PySCF grids were used with P-junction screening turned off. For a given sub-database in GMTKN55, restricted DFT was used if all systems in that sub-database had $S_z=0$; otherwise, unrestricted DFT was used. The exception was the \cf{PCl_3} transition state in the INV24 database, for which spin symmetry breaking occurs for some functionals. This system was calculated with unrestricted DFT as well. For SCF calculations, the nonlocal features and XC energy were computed on level 3 PySCF grids.

For benchmarking comparison, the GMTKN55 database was evaluated with the above settings using the PBE~\cite{Perdew1996}, revPBE~\cite{Zhang1998}, r$^2$SCAN~\cite{Furness2020}, PBE0~\cite{Adamo1999}, B3LYP~\cite{Stephens1994}, and PW6B95~\cite{Zhao2005} functionals, all with D4 dispersion corrections~\cite{Caldeweyher2019}, as well as with the B97M-V~\cite{Mardirossian2015}, $\omega$B97X-V~\cite{Mardirossian2014}, and $\omega$B97M-V~\cite{Mardirossian2016} functionals.

To collect the exact exchange energy training data, the ground-state orbitals from the PBE calculations were used to compute the exact exchange energy $E_\text{x}^\text{exact}$ via Eq.\ \ref{eq:exact_exchange_energy}, and the feature vectors were computed from the corresponding PBE densities. The features were computed on level 1 PySCF grids. For training set feature generation only (not the SCF calculations), the integral in Eq.\ \ref{eq:cider_feature} was computed numerically with the $\mathbf{r}_2$ coordinate evaluated on level 3 PySCF grids without pruning. This dense grid is necessary for numerical stability and precision in the core region.

\subsection{Solid-State Calculations and Reference Data Generation}\label{sec:ssrefgpaw}

The electron densities and ground-state energies of the SOL62 dataset were computed in GPAW~\cite{Mortensen2005,Enkovaara2010} using the PBE geometries obtained by Trepte and Voss~\cite{Trepte2022}, with an energy cutoff of 1000 eV and the same k-point meshes as Trepte and Voss. The $h$ parameter in GPAW was set to the minimum value required to avoid aliasing of the density, i.e. 0.096 \AA. Fermi-Dirac smearing with a width of 0.01 eV was used for periodic systems. For isolated atoms, fixed occupations were used unless numerical instabilities occurred as a result, which was the case for six atoms (Pb-2, Pd-2, Pt-2, Sn-2, Ta-5, and V-5, where the number is the ground-state magnetic moment $2S$). In these cases, Fermi-Dirac smearing with a width of 0.002 eV was used. The default Davidson solver was used for all solids except the alkaline earth metals, which were solved with the conjugate gradient (CG) method. The CG method was used for all isolated atomic systems. To compute cohesive energies for training, the experimental ground state magnetic moments were used for the isolated atom references. The convergence threshold was set to $10^{-6}$ eV per valence electron for both isolated atoms and periodic systems. The density convergence threshold was set to $10^{-4}$ per electron for periodic systems and $10^{-3}$ per electron for isolated atoms. The eigenstates convergence threshold was set to $4\times 10^{-8}$ eV$^2$/electron for periodic systems and $10^{-5}$ eV$^2$/electron for isolated atoms. The cell size for the isolated atoms was chosen to be about 11.7 \AA\ across, with perturbations included to break symmetry. Compared to a larger box size of 17.5 \AA, the PBE atomic energies had a root mean square deviation of 0.012 eV and maximum deviation of 0.080 eV (for Zr). Because of the small average deviation, and the fact that the train and test reference values were computed on the atom-in-a-box systems rather than determined by experiment, the smaller boxes were used to lower the computational cost and memory requirements of training and testing. Symmetry constraints were also turned off for the isolated atom calculations.

Training the CIDER models to the solid-state data required computing the exact exchange energy and the nonlocal features for each system. To obtain exact exchange energies, the hybrid module of GPAW was used to compute the non-self-consistent exact exchange energy (EXX) from the PBE orbitals obtained in the above calculations. However, because the k-point meshes above are too dense for evaluation of EXX to be computationally feasible, the periodic systems were first recomputed with a coarser k-point mesh (for each system, the smallest even, $\Gamma$-centered mesh with a linear k-point density of at least 4.5 k-points per \AA$^{-1}$), and the EXX was computed on this mesh. All nonlocal density features were evaluated on the density and orbitals obtained from the denser k-point mesh. They were evaluated with the internal CIDER settings $q_{\text{max}}=300$ and $\lambda=1.8$ (see Section \ref{sec:cider_settings} for details).

The PBE0 cohesive energies were computed using the same procedure above, with the PBE0 energies being defined as
\begin{equation}
    E^{\text{PBE0}}=E^{\text{PBE,dense}}-E^{\text{PBE,coarse}}+E^{\text{PBE0,coarse}},
\end{equation}
with \emph{dense} denoting the calculation done with the dense k-point mesh and \emph{coarse} denoting the coarser k-point mesh discussed above.

\subsection{Construction of Training and Validation Sets}\label{sec:trval_selection}

The total exchange energy of a molecule is highly basis-set and code-dependent, and it is dominated by the core contributions for large atoms. In practice, only relative exchange energies between systems is relevant for most chemistry and materials science applications. Therefore, rather than training the model to match total exchange energies, we trained to match the exchange energy differences between systems. For example, the target and predicted values for the atomization energy of water would be
\begin{align}
    E_\text{x}^\text{target} &= E_\text{x}^\text{ex}[n_\text{\cf{O}}] + 2 E_\text{x}^\text{ex}[n_\text{\cf{H}}] - E_\text{x}^\text{ex}[n_\text{\cf{H2O}}] \\
    E_\text{x}^\text{pred} &= \hat{E}_\text{x}[n_\text{\cf{O}}] + 2 \hat{E}_\text{x}[n_\text{\cf{H}}] - \hat{E}_\text{x}[n_\text{\cf{H2O}}],
\end{align}
where $E_\text{x}^\text{ex}[n]$ and $\hat{E}_\text{x}[n]$ are the exact and model exchange energies for density distribution $n(\mathbf{r})$, respectively.

The training and validation sets were constructed from four sources:

\textbf{Molecules}: The W4-11 (atomization energies), BH76 (barrier heights), BH76RC (reaction energies for BH76 reactions), MB16-43 (``mindless benchmarking'' of reaction energies), S22 (noncovalently bound dimers), G21IP (ionization potentials), PA26 (proton affinities), RG18 (noble gas dimers), ACONF (alkane conformers), ALKBDE10 (alkali/alkaline earth diatomic bond energies), HEAVYSB11 (dissociation energies for molecules with heavy atoms), and SIE4x4 (self-interaction problems) subsets from the GMTKN55 database were used for training and validation. See the original GMTKN55 paper~\cite{Goerigk2017} and references therein for more details on the datasets. Each set was randomly split into two-thirds training and one-third validation, except for the BH76 and BH76RC subsets. Because the data points in these subsets are based on the same systems, and also because they include the forward and backward reactions for the same chemical equation, a different splitting procedure was chosen to maintain train-validation independence. First, the 34 unique transition states in BH76 were identified and split half-and-half into the training and validation sets. All 40 reactions with the training transition states were used for training, and all 36 reactions with the validation transition states were used for validation. All 17 BH76RC reactions with systems in the BH76 training partition were used for training, and the other 13 reactions were used for validation. All data not in the training or validation set was classified as the GMTKN55 test set. This data partitioning scheme resulted in 280 train data, 163 validation data, and 1062 test data in GMTKN55.

\textbf{Bulk solids}: Half of the SOL62 cohesive energy database (11 randomly selected elemental metals and 20 randomly selected non-metals) was used for training and validation. Nine systems (C-diamond, Li, GaAs, MgO, LiF, ZrC, TiC, Pd, and Pt) were hand-selected for the training set to cover a range of chemical space, while the other 22 systems were used for validation. All data not in the training or validation set was classified as the SOL62 test set.

\textbf{Nanoclusters}: Six nanocluster atomization energies were added to the training set as a ``bridge'' between the isolated and extended systems in the study. These were \cf{Ca13} and \cf{Zn26}, with the initial structure obtained based on the Materials Project~\cite{Jain2013} bulk lattice constant and structure optimization at the PBE/def2-TZVP level in Orca~\cite{Neese2012,Neese2018,Neese2020}; \cf{C35H36} (hydrogen-capped diamond), obtained from Refs.\ \cite{Barnard2013,Barnard2014} and also structurally optimized in Orca at the PBE/def2-TZVP level; and \cf{Na4Cl4}, \cf{[Na13Cl14]-}, and \cf{[Na14Cl13]+}, all using the Materials Project bulk lattice constant without further structure optimization.

\textbf{Atoms}: Absolute and relative energies of the isolated atoms H-Kr and a few excited spin states were included in the training set. Specifically, the training points consisted of the absolute energies of the H atom and the noble gases He, Ne, Ar, and Kr; the relative energies of the elements with nuclear charge $Z$ and $Z+1$ for $Z=\{1,...,35\}$, and the relative energies of the following ground state/excited state spin pairs: Sc 1/3, Ti 2/4, V 3/5, Cr 6/4, and Ni 2/0. The preceding numbers correspond to the magnetic moment $2S$.

In addition to fitting to the above training data, the uniform electron gas constraint was enforced as described in Section \ref{sec:feature_defs}, and the uniform scaling constraint was enforced by the scale-invariance of the input features. The ground-state total energies of the SOL62 solids and isolated atoms were computed with GPAW as described in Section \ref{sec:ssrefgpaw}. The ground-state energies for all other training data were evaluated with PySCF as described in Section \ref{sec:molcalc_settings}. The selection of train and validation databases from GMTKN55 includes a wide variety of chemistries and properties. However, because training data is sampled disproportionately from small-molecule properties, there is a large distribution shift between the train/validation and test sets, with the test sets containing more large molecules, isomerization reactions, and barrier heights. The use of this train/test partition therefore serves as a strong test of transferability. The accuracy on SOL62 also strongly reflects the transferability  of a model because only 9 solid-state systems were included in the training set, and SOL62 covers a wide variety of chemistries.

\subsection{A More Diverse Training Set for Numerical Stability} \label{sec:dtr_method}

To design a more diverse training set for the NL-MGGA-DTR functional, a kernel function was defined for the ``covariance'' between two training points in GMTKN55:
\begin{equation}
    k_\text{GMTKN}(\mathbf{x}, \mathbf{x}') = \exp\left(-\frac{|\mathbf{x}-\mathbf{x}'|^2}{328}\right),\label{eq:dtr_kernel_func}
\end{equation}
where $\mathbf{x}$ is the vector of errors for 82 functionals benchmarked on the GMTKN55 database by Goerigk \emph{et al.}~\cite{Goerigk2017} with D3(BJ) dispersion corrections~\cite{Becke2005,Johnson2006,Grimme2010}, normalized so that the mean square error across all functionals for that individual data point is 1. This kernel essentially provides a larger similarity score between data points for which the (normalized) errors tend to be similar for a given functional. The kernel matrix was constructed for the entire GMTKN55 database and then factorized using a method called Cholesky decomposition with complete pivoting~\cite{Higham2002,HARBRECHT2012428}, which factors a given matrix $\mathbf{A}$ as follows:
\begin{equation}
    \mathbf{P}^\top \mathbf{A} \mathbf{P} = \mathbf{L} \mathbf{L}^\top, \label{eq:pivchol}
\end{equation}
In Eq.\ \ref{eq:pivchol}, $\mathbf{P}$ is a permutation matrix that orders the vectors $\mathbf{x}$ in Equation~\ref{eq:dtr_kernel_func} in decreasing order of their variance, conditioned on the preceding feature vectors, \emph{in the space implicitly spanned by the kernel function}. Intuitively, this means that later vectors, as ordered by $\mathbf{P}$, are likely to be nearly linearly dependent (in the kernel space) on the earlier vectors. Therefore, by picking the first $N$ training points as ordered by $\mathbf{P}$ and using the remaining data for testing, we can get a training set whose errors correlate as much as possible with all the test data, effectively making the training set more diverse and potentially preventing out-of-distribution error of the ML model.

We follow this procedure and select the first 225 pivots for the training set, the next 75 pivots for the validation set, and the remaining 1205 pivots for the test set. With this dataset partitioning, more than 99\% of the GMTKN55 test set data points contain at least one system not present in the train or validation sets, so the test set is still sufficiently independent to assess the transferability of the functionals to new systems. All training points not contained in GMTKN55 were kept the same as in Appendix \ref{sec:trval_selection}.

\subsection{Control Point Selection for Sparse Gaussian processes}\label{sec:control_points}

In Eqs.\ \ref{eq:sparse_cov} and \ref{eq:sparse_cross_cov}, a set of control points $\{\mathbf{\tilde{x}}_a\}$ is required to evaluate the Gaussian process covariance kernel in the sparse approximation. These control points were initially selected by randomly sampling feature vectors from the PySCF grids for the training set (Appendix \ref{sec:molcalc_settings}). All training data was sampled except the SOL62 data, since the features for these systems were evaluated in GPAW.

Due to the size of the numerical grids from which the features were sampled, there were a large number of potential control points for the sparse Gaussian process, many of which were nearly linearly dependent in the implicit feature space of the covariance kernel function. To reduce the size of the control point set, a Cholesky decomposition with pivoting (Eq.\ \ref{eq:pivchol}) was performed on an initial sample size of roughly $10^4$ feature vector samples with a tolerance of $10^{-5}$, and only control points corresponding to signficant pivot indices were kept. This reduced the number of control points from roughly $10^4$ to $10^2$.

\subsection{Selection of Noise Hyperparameters, Training Loss, and Validation Loss}\label{sec:noise_hparam}

Chemical and material properties have a wide range of magnitudes, from less than 1 kcal/mol for some van der Waals binding energies to greater than 100 kcal/mol for many covalent bond dissociation energies. Correspondingly, different magnitudes of XC errors are expected and acceptable for these properties; an error of 1 kcal/mol might make a van der Waals binding energy prediction useless but be acceptable for an atomization energy. To train the exchange functionals, it was therefore necessary to weight the noise hyperparameter of each training point to reflect the feasible and desired precision for the property of interest. To do so, we used the data provided with GMTKN55~\cite{Goerigk2017} to compute the MAD of four hybrid functionals (PBE0-D3(BJ), B3LYP-D3(BJ), PW6B95-D3(BJ), and $\omega$B97X-V) compared to the reference data on each sub-database (denoted $\text{MAD}^\text{hybrid}_{\text{DB}}$, where DB is the sub-database of interest), and then created characteristic uncertainties for each sub-database as follows (in Hartree atomic units):
\begin{equation}
    \tilde{\sigma}_{\text{DB}}=\text{MAD}^\text{hybrid}_{\text{DB}} \left(\frac{\sigma_0}{\text{MAD}^\text{hybrid}_{\text{W4-11}}}\right),
\end{equation}
where W4-11 is a small-molecule atomization energy sub-database in GMTKN55 and $\sigma_0=0.03$ Eh. One drawback of the above approach is that it favors high accuracy on systems that are already described well by conventional, dispersion-corrected DFT (like simple van der Waals dimers) and assumes a large uncertainty on systems like self-interaction-dominated bond energies, for which even hybrid functionals with small exchange mixing perform poorly. To remedy this, the noise parameter was augmented with a constant weighting for every system:
\begin{equation}
    \sigma_{\text{DB}}=\sqrt{\frac{2}{1/\sigma_0^2+1/\tilde{\sigma}_\text{DB}^2}}.
\end{equation}
With this definition, the matrix $\boldsymbol{\Sigma}_{\text{noise}}$ in Eq.\ \ref{eq:gp_predictive} is a diagonal matrix with $\left(\boldsymbol{\Sigma}_{\text{noise}}\right)_{ii}$ being the $\sigma_{\text{DB}}^2$ for the database to which training point $i$ belongs. Constructing the Gaussian process for this noise parameter is equivalent to regression with the loss function
\begin{equation}
    \mathcal{L}=\frac{1}{2}\sum_{\text{DB}} N_\text{DB} \text{MSD}_\text{DB} \left(\frac{1}{\sigma_0^2} + \frac{1}{\tilde{\sigma}_\text{DB}^2}\right), \label{eq:trloss}
\end{equation}
where $\text{MSD}_\text{DB}$ stands for the mean squared deviation of a given GMTKN55 sub-database labeled by ``DB,'' and $N_\text{DB}$ is the number of points in the sub-database. This weighting ensures that training points with small expected errors are appropriately weighted while those with larger expected errors for conventional functionals do not become under-weighted. For systems outside the GMTKN55 database that were in the training set, $\tilde{\sigma}_{\text{DB}}$ was set by hand.

\subsection{Hyperparameter Selection and Validation Procedure} \label{sec:hparam_sweep}

The models tested in this work have several discrete and continuous hyperparameters to be optimized, which are listed below.
\begin{itemize}
    \item The covariance prefactor hyperparameter $\Sigma_{\text{Cov}}$ in eqs~\ref{eq:slgga_kernel}--\ref{eq:nlmgga_kernel}.
    \item The kernel length-scale ($l_i$, where $i$ is the feature index) in eq~\ref{eq:basekernel}.
    \item For the nonlocal functionals only, the feature exponent parameters $B_i$ and $C_i$ in Eqs.\ \ref{eq:anl_new}--\ref{eq:abnl_mgga_new}.
    \item The baseline exchange functional on top of which the CIDER model is trained via $\Delta$-learning. We tested PBE~\cite{Perdew1996} and the Chachiyo GGA~\cite{Chachiyo2020}.
\end{itemize}
It would be quite challenging and expensive to implement differentiation of the marginal likelihood with respect to the second parameter, and nearly impossible for the third since backpropagation would need to be performed through the feature evaluation step. Therefore, we used various combinations of hyperparameters to train the models, and then for each functional type, we selected the best-performing model on the validation set as the final candidate functional.

To obtain a discrete set of exponent hyperparameters to test for the nonlocal functionals, we developed a reduced number of parameters to specify the feature exponents. Similarly to the previous CIDER work~\cite{Bystrom2022}, a control parameter $A$ tunes $B_2$ and $C_2$ simultaneously, in two different schemes. The first scheme (S1) is
\begin{align}
    B_2&=A\notag\\
    C_2&=\frac{A}{32}\frac{6}{5\pi}(6\pi^2)^{2/3}\approx 0.18A.\label{eq:exp_coef_s1}
\end{align}
The second scheme (S2) provides a different mixture of density-based and gradient/kinetic energy-based terms:
\begin{align}
    B_2&=\frac{A}{16}\frac{6}{5\pi}(6\pi^2)^{2/3}\approx 0.36A\notag\\
    C_2&=B_2.\label{eq:exp_coef_s2}
\end{align}
In both schemes, the other constants are determined by $B_2$ and $C_2$:
\begin{align}
    &B_1=\frac{1}{2}B_2,\,\,B_3=2B_2,\,\,B_0=\frac{D}{A}B_2\\
    &C_1=\frac{1}{2}C_2,\,\,C_3=2C_2,\,\,C_0=\frac{D}{A}C_2,
\end{align}
where $D$ is an additional parameter to tune the $B_0$ and $C_0$ coefficients separately from the other coefficients. With these schemes, all 8 parameters $\{B_i\},\{C_i\}$ that define the three-feature CIDER model were reduced to two parameters per scheme. For each scheme, we performed a hyperparameter search over the grid $A=\{1/2,1,2,4,8\}$ and $D=\{1/2,1,2,4,8\}$, resulting in a total of 50 feature constructions to test.

Because there were already 50 features from which to select the nonlocal functionals, we set the covariance and feature length-scales for these models heuristically:
\begin{align}
    \Sigma_{\text{Cov}} &= R_1 \expval{\Delta F_\text{x}^2}_{\text{train}} \\
    l_i &= R_2 \sqrt{\expval{x_i^2}_{\text{train}}},
\end{align}
where $\expval{}_{\text{train}}$ is an expectation value over all the grid points in the exchange energy training set, $\Delta F_\text{x}$ is the deviation of the exact exchange enhancement factor at a given point from the baseline functional enhancement factor, $x_i$ is the $i$-th regularized feature (see Section \ref{sec:feature_defs}), and $R_1$ and $R_2$ are constants. $R_1$ and $R_2$ were set based on the criteria that the resulting functionals converge consistently in both GTO and plane-wave calculations and exhibit a small total energy deviation between the PAW and all-electron GTO atomization energies for a few small molecules. $R_2=1$ worked best for both the nonlocal GGA and meta-GGA, while the most practical values of $R_1$ depended on the baseline functional and model type, as shown in Table \ref{tab:r1table}. The meta-GGA models required a smaller covariance prefactor to achieve good stability than the GGA models. We set $R_1$ to be larger by a factor of 20 for functionals with the Chachiyo baseline compared to the PBE baseline because $\expval{\Delta F_\text{x}^2}_{\text{train}}$ was roughly 20 times larger for PBE than Chachiyo.

\begin{table}
    \caption{$R_1$ values for each functional type and baseline functional.}
    \begin{ruledtabular}
    \begin{tabular}{cd}
        \textrm{Type/Baseline}&
        \multicolumn{1}{c}{\textrm{$R_1$}} \\
        \colrule
        NL-GGA/PBE & 1.0 \\
        NL-GGA/Chachiyo & 20.0 \\
        NL-MGGA/PBE & 0.05 \\
        NL-MGGA/Chachiyo & 1.0
    \end{tabular}
    \end{ruledtabular}
    \label{tab:r1table}
\end{table}

For the semilocal functionals, there was no need to validate over the possible feature constructions, leaving more flexibility to test $R_1$ and $R_2$ systematically. For the functionals with Chachiyo baseline, we tested $R_1=\left\{1,2,4,8,16,32,64,128\right\}$, while for the PBE baseline each $R_1$ was divided by 20 to give a roughly equivalent $\Sigma_{\text{Cov}}$. For the semilocal functionals, $R_2=\left\{1,\frac{1}{2},\frac{1}{4}\right\}$ were tested, resulting in 24 functionals to validate. Note that for NL-MGGA-DTR, we determined from validation set SCF calculations that $R_1=0.1$ with the PBE baseline resulted in sufficiently stable models, so this larger value was used. In addition, we did not search over feature length-scale for NL-MGGA-DTR, and instead used the features determined to be most accurate for NL-MGGA and NL-MGGA-PBE.

After the functionals were trained, we selected the top 10 functionals of each type and baseline functional, as classified by non-self-consistent performance on the validation set via Eq.\ \ref{eq:trloss}. To validate these functionals, the GMTKN55 train and validation set reaction energies were predicted using self-consistent calculations with a convergence threshold of $10^{-6}$ Eh and a gradient threshold of $2\times 10^{-3}$ Eh. For a given functional, if any calculation failed to converge when starting from both the default initial guess of PySCF and from the PBE orbitals, the functional was deemed insufficiently numerically stable and disqualified from further consideration. The non-self-consistent cohesive energies of the SOL62 validation set were also computed. Of the functionals for which all calculations converged, the validation loss consisted of
\begin{align}
    \mathcal{L}_{\text{val}} &= \mathcal{L}_{\text{MOL}} + \mathcal{L}_{\text{SOL}} \label{eq:val_loss} \\
    \mathcal{L}_{\text{MOL}} &= \frac{1}{2}\sum_{\text{DB}} N_\text{DB/VAL} \times \text{MSD}_\text{DB/VAL}^{\text{SCF}} \times \left(1 + \frac{\sigma_0^2}{\tilde{\sigma}_\text{DB}^2}\right) \\
    \mathcal{L}_{\text{SOL}} &= \text{MSD}_{\text{SOL62/VAL22}}^{\text{NSCF}},
\end{align}
where $\text{MSD}_\text{DB/VAL}^{\text{SCF}}$ is the self-consistent mean square deviation on the validation partition of GMTKN55 subset DB, $N_\text{DB/VAL}$ is the number of validation data in DB, and $\text{MSD}_{\text{SOL62/VAL22}}^{\text{NSCF}}$ is the non-self-consistent mean square deviation of the 22 validation-set cohesive energies per atom. The final functional of each type was selected based on which candidate had the lowest loss for Eq.\ \ref{eq:val_loss}. The validation results and selected hyperparameters for the functionals are provided in the Supplemental Material (Section S4).

\subsection{Difficult Convergence Cases} \label{sec:convergence_issues}

The default convergence tolerance for the calculations performed with CIDER functionals was $10^{-8}$ Eh, except for the NL-MGGA-PBE and NL-MGGA-DTR calculations, which were converged to $10^{-9}$ Eh when possible. Some CIDER calculations did not converge to this tolerance within the default PySCF settings. There were two different classes of convergence problems. First, for a few systems, some CIDER calculations did not converge from the PySCF starting guess, but the convergence issues were resolved by starting from PBE ground-state orbitals. Second, for some  systems (especially isolated atoms and ions, as well as alkali and alkaline earth-containing systems), a numerical stability issue was observed for some CIDER calculations, in which the energy converged to a reasonable tolerance (less than $10^{-7}$ Eh), but the orbital gradients oscillated around $10^{-3}$ Eh without converging fully. In these cases, the energy tolerance was set to $10^{-7}$ Eh, and the gradient tolerance was set to the (rather large) $2\times 10^{-3}$ Eh. All CIDER calculations converged to these thresholds. PBE starting orbitals were also used for these loose-threshold calculations. Even with these looser thresholds, the achieved energy and orbital gradient convergence is still sufficient to compare thermochemical data in the GMTKN55 database.

\subsection{CIDER Functional Settings} \label{sec:cider_settings}

The CIDER functionals require a few additional settings to specify the precision of the accelerated feature calculations. For the molecular version, these settings are $l_{\text{max}}$, the maximum spherical harmonic order for auxiliary basis expansions in Section \ref{sec:theory_cidermol}; $\lambda$, in Eq.\ \ref{eq:qlambd}; $\beta^{\text{aux}}$, in Eq.\ \ref{eq:auxiliary_etb}; and $q_{\text{max}}$, the maximum value of $q_{\alpha}$ (Eq.\ \ref{eq:qlambd}). The $\xi$ basis in Eq.\ \ref{eq:xi_proj} uses the same parameter $\lambda$ for the even-tempered basis as Eq.\ \ref{eq:qlambd}. We chose $l_{\text{max}}=10$, $\lambda=\beta^{\text{aux}}=1.6$, and $q_{\text{max}}=\frac{1000}{36}Z_{\text{max}}^2$, where $Z_{\text{max}}$ is the minimum of 36 and the charge of the largest nucleus in the given GMTKN55 sub-database. These settings were found to be in good agreement with numerical integration of the features. The minimum control point $q_0$ was set automatically based on the minimum allowed value of the kernel exponent $a(\mathbf{r})$.

For the plane-wave implementation, only $\lambda=1.8$ and $q_{\text{max}}=300$ needed to be specified. A larger $q_{\text{max,atom}}=10Z^2$ was used for each atom on the radial support grids, with $Z$ being the atomic number. In both the Gaussian-type orbital and plane-wave implementations, all other CIDER-specific parameters (like the minimum exponents for the various auxiliary bases) were set automatically based on the chosen basis set and nonlocal feature parameters (i.e. the parameters in Appendix \ref{sec:hparam_sweep}).

\subsection{Band Gap Calculations} \label{sec:band_gap_methods}

Band gaps for the database of Borlido \emph{et al.}~\cite{Borlido2019} were computed in GPAW~\cite{Mortensen2005,Enkovaara2010} as the difference between the conduction band minimum and valence band maximum for a ground-state SCF calculation. The energy cutoff for these calculations was 520 eV, and the grid spacing was set to $h=0.134$ \AA\ to avoid any aliasing of the density. Fermi-Dirac k-point smearing was used with a width of 0.01 eV, except for the noble gas systems, for which no k-point smearing was used. The k-point mesh for each system was selected to be the smallest even, $\Gamma$-centered mesh with a linear k-point density of at least 6 k-points per \AA$^{-1}$. For each system, if the PBE band gap we obtained was more than 0.1 eV larger than the PBE band gap obtained by Borlido \emph{et al.}~\cite{Borlido2019}, we inferred that the k-point mesh was too coarse and reran the band gap calculation with a linear k-point density of 10 k-points per \AA$^{-1}$. Using this methodology, we reproduced the PBE band gaps of Borlido \emph{et al.} with a mean, mean absolute, and maximum error of -0.01, 0.03, amd 0.17 eV, respectively, and the SCAN results with a mean, mean absolute, and maximum error of 0.00, 0.03, and 0.15 eV, respectively.

For PBE0/NL-MGGA-DTR and PBE0(0.35)/NL-MGGA-DTR, a few systems did not converge with the default CIDER PAW corrections, and a denser core integration grid and smaller set of feature projectors (i.e. the $g_i^A$ and $p_i^A$ functions of Section \ref{sec:theory_ciderpaw}) were used. Even with these settings, one system (\cf{ReSe2}) did not converge with PBE0(0.35)/NL-MGGA-DTR, and it had to be run with SG15 norm-conserving pseudopotentials (NCPP)~\cite{Hamann2013}. For the 255 calculations in the main text, the mean, mean absolute, and maximum deviations between PBE0/NL-MGGA-DTR with NCPP and PAW were 0.03, 0.12, and 1.12 eV, respectively. In addition, the band gap for \cf{ReSe2} with PBE0/NL-MGGA-DTR was 1.29 eV with PAW and 1.30 eV with NCPP, so the use of pseudopotentials appears reasonable for this particular case. We think that the convergence issues are due to numerical problems with the PAW corrections for nonlocal features, and improving the numerical stability of these corrections will be a subject of future work.

For the band gaps in Table \ref{tab:bgsubset}, the same procedure was used, except with denser k-point meshes containing 12 k-points per \AA$^{-1}$. The HSE06 band gaps were computed using VASP~\cite{Kresse1993}, with the k-point mesh set to contain 4000 k-points divided by the number of atoms in the unit cell. This is a coarser mesh than was used for the semilocal and CIDER calculations, but we found that the band gap difference compared to larger k-point densities was less than 0.01 eV.

\subsection{Supercell and Neutral Defect Performance Benchmark} \label{sec:ss_performance}

For the neutral defect calculations, the atomic structure of the 215-atom single vacancy cells of diamond and silicon were obtained using VASP~\cite{Kresse1993,Kresse1996,Kresse1996a} with PAW datasets~\cite{Kresse1999a,Blochl1994}, an energy cutoff of 520 eV, the ``Accurate'' precision setting, and a $2\times 2\times 2$ Monkhorst-Pack k-point mesh~\cite{Monkhorst1976}, starting with the PBE lattice constant obtained in VASP using the same settings with a $14 \times 14 \times 14$ $\Gamma$-centered k-point mesh. Using these structures, SCF ground-state calculations with static atomic positions were performed in GPAW~\cite{Mortensen2005,Enkovaara2010} for the 216-atom bulk and 215-atom vacancy cells for each functional. The formation energy of the neutral defect was computed as
\begin{equation}
    E^{\text{formen}}=E^{\text{defect}}-\frac{215}{216}E^{\text{bulk}},
\end{equation}
where the bulk and defect energies were computed with 216 and 215-atom supercells, respectively, with a $2\times 2\times 2$ Monkhorst-Pack k-point mesh (resulting in 1 reduced k-point for the bulk supercell and 4 reduced k-points for the defect supercell), 520 eV plane-wave cutoff, 0.134 \AA\ grid spacing, and $10^{-5}$ eV convergence threshold. 523 bands were computed for the bulk, and 520 bands were computed for the defect cell. The time to evaluate the formation energy for each functional was computed as the sum of the time taken for the bulk and vacancy cell calculations, as obtained from the GPAW internal calculation timer. The HSE06 and PBE0 formation energies were obtained using VASP with the same structures, k-point grid, and energy cutoff.

For the $\Gamma$-point calculations used in Fig.~\ref{fig:perf_bulk}, GPAW was also used for the semilocal and ML functionals. For the hybrid DFT calculations, Quantum ESPRESSO~\cite{QE-2009,QE-2017} was used with the adaptively compressed exchange~\cite{Lin2016} algorithm for exchange, a 38 Rydberg plane-wave cutoff for the wave functions, a 152 Rydberg cutoff for the charge density and potential, a $10^{-6}$ Rydberg convergence threshold, and SG15 norm-conserving pseudopotentials~\cite{Hamann2013}. The use of a 38 Rydberg (517 eV) plane-wave cutoff might be too small when using norm-conserving pseudopotentials, but for performance benchmarks, it provides the most fair comparison to the 520 eV cutoff used with the PAW setups in GPAW. The PBE calculation with Quantum ESPRESSO used the same settings but with PBE in place of PBE0. The GPAW calculations used 523 bands, and the Quantum ESPRESSO calculations used 432 bands.

\subsection{Charged Defect Calculations} \label{sec:charged_defects}

The charged defect levels were computed with PBE, PBE0/NL-MGGA-DTR, and PBE0(0.3)/NL-MGGA-DTR. The mixing parameter $\alpha=0.3$ of the last functional was selected to match the zero-temperature experimental band gap of silicon, 1.17 eV. For each functional, the lattice constant, band edges, and band gap were obtained from a unit cell relaxation with 520 eV cutoff, $h=0.134$ \AA, and a $24 \times 24 \times 24$ k-point mesh in GPAW~\cite{Mortensen2005,Enkovaara2010}. Then, a 512-atom supercell was created for each defect, with perturbed structures used for the vacancy to break symmetry. For each charge state and defect, the geometry was relaxed in GPAW using the BFGS with Line Search algorithm to a tolerance of 0.05 eV/\AA\ with a $\Gamma$-point only calculation. This geometry was then used as the starting point for the final geometry relaxation, which was performed with a $2 \times 2 \times 2$ $\Gamma$-centered k-point mesh and a force convergence tolerance of 0.01 eV/\AA. Because PBE0(0.3)/NL-MGGA-DTR is so similar to PBE0/NL-MGGA-DTR, and because the lattice constants are similar, we saved time for this functional by starting the final relaxation from the fractional coordinates of the PBE0/NL-MGGA-DTR calculations.

The formation energies of the defects were calculated as~\cite{Freysoldt2014}
\begin{align}
    E^f[X^q] =& E_\text{tot}[X^q] - E_\text{tot}[\text{bulk}] \notag\\ &- \sum_i n_i\mu_i + qE_F + E_\text{corr}[X^q],
\end{align}
where $E_\text{tot}$ is the total ground-state energy of the supercell DFT calculation, $X^q$ is defect $X$ in charge state $q$, $E_F$ is the Fermi level, and $E_\text{corr}[X^q]$ is a finite-size energy correction for the defect. The stoichiometry changes and chemical potentials $n_i$ and $\mu_i$ were neglected because they do not impact electronic transition levels. We used the Freysoldt finite-size correction-scheme for $E_\text{corr}[X^q]$~\cite{Freysoldt2009}, which combines an electrostatic and potential alignment correction. This correction requires an estimate of the dielectric constant $\epsilon$, for which we used the experimental value of 11.9~\cite{Sze2006}.

Because some of the defects studied in this work are shallow donors and acceptors, and because electrostatic corrections like those developed by Freysoldt \emph{et al.}~\cite{Freysoldt2009} and Kumagai \emph{et al.}~\cite{Kumagai2014} assume localized charges, these corrections might be inaccurate. Therefore, for comparison, we also computed the transition levels using the potential alignment correction only, i.e.
\begin{equation}
    E_\text{corr}[X^q] = q \left( \Bar{V}_\text{esp}^\infty[X^q] - \Bar{V}_\text{esp}^\infty[\text{bulk}] \right),
\end{equation}
where $\Bar{V}_\text{esp}^\infty$ is the average electrostatic potential in a region far from the defect site. This is equivalent to the full Freysoldt correction with $\epsilon\rightarrow\infty$. We used pymatgen~\cite{Ong2013} to perform the corrections.

\section{Construction of Feature Projector Functions for CIDER within PAW}\label{app:feat_proj}

\subsection{Construction of Feature Smoothing Projectors}\label{app:feat_smooth_const}

We solve for the coefficients $D_{j\beta}$ and $C_{i\alpha}$ in Eqs.\ \ref{eq:djb} and \ref{eq:cia} by minimizing the loss function
\begin{equation}
    \mathcal{L} = \sum_\beta w_\beta \int \dd[3]\mathbf{r} \left[\hat{y}_{A,\beta}(\mathbf{r}) - \Delta F_{A,\beta}(\mathbf{r})\right]^2. \label{eq:pasdw_loss}
\end{equation}
For the current implementation, $w_\beta=1$, but this could be changed if needed to encourage better smoothness or numerical precision. We start with the following set of functions:
\begin{align}
    g_{i\leftarrow nlm}(\mathbf{r}) &= \begin{cases}
    Y_{lm}(\mathbf{\hat{r}}) P(2n+l, r/r_c) & r < r_c\\
    0 & r \ge r_c
    \end{cases}\\
    P(n, x) &= \begin{cases}
        \frac{1}{2}(1 + \cos \pi x) & n = 0\\
        Q(n-1,x)  & n \in \mathbb{O}\\
        R(n-2,x) & n \in \mathbb{E}
    \end{cases}\\
    Q(m,x) &= 5\left(\sin\frac{\pi x}{2}-\frac{1}{2}+\frac{1}{2}\cos \pi x\right) x^{m} e^{-mx^2} \\
    R(m,x) &= \frac{1}{2}\left(1-\cos(2\pi x)\right) x^m e^{-mx^2},
\end{align}
where $\mathbb{O}$ and $\mathbb{E}$ are the sets of positive odd and even integers, respectively. Then we introduce the matrix $\mathbf{P}$
\begin{equation}
    \mathbf{P} = \begin{pmatrix}
    \mathbf{P}^{11} & \mathbf{P}^{12} \\ (\mathbf{P}^{12})^\top & \mathbf{P}^{22}
    \end{pmatrix}.
\end{equation}
In the above equation, $\mathbf{P}^{11}$ is an $N_{h} N_{a} \times N_{h} N_{a}$ matrix,
\begin{equation}
    P^{11}_{j\beta,j'\beta'} = \delta_{\beta\beta'} \braket{h_j^A}{h_{j'}} + \delta_{jj'}\delta_{\beta\beta'} \sigma_{11},
\end{equation}
where $N_{h}$ is the number of $h_j^A$ functions and $N_a$ is the number of control points for kernel interpolation. Similarly, $\mathbf{P}^{22}$ is an $N_g N_{a} \times N_g N_{a}$ matrix,
\begin{align}
    P^{22}_{i\alpha,i'\alpha'} &= \sum_\beta w_\beta I_{\beta,i\alpha,i'\alpha'} + \delta_{ii'}\delta_{\alpha\alpha'}\sigma_{22}' \\
    I_{\beta,i\alpha,i'\alpha'} &= \int \dd[3]\mathbf{r}\, \phi_{i\alpha\beta}(\mathbf{r}) \phi_{i'\alpha'\beta}(\mathbf{r}) \left(1 + \sigma_{22}f_{cut}(\mathbf{r}) \right)\\
    \phi_{i\alpha\beta}(\mathbf{r}) &= \int \dd[3]\mathbf{r}'\, \Phi_{\alpha\beta}(\mathbf{r}'-\mathbf{r}) g_i^A(\mathbf{r}'),
\end{align}
where $N_g$ is the number of $g_i^A$ functions. The constants $\sigma_{11}$, $\sigma_{22}$, and $\sigma_{22}'$ are regularization constants to reduce the magnitude of the feature augmentation correction in the core region, which helps with numerical stability by preventing the pseudo-feature from being unrealistically large in the core.

Lastly, $\mathbf{P}^{12}$ is an $N_h N_{a} \times N_g N_{a}$ matrix:
\begin{equation}
    P^{12}_{j\beta,i\alpha} = w_\beta \int \dd[3]\mathbf{r}\, h_{j}(\mathbf{r}) \phi_{i\alpha\beta}(\mathbf{r}).
\end{equation}
The coefficients can be solved by minimizing the loss Eq.\ \ref{eq:pasdw_loss}, which results in the linear system
\begin{equation}
    \begin{pmatrix} \mathbf{D} \\ \mathbf{C} \end{pmatrix} =
    \begin{pmatrix}
    \mathbf{P}^{11} & \mathbf{P}^{12} \\ (\mathbf{P}^{12})^\top & \mathbf{P}^{22}
    \end{pmatrix}^{-1}
    \begin{pmatrix} \mathbf{b}^{1} \\ \mathbf{b}^{2} \end{pmatrix},
\end{equation}
where
\begin{align}
    b^{1}_{j\beta} &= w_\beta \int \dd[3]\mathbf{k}\, \Delta F_{A,\beta}(\mathbf{k}) h_j^A (\mathbf{k}) \label{eq:paw_b1l} \\
    b^{2}_{i\alpha} &= \sum_\beta w_\beta \int \dd[3]\mathbf{k}\, \Delta F_{A,\beta}(\mathbf{k}) \phi_{i\alpha\beta}(\mathbf{k}). \label{eq:paw_b2l}
\end{align}
The matrix $\mathbf{P}$ is block-diagonal in the angular momentum channels due to the orthogonality of spherical harmonics, so each angular momentum channel can be solved separately. This makes the whole routine fairly efficient, and the number of operations per SCF cycle is linear-scaling in the number of atoms. The Fourier transforms of the $g_i^A$ and $h_j^A$ functions needed for Eqs.\ \ref{eq:paw_b1l} and \ref{eq:paw_b2l} are obtained using the NUMSBT algorithm~\cite{Talman2009}, and the $\Delta F_{A,\beta}(\mathbf{k})$ is decomposed into angular momentum channels via the $\Delta F_{A,L,\beta}(k)$ term obtained in Eq.\ \ref{eq:dfalbk}. Also, note that there are a larger number of $\alpha$ and $\beta$ indices in the atomic augmentation regions than on the FFT grid. The above fitting procedure is only used for indices going up to the maximum $\alpha$ index on the FFT grid. The higher indices (corresponding to larger values of $q_\alpha$ that are only relevant in the core regions) are solved by simply projecting onto the $h_j^A(\mathbf{r})$ basis, i.e. setting $\mathbf{D} = \left(\mathbf{P}^{11}\right)^{-1} \mathbf{b}^1$ in the subspace of core region-only $\alpha$ indices.
This procedure is equivalent to assuming that the larger $q_\alpha$ contributions to $\theta_\alpha$ and $F_\beta$ are localized in the core region, which is reasonable because these contributions correspond to high densities and short length-scales for $\Phi_{\alpha\beta}(r)$ (Eq.\ \ref{eq:cider_kernel}).

\subsection{Construction of the Feature Augmentation Projectors}\label{app:feat_aug_const}

The feature augmentation projectors used in Eq.\ \ref{eq:feat_paw_aug} are constructed as
\begin{align}
    p_{i\leftarrow nlm}(\mathbf{r}) &= Y_{lm}(\hat{\mathbf{r}}) P(2n+l, r/r_0) \\
    P(n, x) &= \begin{cases} x^n (x^2-1)^2 & x < 1 \\ 0 & x \ge 1 \end{cases}.
\end{align}
The on-site function sets $\tilde{f}_i^A(\mathbf{r})$ and $\tilde{p}_j^A(\mathbf{r})$ are then computed as
\begin{align}
    \tilde{p}_i(\mathbf{r}) &= p_i(\mathbf{r}) f_\text{cut}(\mathbf{r}) \\
    \tilde{f}_i(\mathbf{r}) &= \sum_j \left(\mathbf{S}^{-1}\right)_{ij} p_j(\mathbf{r}) \\
    S_{ij} &= \int \dd[3]\mathbf{r} \, p_i(\mathbf{r}) p_j(\mathbf{r}) f_\text{cut}(\mathbf{r}) \\
    f_\text{cut}(\mathbf{r}) &= \begin{cases}
    \frac{1}{2} \left(1 + \cos \frac{\pi r}{r_c}\right) & r<r_c \\
    0 & r\ge r_c
    \end{cases}.
\end{align}
The construction localizes $\tilde{p}_j^A(\mathbf{r})$ inside the core region and satisfies the condition that $\braket{\tilde{f}_i^A}{\tilde{p}_j^A}=\delta_{ij}$.

\end{appendix}




\bibliography{references}

\end{document}


\preprint{APS/123-QED}

\title{Nonlocal Machine-Learned Exchange Functional for Molecules and Solids}

\author{Kyle Bystrom}
\email{kylebystrom@g.harvard.edu}
\affiliation{Harvard John A. Paulson School of Engineering and Applied Sciences}
\author{Boris Kozinsky}
\email{bkoz@g.harvard.edu}
\affiliation{Harvard John A. Paulson School of Engineering and Applied Sciences}
\affiliation{Robert Bosch LLC Research and Technology Center, Cambridge, MA, USA}

\date{\today}


\onecolumngrid

\setcounter{page}{1}

\MakeTitle{Supplementary Information for ``Nonlocal Machine-Learned Exchange Functional for Molecules and Solids''}{Kyle Bystrom, Boris Kozinsky}

\beginsupplement

\section{Table of all GMTKN55 Subset Error Metrics}

This section consists of a large table (shown below) with detailed GMTKN55 error metrics for each functional presented in this work. All entries contain the mean absolute deviation (MAD) relative to GMTKN55 reference values, except for the aggregate errors WT2 (WTMAD-2 as defined by Goerigk \emph{et al.}~\cite{Goerigk2017}) and MOM (mean of means, as defined in Section III A). The ``DTR tr,'' ``DTR val,'' and ``DTR test'' labels denote the train/test partition used to train NL-MGGA-DTR on GMTKN55, while ``tr/val'' and ``test'' refer to the train/test partition used for all other functionals. All units are in kcal/mol.

\begin{scriptsize}
\begin{ruledtabular}
\begin{tabular}{lrrrrrrrrr}
{} &  wB97M-V &  wB97X-V &  B97M-V &   PBE &  revPBE &  r$^2$SCAN &  PW6B95 &  PBE0 &  B3LYP \\
\colrule
W4-11         &     2.11 &     2.84 &    2.88 & 15.67 &    7.81 &    3.87 &    2.37 &  3.47 &   3.24 \\
G21EA         &     2.24 &     2.54 &    2.06 &  2.34 &    2.35 &    3.83 &    1.98 &  2.81 &   2.29 \\
G21IP         &     2.90 &     3.06 &    3.04 &  3.73 &    3.91 &    4.66 &    2.78 &  3.64 &   3.48 \\
DIPCS10       &     5.09 &     4.06 &    4.60 &  4.23 &    3.71 &    5.14 &    2.79 &  2.93 &   4.28 \\
PA26          &     1.50 &     2.54 &    3.24 &  1.82 &    2.79 &    2.40 &    2.04 &  2.29 &   1.91 \\
SIE4x4        &    10.62 &    11.38 &   16.15 & 23.42 &   23.07 &   18.02 &   15.23 & 14.18 &  17.76 \\
ALKBDE10      &     3.73 &     4.06 &    4.04 &  6.45 &    5.14 &    5.02 &    3.82 &  5.55 &   4.38 \\
YBDE18        &     2.98 &     2.06 &    4.41 &  4.92 &    3.93 &    3.37 &    2.97 &  0.96 &   4.96 \\
AL2X6         &     1.31 &     1.21 &    1.53 &  1.60 &    2.06 &    1.58 &    0.60 &  1.32 &   3.41 \\
HEAVYSB11     &     2.70 &     1.44 &    2.18 &  3.37 &    3.21 &    3.31 &    2.26 &  1.75 &   4.28 \\
NBPRC         &     0.94 &     1.49 &    1.78 &  2.37 &    2.11 &    1.52 &    1.41 &  3.07 &   2.07 \\
ALK8          &     2.50 &     0.92 &    2.65 &  2.26 &    2.33 &    2.87 &    2.43 &  2.41 &   4.92 \\
RC21          &     1.82 &     3.50 &    3.47 &  6.68 &    5.32 &    4.98 &    3.13 &  5.48 &   2.35 \\
G2RC          &     2.02 &     3.85 &    4.44 &  6.77 &    6.44 &    5.55 &    3.10 &  6.62 &   2.56 \\
BH76RC        &     0.86 &     1.50 &    2.02 &  4.20 &    2.86 &    2.98 &    1.19 &  2.18 &   1.98 \\
FH51          &     0.95 &     2.22 &    2.17 &  3.06 &    3.18 &    2.16 &    1.48 &  2.71 &   2.56 \\
TAUT15        &     0.33 &     0.71 &    0.89 &  1.80 &    1.56 &    1.57 &    0.92 &  1.13 &   1.13 \\
DC13          &     5.30 &     6.47 &    5.20 &  8.99 &    8.84 &    7.71 &    6.40 &  8.32 &   9.60 \\
MB16-43       &    15.35 &    33.98 &   36.76 & 24.88 &   28.40 &   14.12 &    9.30 & 15.75 &  29.74 \\
DARC          &     0.71 &     4.31 &    3.50 &  3.13 &    2.21 &    2.70 &    2.73 &  3.93 &   7.92 \\
RSE43         &     0.79 &     0.99 &    2.13 &  3.00 &    2.40 &    1.54 &    2.16 &  1.50 &   1.80 \\
BSR36         &     0.43 &     2.02 &    0.21 &  2.30 &    0.99 &    0.46 &    2.07 &  2.76 &   2.51 \\
CDIE20        &     0.59 &     0.62 &    1.62 &  1.69 &    1.51 &    1.61 &    1.11 &  1.28 &   1.07 \\
ISO34         &     0.62 &     1.16 &    1.45 &  1.48 &    1.31 &    1.29 &    1.19 &  1.43 &   1.77 \\
ISOL24        &     1.59 &     2.97 &    4.09 &  4.17 &    3.28 &    4.02 &    3.22 &  2.07 &   5.45 \\
C60ISO        &    11.82 &    13.65 &    4.89 & 11.35 &   12.08 &    5.65 &    2.30 &  2.32 &   2.74 \\
PArel         &     0.59 &     0.63 &    1.36 &  1.77 &    1.52 &    1.54 &    0.94 &  1.19 &   1.14 \\
BH76          &     1.33 &     1.68 &    4.16 &  9.28 &    8.16 &    6.99 &    3.31 &  4.19 &   5.04 \\
BHPERI        &     1.14 &     2.11 &    1.16 &  6.79 &    7.96 &    4.65 &    1.31 &  3.27 &   1.11 \\
BHDIV10       &     1.28 &     0.87 &    2.90 &  8.89 &    8.40 &    6.11 &    2.84 &  4.81 &   3.21 \\
INV24         &     1.24 &     1.63 &    1.87 &  2.95 &    2.57 &    2.12 &    1.28 &  1.13 &   1.02 \\
BHROT27       &     0.22 &     0.31 &    0.69 &  0.47 &    0.42 &    0.76 &    0.57 &  0.58 &   0.41 \\
PX13          &     1.74 &     2.43 &    1.15 & 11.78 &    9.07 &    8.81 &    1.53 &  6.38 &   4.10 \\
WCPT18        &     1.23 &     1.58 &    1.42 &  9.08 &    7.43 &    5.99 &    1.50 &  4.17 &   2.02 \\
RG18          &     0.06 &     0.07 &    0.07 &  0.23 &    0.08 &    0.16 &    0.22 &  0.07 &   0.17 \\
ADIM6         &     0.11 &     0.13 &    0.16 &  0.10 &    0.41 &    0.34 &    0.36 &  0.16 &   0.29 \\
S22           &     0.22 &     0.21 &    0.24 &  0.37 &    0.40 &    0.24 &    0.22 &  0.38 &   0.40 \\
S66           &     0.09 &     0.09 &    0.14 &  0.32 &    0.34 &    0.26 &    0.14 &  0.31 &   0.29 \\
HEAVY28       &     0.18 &     0.17 &    0.21 &  0.23 &    0.45 &    0.33 &    0.29 &  0.18 &   0.33 \\
WATER27       &     0.57 &     1.04 &    1.22 &  7.59 &    3.11 &    6.30 &    1.58 &  4.92 &   2.64 \\
CARBHB12      &     0.19 &     0.30 &    0.28 &  1.70 &    0.86 &    1.06 &    0.33 &  1.26 &   0.60 \\
PNICO23       &     0.23 &     0.19 &    0.21 &  1.24 &    0.94 &    0.76 &    0.31 &  0.82 &   0.28 \\
HAL59         &     0.28 &     0.30 &    0.46 &  1.19 &    0.78 &    0.80 &    0.35 &  0.55 &   0.44 \\
AHB21         &     0.31 &     0.34 &    0.57 &  1.19 &    0.91 &    1.27 &    0.35 &  1.28 &   0.37 \\
CHB6          &     0.78 &     0.73 &    0.90 &  0.53 &    1.94 &    0.52 &    1.14 &  0.98 &   0.96 \\
IL16          &     0.89 &     1.00 &    0.50 &  0.90 &    0.96 &    0.64 &    0.72 &  0.47 &   0.40 \\
IDISP         &     1.63 &     2.59 &    3.19 &  2.55 &    2.09 &    2.59 &    2.22 &  1.51 &   3.05 \\
ICONF         &     0.14 &     0.26 &    0.29 &  0.31 &    0.25 &    0.29 &    0.20 &  0.27 &   0.28 \\
ACONF         &     0.05 &     0.02 &    0.14 &  0.08 &    0.25 &    0.18 &    0.22 &  0.06 &   0.06 \\
Amino20x4     &     0.19 &     0.18 &    0.23 &  0.33 &    0.32 &    0.19 &    0.29 &  0.25 &   0.20 \\
PCONF21       &     0.58 &     0.29 &    0.80 &  1.01 &    0.69 &    0.41 &    0.53 &  0.75 &   0.35 \\
MCONF         &     0.35 &     0.23 &    0.34 &  0.48 &    0.46 &    0.45 &    0.39 &  0.27 &   0.24 \\
SCONF         &     0.13 &     0.16 &    0.21 &  0.82 &    0.78 &    0.51 &    0.26 &  0.28 &   0.29 \\
UPU23         &     0.45 &     0.53 &    0.45 &  0.53 &    0.50 &    0.39 &    0.72 &  0.54 &   0.60 \\
BUT14DIOL     &     0.04 &     0.04 &    0.19 &  0.50 &    0.28 &    0.21 &    0.28 &  0.25 &   0.43 \\
WT2-tr/val    &     2.43 &     3.02 &    4.94 &  9.84 &    8.13 &    7.00 &    4.60 &  4.62 &   5.85 \\
WT2-test      &     3.42 &     4.24 &    5.82 & 10.10 &    8.60 &    7.27 &    5.72 &  6.80 &   6.40 \\
WT2-DTR tr    &     2.53 &     3.06 &    3.69 &  4.18 &    4.43 &    3.88 &    2.83 &  2.87 &   3.72 \\
WT2-DTR val   &     2.62 &     3.49 &    3.59 &  5.33 &    4.99 &    4.33 &    3.20 &  2.96 &   3.36 \\
WT2-DTR test  &     3.27 &     4.06 &    6.04 & 11.41 &    9.43 &    7.98 &    6.01 &  6.97 &   6.89 \\
WT2-small     &     2.32 &     3.24 &    3.85 &  6.68 &    5.70 &    4.98 &    3.29 &  4.38 &   4.28 \\
WT2-big       &     4.02 &     6.68 &    9.42 & 12.08 &    9.78 &    8.26 &    8.30 &  8.34 &  10.11 \\
WT2-barrier   &     3.08 &     4.17 &    7.33 & 18.42 &   16.75 &   14.02 &    6.13 &  9.00 &   8.30 \\
WT2-inter nci &     2.66 &     2.76 &    3.48 &  8.89 &    7.22 &    7.01 &    4.69 &  5.44 &   5.25 \\
WT2-intra nci &     4.22 &     3.55 &    6.13 &  9.32 &    7.63 &    5.51 &    6.63 &  6.08 &   5.84 \\
WT2-all       &     3.13 &     3.88 &    5.56 & 10.02 &    8.46 &    7.19 &    5.39 &  6.16 &   6.24 \\
MOM-tr/val    &     3.45 &     5.23 &    6.24 &  7.79 &    7.17 &    5.16 &    3.58 &  4.46 &   6.04 \\
MOM-test      &     1.32 &     1.69 &    1.67 &  3.15 &    2.77 &    2.44 &    1.45 &  2.08 &   2.06 \\
MOM-small     &     2.77 &     3.10 &    3.71 &  5.76 &    5.04 &    4.47 &    3.16 &  3.93 &   4.29 \\
MOM-big       &     3.61 &     6.70 &    6.22 &  5.97 &    5.97 &    3.66 &    2.78 &  3.58 &   6.02 \\
MOM-barrier   &     1.17 &     1.52 &    1.91 &  7.03 &    6.29 &    5.06 &    1.76 &  3.50 &   2.42 \\
MOM-inter nci &     0.33 &     0.38 &    0.41 &  1.30 &    0.93 &    1.06 &    0.50 &  0.95 &   0.60 \\
MOM-intra nci &     0.40 &     0.48 &    0.65 &  0.73 &    0.63 &    0.58 &    0.57 &  0.46 &   0.61 \\
MOM-all       &     1.78 &     2.47 &    2.67 &  4.16 &    3.73 &    3.03 &    1.92 &  2.60 &   2.93 \\
MAD-tr/val    &     3.32 &     5.54 &    6.51 & 10.76 &    8.36 &    5.38 &    3.37 &  4.63 &   6.18 \\
MAD-test      &     0.88 &     1.24 &    1.30 &  2.49 &    2.11 &    1.88 &    1.18 &  1.70 &   1.55 \\
MAD-DTR tr    &     1.52 &     2.11 &    2.25 &  2.36 &    2.59 &    1.94 &    1.18 &  1.49 &   2.18 \\
MAD-DTR val   &     1.67 &     2.62 &    3.26 &  2.67 &    3.40 &    2.12 &    1.39 &  2.15 &   2.61 \\
MAD-DTR test  &     1.60 &     2.57 &    2.91 &  5.54 &    4.24 &    3.14 &    1.97 &  2.79 &   3.07 \\
MAD-small     &     2.35 &     2.98 &    3.35 &  8.10 &    5.59 &    4.17 &    2.78 &  3.74 &   3.64 \\
MAD-big       &     3.74 &     7.80 &    8.15 &  6.78 &    6.93 &    4.04 &    3.23 &  4.38 &   7.48 \\
MAD-barrier   &     1.16 &     1.54 &    2.47 &  7.06 &    6.37 &    5.19 &    2.10 &  3.36 &   2.93 \\
MAD-inter nci &     0.27 &     0.33 &    0.38 &  1.34 &    0.84 &    1.07 &    0.42 &  0.91 &   0.57 \\
MAD-intra nci &     0.25 &     0.24 &    0.35 &  0.51 &    0.43 &    0.34 &    0.38 &  0.33 &   0.36 \\
MAD-all       &     1.59 &     2.51 &    2.83 &  4.92 &    3.95 &    2.91 &    1.83 &  2.56 &   2.91 \\
\end{tabular}

\begin{tabular}{lrrrrr}
{} &  PBE0/SL-GGA &  PBE0/NL-GGA &  PBE0/SL-MGGA &  PW6B95/SL-MGGA &  PBE0/NL-MGGA \\
\colrule
W4-11         &         5.63 &         4.04 &          3.79 &            2.93 &          3.42 \\
G21EA         &         2.49 &         2.71 &          3.53 &            2.72 &          3.45 \\
G21IP         &         4.09 &         3.59 &          4.51 &            3.20 &          3.88 \\
DIPCS10       &         6.30 &         3.81 &          5.29 &            4.36 &          3.22 \\
PA26          &         2.14 &         3.11 &          4.43 &            4.51 &          2.83 \\
SIE4x4        &        22.26 &        17.66 &         17.07 &           18.56 &         15.83 \\
ALKBDE10      &         5.08 &         6.52 &          6.70 &            5.23 &          5.85 \\
YBDE18        &         4.59 &         2.92 &          3.17 &            5.59 &          1.04 \\
AL2X6         &         1.26 &         2.05 &          1.38 &            1.41 &          1.48 \\
HEAVYSB11     &         4.41 &         2.64 &          2.23 &            2.39 &          2.01 \\
NBPRC         &         2.55 &         3.31 &          2.17 &            2.03 &          2.97 \\
ALK8          &         3.83 &         3.17 &          5.06 &            5.15 &          3.22 \\
RC21          &         5.66 &         4.93 &          5.70 &            3.41 &          5.86 \\
G2RC          &         5.81 &         6.47 &          6.88 &            3.40 &          6.69 \\
BH76RC        &         3.03 &         2.40 &          2.78 &            1.90 &          2.27 \\
FH51          &         3.10 &         2.93 &          2.83 &            1.72 &          2.95 \\
TAUT15        &         1.68 &         2.31 &          1.64 &            1.41 &          1.49 \\
DC13          &         8.06 &         6.65 &          8.14 &            6.42 &          9.84 \\
MB16-43       &        22.43 &        17.04 &         16.53 &            9.76 &         17.29 \\
DARC          &         3.23 &         1.43 &          3.36 &            3.19 &          4.18 \\
RSE43         &         2.74 &         2.10 &          1.38 &            2.06 &          1.50 \\
BSR36         &         2.74 &        11.95 &          4.08 &            3.80 &          0.65 \\
CDIE20        &         1.45 &         1.51 &          1.67 &            1.56 &          1.53 \\
ISO34         &         1.46 &         1.79 &          1.53 &            1.24 &          1.49 \\
ISOL24        &         4.40 &         3.21 &          2.62 &            4.07 &          2.89 \\
C60ISO        &        12.46 &         5.53 &          6.79 &            7.42 &          5.35 \\
PArel         &         1.73 &         1.79 &          1.58 &            1.36 &          1.47 \\
BH76          &         9.84 &         6.34 &          6.28 &            5.72 &          4.68 \\
BHPERI        &         7.97 &         4.20 &          5.44 &            3.46 &          3.02 \\
BHDIV10       &         9.81 &         8.33 &          7.06 &            5.34 &          4.41 \\
INV24         &         3.53 &         2.99 &          2.47 &            2.35 &          2.16 \\
BHROT27       &         0.39 &         0.68 &          0.64 &            0.64 &          0.94 \\
PX13          &        12.58 &        12.44 &          9.36 &            4.85 &          1.71 \\
WCPT18        &         9.15 &         7.31 &          6.60 &            4.17 &          3.20 \\
RG18          &         0.35 &         0.17 &          0.14 &            0.21 &          0.12 \\
ADIM6         &         2.12 &         2.35 &          0.99 &            1.47 &          0.47 \\
S22           &         0.63 &         0.52 &          0.40 &            0.44 &          0.44 \\
S66           &         0.80 &         0.75 &          0.42 &            0.54 &          0.42 \\
HEAVY28       &         0.36 &         0.44 &          0.49 &            0.73 &          0.28 \\
WATER27       &        10.63 &         4.96 &          4.26 &            2.96 &          4.30 \\
CARBHB12      &         1.85 &         1.12 &          1.12 &            0.63 &          0.99 \\
PNICO23       &         1.45 &         1.12 &          0.60 &            0.79 &          0.75 \\
HAL59         &         1.43 &         1.25 &          0.82 &            0.82 &          0.73 \\
AHB21         &         1.52 &         0.99 &          1.32 &            0.65 &          0.99 \\
CHB6          &         0.93 &         1.09 &          1.03 &            1.15 &          0.84 \\
IL16          &         1.21 &         0.69 &          0.70 &            0.70 &          0.84 \\
IDISP         &         2.68 &         1.42 &          1.96 &            3.29 &          0.88 \\
ICONF         &         0.34 &         0.48 &          0.39 &            0.36 &          0.23 \\
ACONF         &         0.08 &         0.04 &          0.05 &            0.26 &          0.03 \\
Amino20x4     &         0.38 &         0.42 &          0.31 &            0.37 &          0.29 \\
PCONF21       &         1.22 &         0.86 &          0.66 &            0.50 &          0.56 \\
MCONF         &         0.40 &         0.70 &          0.35 &            0.43 &          0.41 \\
SCONF         &         0.78 &         0.50 &          0.24 &            0.27 &          0.16 \\
UPU23         &         0.61 &         1.24 &          0.51 &            0.51 &          0.38 \\
BUT14DIOL     &         0.51 &         0.16 &          0.18 &            0.40 &          0.25 \\
WT2-tr/val    &         9.80 &         6.56 &          6.41 &            6.39 &          5.26 \\
WT2-test      &        11.24 &        11.26 &          8.46 &            8.65 &          7.14 \\
WT2-DTR tr    &         5.25 &         5.73 &          4.50 &            4.55 &          3.71 \\
WT2-DTR val   &         5.91 &         6.26 &          4.75 &            4.95 &          3.41 \\
WT2-DTR test  &        12.17 &        10.88 &          8.68 &            8.82 &          7.32 \\
WT2-small     &         5.68 &         5.48 &          5.26 &            4.36 &          4.93 \\
WT2-big       &        11.66 &        15.24 &          9.98 &           10.45 &          8.17 \\
WT2-barrier   &        19.74 &        13.93 &         13.60 &           11.16 &          9.48 \\
WT2-inter nci &        13.19 &        10.89 &          8.05 &            9.76 &          6.74 \\
WT2-intra nci &        10.06 &         8.81 &          6.25 &            7.86 &          5.88 \\
WT2-all       &        10.82 &         9.88 &          7.86 &            7.99 &          6.59 \\
MOM-tr/val    &         6.66 &         5.34 &          5.41 &            4.59 &          4.89 \\
MOM-test      &         3.45 &         2.95 &          2.71 &            2.32 &          2.10 \\
MOM-small     &         5.11 &         4.51 &          4.85 &            4.24 &          4.35 \\
MOM-big       &         5.85 &         5.15 &          4.39 &            3.83 &          4.04 \\
MOM-barrier   &         7.61 &         6.04 &          5.41 &            3.79 &          2.87 \\
MOM-inter nci &         1.94 &         1.29 &          1.02 &            0.92 &          0.93 \\
MOM-intra nci &         0.78 &         0.65 &          0.52 &            0.71 &          0.35 \\
MOM-all       &         4.15 &         3.48 &          3.30 &            2.81 &          2.71 \\
MAD-tr/val    &         7.38 &         5.54 &          5.54 &            4.40 &          4.98 \\
MAD-test      &         2.70 &         2.50 &          2.11 &            1.79 &          1.70 \\
MAD-DTR tr    &         2.36 &         2.13 &          2.01 &            1.79 &          1.87 \\
MAD-DTR val   &         3.05 &         2.51 &          2.67 &            1.72 &          2.47 \\
MAD-DTR test  &         4.46 &         3.68 &          3.36 &            2.75 &          2.83 \\
MAD-small     &         5.00 &         4.22 &          4.43 &            3.62 &          4.00 \\
MAD-big       &         6.41 &         6.29 &          4.96 &            3.93 &          4.60 \\
MAD-barrier   &         7.61 &         5.45 &          5.19 &            4.07 &          3.27 \\
MAD-inter nci &         1.91 &         1.24 &          0.97 &            0.87 &          0.91 \\
MAD-intra nci &         0.54 &         0.51 &          0.35 &            0.45 &          0.31 \\
MAD-all       &         4.08 &         3.39 &          3.12 &            2.56 &          2.67 \\
\end{tabular}

\begin{tabular}{lrrrr}
{} &  PBE0/NL-MGGA-PBE &  PW6B95/NL-MGGA-PBE &  PBE0/NL-MGGA-DTR &  PW6B95/NL-MGGA-DTR \\
\colrule
W4-11         &              3.44 &                2.20 &              3.97 &                3.39 \\
G21EA         &              3.44 &                2.44 &              3.32 &                2.46 \\
G21IP         &              3.87 &                2.77 &              4.01 &                2.75 \\
DIPCS10       &              3.21 &                3.06 &              3.81 &                3.26 \\
PA26          &              2.78 &                2.61 &              3.60 &                3.54 \\
SIE4x4        &             15.81 &               17.15 &             17.17 &               18.66 \\
ALKBDE10      &              5.83 &                4.18 &              5.59 &                4.03 \\
YBDE18        &              1.04 &                2.76 &              1.55 &                3.64 \\
AL2X6         &              1.45 &                1.01 &              1.32 &                0.94 \\
HEAVYSB11     &              1.98 &                2.30 &              1.72 &                2.21 \\
NBPRC         &              2.95 &                1.80 &              3.09 &                1.60 \\
ALK8          &              3.18 &                2.79 &              2.92 &                2.42 \\
RC21          &              5.85 &                3.56 &              6.65 &                4.43 \\
G2RC          &              6.70 &                3.21 &              7.54 &                4.01 \\
BH76RC        &              2.27 &                1.26 &              2.77 &                1.59 \\
FH51          &              2.94 &                1.48 &              3.23 &                1.81 \\
TAUT15        &              1.49 &                1.43 &              1.52 &                1.38 \\
DC13          &              9.78 &                7.37 &              8.10 &                6.25 \\
MB16-43       &             17.29 &               10.41 &             18.88 &               12.00 \\
DARC          &              4.12 &                3.10 &              3.94 &                3.23 \\
RSE43         &              1.53 &                2.24 &              1.60 &                2.34 \\
BSR36         &              0.86 &                1.79 &              2.59 &                2.03 \\
CDIE20        &              1.52 &                1.38 &              1.68 &                1.56 \\
ISO34         &              1.47 &                1.36 &              1.53 &                1.27 \\
ISOL24        &              2.85 &                3.68 &              2.57 &                3.94 \\
C60ISO        &              5.37 &                5.83 &              5.23 &                5.60 \\
PArel         &              1.46 &                1.44 &              1.50 &                1.25 \\
BH76          &              4.67 &                3.98 &              5.53 &                4.91 \\
BHPERI        &              2.97 &                1.27 &              4.44 &                2.35 \\
BHDIV10       &              4.41 &                2.49 &              5.65 &                3.82 \\
INV24         &              2.17 &                1.92 &              2.01 &                1.82 \\
BHROT27       &              0.93 &                0.95 &              0.80 &                0.82 \\
PX13          &              1.67 &                4.91 &              4.90 &                1.44 \\
WCPT18        &              3.11 &                1.42 &              4.80 &                2.22 \\
RG18          &              0.13 &                0.29 &              0.08 &                0.17 \\
ADIM6         &              0.51 &                1.04 &              0.16 &                0.49 \\
S22           &              0.45 &                0.20 &              0.43 &                0.22 \\
S66           &              0.43 &                0.27 &              0.36 &                0.15 \\
HEAVY28       &              0.28 &                0.50 &              0.20 &                0.33 \\
WATER27       &              4.21 &                1.72 &              5.67 &                1.19 \\
CARBHB12      &              0.99 &                0.34 &              1.20 &                0.38 \\
PNICO23       &              0.75 &                0.32 &              0.79 &                0.35 \\
HAL59         &              0.74 &                0.51 &              0.87 &                0.62 \\
AHB21         &              0.99 &                0.30 &              1.18 &                0.36 \\
CHB6          &              0.84 &                0.97 &              0.99 &                1.14 \\
IL16          &              0.86 &                0.37 &              0.69 &                0.51 \\
IDISP         &              0.83 &                4.12 &              1.78 &                2.29 \\
ICONF         &              0.23 &                0.20 &              0.23 &                0.17 \\
ACONF         &              0.03 &                0.25 &              0.11 &                0.15 \\
Amino20x4     &              0.29 &                0.37 &              0.28 &                0.29 \\
PCONF21       &              0.57 &                0.62 &              0.74 &                0.45 \\
MCONF         &              0.40 &                0.53 &              0.39 &                0.50 \\
SCONF         &              0.16 &                0.17 &              0.27 &                0.24 \\
UPU23         &              0.37 &                0.50 &              0.39 &                0.57 \\
BUT14DIOL     &              0.26 &                0.24 &              0.34 &                0.18 \\
WT2-tr/val    &              5.28 &                5.39 &              5.83 &                5.56 \\
WT2-test      &              7.18 &                7.07 &              7.94 &                6.60 \\
WT2-DTR tr    &              3.72 &                4.02 &              3.36 &                3.23 \\
WT2-DTR val   &              3.40 &                3.39 &              3.27 &                3.00 \\
WT2-DTR test  &              7.37 &                7.25 &              8.31 &                7.07 \\
WT2-small     &              4.92 &                3.86 &              5.27 &                4.21 \\
WT2-big       &              8.27 &                9.47 &              9.49 &                9.79 \\
WT2-barrier   &              9.42 &                7.92 &             11.53 &                9.07 \\
WT2-inter nci &              6.85 &                6.83 &              6.48 &                5.27 \\
WT2-intra nci &              5.92 &                7.43 &              6.89 &                5.97 \\
WT2-all       &              6.62 &                6.58 &              7.32 &                6.29 \\
MOM-tr/val    &              4.88 &                3.97 &              5.32 &                4.47 \\
MOM-test      &              2.10 &                1.81 &              2.39 &                1.77 \\
MOM-small     &              4.33 &                3.52 &              4.55 &                3.80 \\
MOM-big       &              4.05 &                3.47 &              4.39 &                3.69 \\
MOM-barrier   &              2.85 &                2.42 &              4.02 &                2.48 \\
MOM-inter nci &              0.93 &                0.57 &              1.05 &                0.49 \\
MOM-intra nci &              0.35 &                0.78 &              0.50 &                0.54 \\
MOM-all       &              2.70 &                2.28 &              3.03 &                2.36 \\
MAD-tr/val    &              4.98 &                3.65 &              5.58 &                4.46 \\
MAD-test      &              1.70 &                1.39 &              1.97 &                1.42 \\
MAD-DTR tr    &              1.86 &                1.53 &              1.72 &                1.38 \\
MAD-DTR val   &              2.47 &                1.46 &              2.40 &                1.33 \\
MAD-DTR test  &              2.83 &                2.19 &              3.31 &                2.55 \\
MAD-small     &              4.00 &                2.95 &              4.36 &                3.53 \\
MAD-big       &              4.63 &                3.69 &              5.16 &                4.03 \\
MAD-barrier   &              3.25 &                2.69 &              4.19 &                3.08 \\
MAD-inter nci &              0.91 &                0.50 &              1.05 &                0.44 \\
MAD-intra nci &              0.31 &                0.44 &              0.37 &                0.36 \\
MAD-all       &              2.67 &                2.06 &              3.03 &                2.31 \\
\end{tabular}
\end{ruledtabular}
\end{scriptsize}

\section{Analysis of GMTKN55 Results with Different Error Averaging Metrics}

\subsection{Results Relative to PBE0} \label{sec:si_pbe0_err_metric}

\begin{figure}[htbp]
    \centering
    \includegraphics[width=0.75\columnwidth]{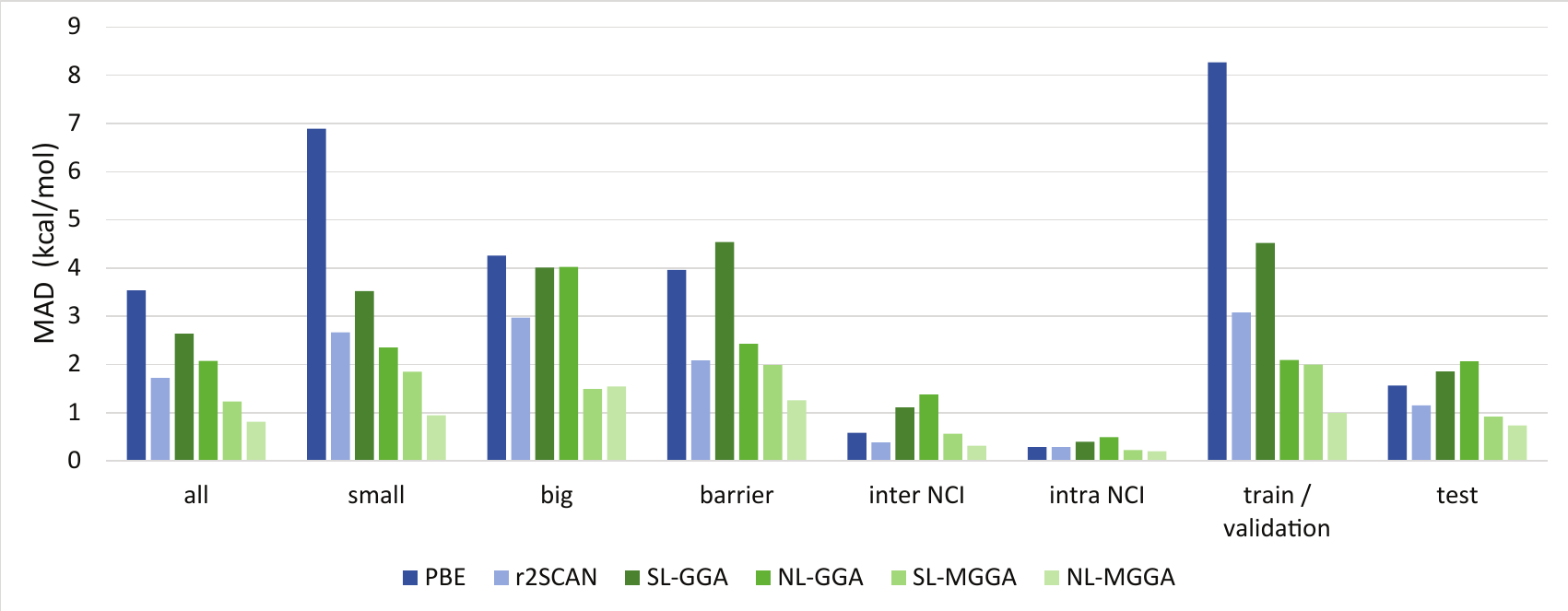}\\
    \includegraphics[width=0.75\columnwidth]{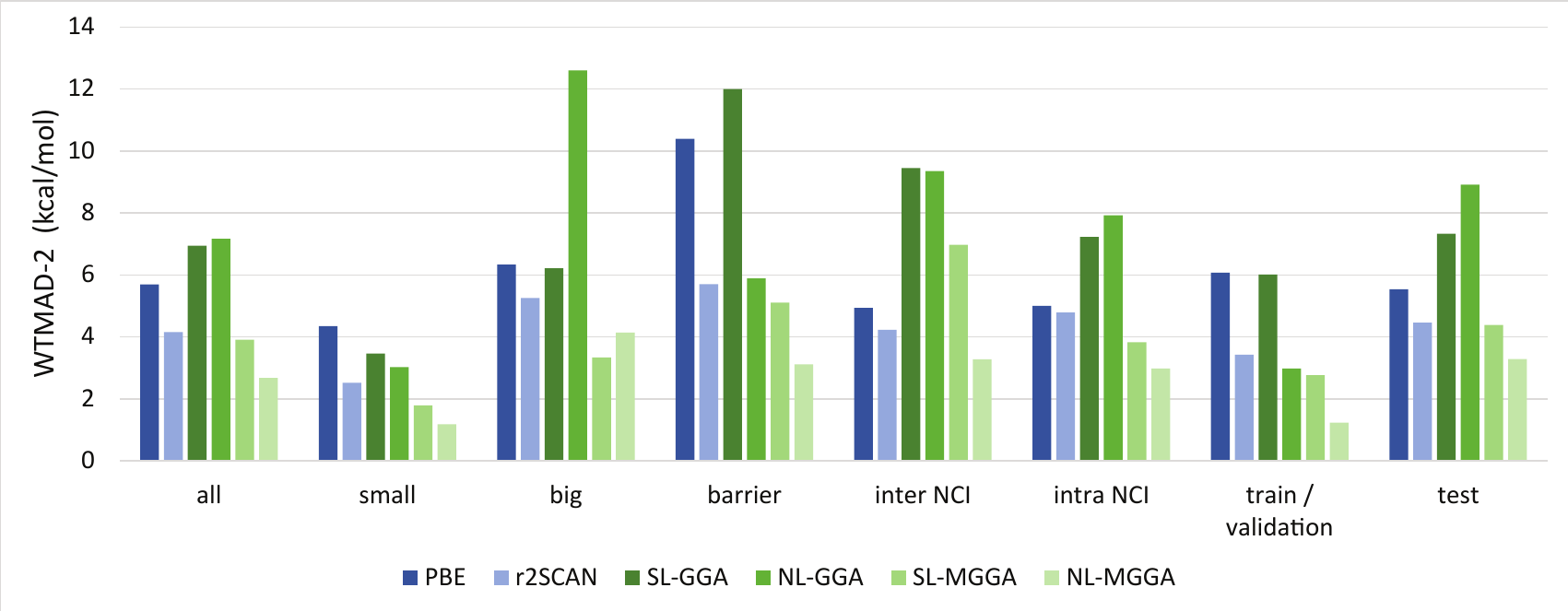}
    \caption{Deviations from PBE0 for semilocal and PBE0/CIDER functionals using the MAD (top) and WTMAD-2 (bottom) error metrics.}
    \label{fig:si_gmtkn55_vs_pbe0}
\end{figure}

As mentioned in Section III A, the analysis of functional accuracy can be sensitive to the averaging scheme for aggregate errors across the entire GMTKN55 database. In Fig.\ \ref{fig:si_gmtkn55_vs_pbe0}, we plot the average deviations relative to PBE0 of semilocal and ML functionals using two alternative metrics to the mean of means (MoM, i.e. the mean of the mean absolute deviations for the sub-databases) used in Section III A: mean absolute deviation (MAD, top) and weighted MAD (specifically, the WTMAD-2 metric used by Goerigk \emph{et al.}~\cite{Goerigk2017}). The general trends for these metrics are similar to those for the MoM; the SL-MGGA is more accurate than the SL-GGA and NL-GGA, and the NL-MGGA is more accurate than the SL-MGGA. However, the transferability issues suffered by the SL-GGA and NL-GGA appear even more severe when using the WTMAD-2 metric, and the NL-GGA is actually less accurate than the SL-GGA overall. The WTMAD-2 metric gives larger weight to sub-databases with smaller reaction energy magnitudes (e.g. noncovalent interactions are weighted more heavily than atomization energies), so this worse performance for WTMAD-2 suggests that the SL-GGA and NL-GGA cannot accurately describe the smaller density variations associated with these reactions. Notably, the SL-MGGA suffers a similar problem on the intermolecular NCI database; as a result, the NL-MGGA is the only functional that is closer to PBE0 than r$^2$SCAN for every sub-database, a testament to its robustness.

\subsection{Results Relative to Reference Values} \label{sec:si_ref_err_metric}

\begin{figure}[htbp]
    \includegraphics[width=0.75\columnwidth]{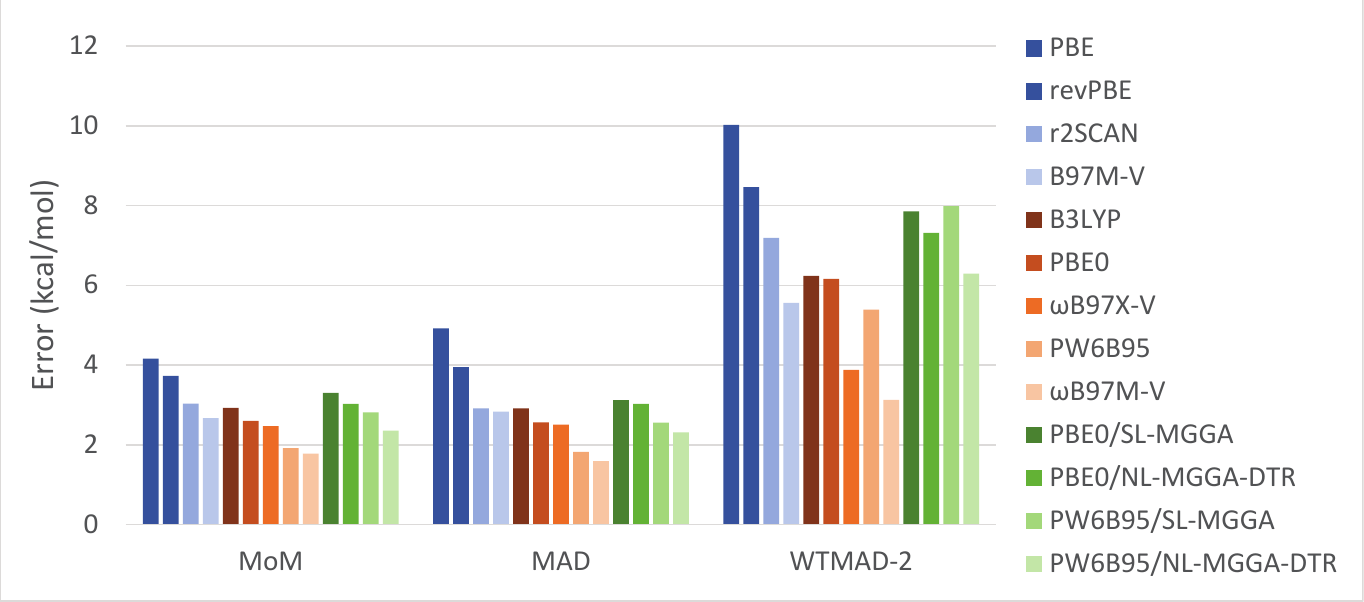}
    \caption{Errors of different semilocal, hybrid, and CIDER functionals on GMTKN55 with different error metrics.}
    \label{fig:SI_REF_GMTKN55}
\end{figure}

\begin{table}[htbp]
    \caption{WTMAD-2 (kcal/mol) for a few CIDER functionals. PW6B95/NL-MGGA was not evaluated on GMTKN55.}
    \begin{ruledtabular}
    \begin{tabular}{lrrrr}
        Model & SL-MGGA & NL-MGGA & NL-MGGA-PBE & NL-MGGA-DTR \\
        \colrule
        PBE0/CIDER & 7.86 & 6.59 & 6.62 & 7.32 \\
        PW6B95/CIDER & 7.99 & -- & 6.58 & 6.29 \\
    \end{tabular}
    \end{ruledtabular}
    \label{tab:comp_xc_si}
\end{table}

Figure \ref{fig:SI_REF_GMTKN55} shows the performance of the semilocal, hybrid, and CIDER functionals on the GMTKN55 database with the three different averaging metrics MoM, MAD, and WTMAD-2 discussed in the previous section. The same dispersion corrections are used as in Section III D. The results with MoM and MAD are very similar, while the performance of CIDER functionals relative to hybrid DFT is notably worse for the WTMAD-2. This again speaks to the difficulty of capturing small relative energies accurately for CIDER functionals, which will need to be a topic of future research. However, it is notable that the SL-MGGA models suffer from this problem more severely than the NL-MGGA-DTR; PW6B95/SL-MGGA has a worse WTMAD-2 than r$^2$SCAN, while PW6B95/NL-MGGA-DTR has a WTMAD-2 that is better than r$^2$SCAN and comparable to B3LYP and PBE0. As shown in Table \ref{tab:comp_xc_si}, the NL-MGGA and NL-MGGA-PBE models also significantly outperform the SL-MGGA models, so the worse performance of SL-MGGA is not an artefact of the different training set used for NL-MGGA-DTR. From this analysis, we conclude that introducing a more flexible feature set like that used for the NL-MGGA and its variants results in models that perform better for a variety of systems and therefore are less sensitive to a particular error metric.

\section{More data on band gaps} \label{sec:si_band_gaps}

\begin{table}[]
\caption{Error statistics for 453 band gaps from the database published by Borlido \emph{et al.}~\cite{Borlido2019}. The ML functionals use the PBE0($\alpha$)/CIDER formalism. $^a$This work. $^b$Borlido \emph{et al.}~\cite{Borlido2019}}
\begin{ruledtabular}
\begin{tabular}{cddddddd}
{}&
\multicolumn{1}{c}{PBE\footnote{This work.}}&
\multicolumn{1}{c}{r$^2$SCAN\footnotemark[1]}&
\multicolumn{1}{c}{SL-MGGA\footnotemark[1]}&
\multicolumn{2}{c}{NL-MGGA-DTR\footnotemark[1]}&
\multicolumn{1}{c}{HSE06\footnote{Borlido \emph{et al.}~\cite{Borlido2019}}}&
\multicolumn{1}{c}{PBE0\footnotemark[2]} \\
{} &  \centering {} &  {} &  \multicolumn{1}{c}{\textrm{$\alpha=0.25$}} &  \multicolumn{1}{c}{\textrm{$\alpha=0.25$}} &  \multicolumn{1}{c}{\textrm{$\alpha=0.35$}} &  {} &  {} \\
\colrule
ME (eV)   & -1.03 &   -0.68 &    -0.61 &        -0.51 &         -0.27 &  -0.10 &  0.47 \\
MAE (eV)  &  1.05 &    0.77 &     0.74 &         0.66 &          0.57 &   0.53 &  0.78 \\
RMSE (eV) &  1.46 &    1.15 &     1.10 &         1.04 &          0.93 &   0.86 &  0.99 \\
MRE       & -0.42 &   -0.22 &    -0.18 &        -0.13 &          0.02 &   0.10 &  0.54 \\
MARE      &  0.47 &    0.36 &     0.35 &         0.32 &          0.33 &   0.31 &  0.62 \\
RMSRE     &  0.55 &    0.52 &     0.51 &         0.51 &          0.62 &   0.67 &  1.29 \\
\end{tabular}
\end{ruledtabular}
\label{tab:band_gaps_si}
\end{table}

Table \ref{tab:band_gaps_si} contains the error metrics for semilocal, PBE0/CIDER, and hybrid functionals for all 453 systems in the database of Borlido \emph{et al.}~\cite{Borlido2019} that do not contain La, Th, or Yb. Note that for several systems, we used SG15 norm-conserving pseudopotentials~\cite{Hamann2013} as described in Appendix A11.

\section{Validation Results and Selected Hyperparameters} \label{sec:si_val_results}

Tables \ref{tab:val_sl_gga}-\ref{tab:val_nl_mgga} show the final validation results for each functional type, ordered by validation loss (Eq.\ A15) and labeled by internal project code name. Table \ref{tab:val_final_hparam} shows the final hyperparameters for the CIDER functionals presented in the main text. For all of these figures, see Appendix A8 for hyperparameter definitions.

\begin{table}[p]
\caption{SL-GGA validation results according to Eq.\ A15, with PBE provided as reference. Loss in kcal/mol.}
\begin{ruledtabular}
\begin{tabular}{llllr}
Code Name & Baseline &   $R_1$ &   $R_2$ &  Loss \\
\colrule
 SL\_GGA\_VC\_000 &  Chachiyo &    8 &  0.5 &   7.035046 \\
 SL\_GGA\_VC\_003 &  Chachiyo &    4 &  0.5 &   7.052635 \\
 SL\_GGA\_VP\_008 &       PBE &  0.1 &  0.5 &   7.079104 \\
 SL\_GGA\_VC\_009 &  Chachiyo &    2 &  0.5 &   7.092393 \\
           PBE &        -- &   -- &   -- &  10.868648 \\
 SL\_GGA\_VP\_000 &       PBE &  0.4 &  0.5 &  16.625300 \\
 SL\_GGA\_VP\_003 &       PBE &  0.2 &  0.5 &  38.890716 \\
\end{tabular}
\end{ruledtabular}
\label{tab:val_sl_gga}
\end{table}

\begin{table}[p]
\caption{NL-GGA validation results according to Eq.\ A15, with PBE provided as reference. Loss in kcal/mol.}
\begin{ruledtabular}
\begin{tabular}{lllllr}
Code Name & Baseline & Scheme &   $A$ &  $D$ &   Loss \\
\colrule
 NL\_GGA\_VC\_004 &  Chachiyo &     S1 &    1 &  0.5 &   3.904467 \\
 NL\_GGA\_VP\_004 &       PBE &     S1 &    1 &  0.5 &   3.928614 \\
 NL\_GGA\_VC\_003 &  Chachiyo &     S1 &    2 &  0.5 &   3.990208 \\
 NL\_GGA\_VP\_003 &       PBE &     S1 &    2 &  0.5 &   4.007883 \\
 NL\_GGA\_VP\_005 &       PBE &     S2 &    2 &  0.5 &   4.057075 \\
 NL\_GGA\_VC\_005 &  Chachiyo &     S2 &    2 &  0.5 &   4.061060 \\
 NL\_GGA\_VC\_008 &  Chachiyo &     S1 &  0.5 &  0.5 &   4.286322 \\
 NL\_GGA\_VP\_008 &       PBE &     S1 &  0.5 &  0.5 &   4.290059 \\
 NL\_GGA\_VP\_001 &       PBE &     S1 &    4 &  0.5 &   4.329216 \\
 NL\_GGA\_VC\_001 &  Chachiyo &     S1 &    4 &  0.5 &   4.342482 \\
 NL\_GGA\_VC\_009 &  Chachiyo &     S2 &  0.5 &    4 &   4.407261 \\
 NL\_GGA\_VP\_009 &       PBE &     S2 &  0.5 &    4 &   4.449937 \\
 NL\_GGA\_VP\_000 &       PBE &     S2 &    4 &  0.5 &   4.568491 \\
 NL\_GGA\_VC\_000 &  Chachiyo &     S2 &    4 &  0.5 &   4.581439 \\
 NL\_GGA\_VC\_006 &  Chachiyo &     S2 &  0.5 &  0.5 &   4.798940 \\
 NL\_GGA\_VP\_007 &       PBE &     S2 &  0.5 &  0.5 &   4.800655 \\
           PBE &        -- &     -- &   -- &   -- &  10.868648 \\
\end{tabular}
\end{ruledtabular}
\label{tab:val_nl_gga}
\end{table}

\begin{table}[p]
\caption{SL-MGGA validation results according to Eq.\ A15, with PBE provided as reference. Loss in kcal/mol.}
\begin{ruledtabular}
\begin{tabular}{llllr}
Code Name & Baseline &    $R_1$ &   $R_2$ &   Loss \\
\colrule
 SL\_MGGA\_VC\_006 &  Chachiyo &    64 &   0.5 &   3.218994 \\
 SL\_MGGA\_VP\_007 &       PBE &   3.2 &   0.5 &   3.220336 \\
 SL\_MGGA\_VP\_009 &       PBE &   1.6 &   0.5 &   3.286767 \\
 SL\_MGGA\_VC\_008 &  Chachiyo &    32 &   0.5 &   3.303520 \\
 SL\_MGGA\_VC\_009 &  Chachiyo &    16 &   0.5 &   3.347484 \\
 SL\_MGGA\_VP\_008 &       PBE &  0.05 &  0.25 &   3.724633 \\
 SL\_MGGA\_VP\_004 &       PBE &   0.1 &  0.25 &   3.771061 \\
 SL\_MGGA\_VP\_002 &       PBE &   0.2 &  0.25 &   3.872119 \\
 SL\_MGGA\_VC\_007 &  Chachiyo &     2 &  0.25 &   3.934145 \\
 SL\_MGGA\_VC\_003 &  Chachiyo &     4 &  0.25 &   3.995785 \\
 SL\_MGGA\_VP\_000 &       PBE &   0.4 &  0.25 &   4.085934 \\
 SL\_MGGA\_VC\_001 &  Chachiyo &     8 &  0.25 &   4.130502 \\
 SL\_MGGA\_VP\_001 &       PBE &   0.8 &  0.25 &   4.370939 \\
            PBE &        -- &    -- &    -- &  10.868648 \\
\end{tabular}
\end{ruledtabular}
\label{tab:val_sl_mgga}
\end{table}

\begin{table}[p]
\caption{NL-MGGA validation results according to Eq.\ A15, with PBE provided as reference. Loss in kcal/mol.}
\begin{ruledtabular}
\begin{tabular}{lllllr}
Code Name & Baseline & Scheme &   $A$ &  $D$ &   Loss \\
\colrule
 NL\_MGGA\_VC\_005 &  Chachiyo &     S1 &    1 &    1 &   1.895010 \\
 NL\_MGGA\_VP\_007 &       PBE &     S1 &    1 &    1 &   1.930917 \\
 NL\_MGGA\_VC\_006 &  Chachiyo &     S1 &  0.5 &    1 &   2.011226 \\
 NL\_MGGA\_VP\_005 &       PBE &     S1 &  0.5 &    1 &   2.039220 \\
 NL\_MGGA\_VC\_000 &  Chachiyo &     S2 &    1 &    1 &   2.039227 \\
 NL\_MGGA\_VP\_000 &       PBE &     S2 &    1 &    1 &   2.044209 \\
 NL\_MGGA\_VP\_003 &       PBE &     S2 &    1 &    2 &   2.063379 \\
 NL\_MGGA\_VC\_003 &  Chachiyo &     S2 &    1 &    2 &   2.063773 \\
 NL\_MGGA\_VP\_006 &       PBE &     S2 &    2 &  0.5 &   2.100342 \\
 NL\_MGGA\_VC\_007 &  Chachiyo &     S2 &    2 &  0.5 &   2.118458 \\
 NL\_MGGA\_VC\_001 &  Chachiyo &     S2 &  0.5 &    2 &   2.284749 \\
 NL\_MGGA\_VP\_002 &       PBE &     S2 &  0.5 &    2 &   2.313989 \\
 NL\_MGGA\_VP\_004 &       PBE &     S2 &  0.5 &    4 &   2.457190 \\
 NL\_MGGA\_VC\_004 &  Chachiyo &     S2 &  0.5 &    4 &   2.469842 \\
 NL\_MGGA\_VP\_009 &       PBE &     S1 &  0.5 &    4 &   2.504403 \\
 NL\_MGGA\_VC\_002 &  Chachiyo &     S2 &  0.5 &    1 &   2.584111 \\
 NL\_MGGA\_VP\_001 &       PBE &     S2 &  0.5 &    1 &   2.604407 \\
 NL\_MGGA\_VC\_009 &  Chachiyo &     S1 &  0.5 &    4 &   2.620007 \\
 NL\_MGGA\_VP\_008 &       PBE &     S2 &    1 &    4 &   2.620640 \\
 NL\_MGGA\_VC\_008 &  Chachiyo &     S2 &    1 &    4 &   2.665951 \\
            PBE &        -- &     -- &   -- &   -- &  10.868648 \\
\end{tabular}
\end{ruledtabular}
\label{tab:val_nl_mgga}
\end{table}

\begin{table}[p]
\caption{Names and hyperparameters for CIDER functionals presented in the main text.}
\begin{ruledtabular}
\begin{tabular}{cccccccc}
Name in Paper & Code Name &  Baseline & $R_1$ & $R_2$ & Scheme & $A$ &  $D$ \\
\colrule
SL-GGA &  SL\_GGA\_VC\_000 &  Chachiyo &    8 &  0.5 &   -- & -- & -- \\
NL-GGA & NL\_GGA\_VC\_004 &  Chachiyo &  20.0 & 1.0 & S1 & 1 &  0.5 \\
SL-MGGA & SL\_MGGA\_VC\_006 &  Chachiyo &    64 &   0.5 & -- & -- & -- \\
NL-MGGA & NL\_MGGA\_VC\_005 &  Chachiyo &  1.0  &  1.0  &   S1 &    1 &    1 \\
NL-MGGA-PBE & NL\_MGGA\_VP\_007 &  PBE & 0.05 & 1.0  &  S1 &    1 &    1 \\
NL-MGGA-DTR & NL\_MGGA\_VP\_DTR & PBE & 0.1 & 1.0 & S1 & 1 & 1 \\
\end{tabular}
\end{ruledtabular}
\label{tab:val_final_hparam}
\end{table}

\section{Density Accuracy}

The density error for a molecule $i$ is measured using numerical integrals as
\begin{equation}
    \Delta_i = \frac{\sum_\sigma \sum_g w_g^i |n^i_\sigma(\mathbf{r}_g) - n^{i,\text{ref}}_\sigma(\mathbf{r}_g^i)|}{\sum_\sigma \sum_g w_g^i n^{i,\text{ref}}_\sigma(\mathbf{r}_g^i)}, \label{eq:onemol_rho_error}
\end{equation}
where $w_g^i$ and $\mathbf{r}_g^i$ are the real-space integration weights and positions for molecule $i$. The aggregate percentage error for a sub-database is
\begin{equation}
    \bar{\Delta}_{\text{DB}} = \frac{100}{N_\text{DB}} \sum_{i\in \text{DB}} \Delta_i. \label{eq:rho_error_av}
\end{equation}
Table \ref{tab:rho_error} shows the density deviations from PBE0 for PBE, r$^2$SCAN, and the ML functionals over several subsets of GMTKN55. The NL-MGGA achieves the best match to PBE0 of the ML functionals, and this improvement is transferable to the test set. r$^2$SCAN, however, matches PBE0 more closely than the ML functionals for most subsets. This suggests there might still be some room for improvement in designing ML functionals with accurate densities. However, NL-MGGA yields smaller deviations from PBE0 densities than PBE does for every subset, so the density distributions get closer to those of PBE0 when the ML exchange is added.

\begin{table}[]
\caption{Average percentage density deviations from PBE0 (Eq.\ \ref{eq:rho_error_av} with PBE0 densities for $n^{i,\text{ref}}$ in Eq.\ \ref{eq:onemol_rho_error}) for semilocal and ML functionals. The ML functionals use the PBE0/CIDER surrogate hybrid form. Whether each sub-database was used for training or testing is indicated.}
\begin{ruledtabular}
\begin{tabular}{lrrrrrrr}
{} &   PBE &  r$^2$SCAN &  SL-GGA &  NL-GGA &  SL-MGGA &  NL-MGGA & NL-MGGA-DTR\footnote{Note that all sub-databases contain both train and test data for NL-MGGA-DTR, so the train-test partition does not apply for this functional.} \\
\colrule
W4-11 (Train)  &  0.86 &    0.32 &    0.80 &    0.72 &     0.49 &     0.34 & 0.32 \\
G21IP (Train)  &  0.72 &    0.41 &    0.75 &    0.69 &     0.89 &     0.65 & 0.59 \\
BH76 (Train)   &  1.02 &    0.44 &    1.06 &    0.77 &     0.68 &     0.52 & 0.51 \\
\colrule
G21EA (Test)   &  0.80 &    0.36 &    0.91 &    0.58 &     0.64 &     0.41 & 0.38 \\
FH51 (Test)    &  0.68 &    0.25 &    0.63 &    0.59 &     0.49 &     0.37 & 0.30 \\
DIPCS10 (Test) &  0.61 &    0.28 &    0.62 &    0.48 &     0.52 &     0.30 & 0.27 \\
INV24 (Test)   &  0.68 &    0.30 &    0.59 &    0.57 &     0.55 &     0.42 & 0.36 \\
PX13 (Test)    &  0.81 &    0.21 &    1.02 &    0.62 &     0.38 &     0.28 & 0.30 \\
\end{tabular}
\end{ruledtabular}
\label{tab:rho_error}
\end{table}

\section{Bond Lengths and Lattice Constants}

To assess the accuracy of CIDER functionals for structure prediction, we performed two benchmarks. First, we computed the bond lengths of six diatomic molecules in PySCF (Table \ref{tab:bond_lengths}) and compared the PBE0/CIDER bond length predictions to those of PBE0. For these calculations, the def2-QZVPPD basis set was used without density fitting, level 4 PySCF grids were used, and the internal nonlocal feature evaluation settings were $q_\text{max}=3000$, $\beta^\text{aux}=1.6$, $\lambda=1.6$, and $l_\text{max}=10$. The results are shown in Table \ref{tab:bond_lengths} and suggest that with increasing model complexity, the bond lengths more closely match PBE0. For reference, the atomization energy predictions for these molecules are shown in Table \ref{tab:atomization_energies} and demonstrate a similar trend. Note that the CIDER models are trained on these systems, so this benchmark serves only as a basic sanity test for CIDER forces, and not as a test of generalizability.

For the second benchmark, we computed the lattice constants of the materials in the SOL62 database~\cite{Zhang2018,Trepte2022} in GPAW (with the same settings as were used for the cohesive energy calculations, c.f. Sections A3 and A10) and compared the lattice constants to their experimental values. We separated the results into the train, validation, and test partitions to evaluate model generalizability. Also, we only computed the lattice constants for the PBE0/NL-MGGA-DTR CIDER functional since it was the most numerically stable. The results are shown in Table \ref{tab:lattice_constants}. No zero-point vibrational corrections were performed, but the mean absolute lattice constant deviation with and without these corrections for PBE is 0.01 \AA, significantly smaller than the typical deviations between the DFT predictions and experiment.

The MAE of PBE0/NL-MGGA-DTR compared to experiment is slightly smaller than that of PBE on the test set, but larger than that of r$^2$SCAN and HSE06. Due to unusually large errors on a few alkali and alkaline earth metals (i.e. K, Rb, and Ba), PBE0/NL-MGGA-DTR also has a slightly larger RMSE than PBE. The results suggest that overall, the ML functional can predict reasonable, GGA-accuracy structures for materials outside the training set, but more work is needed to achieve the accuracy of more advanced functionals like r$^2$SCAN and HSE06 for structure optimization. In the future, we will also need to investigate the cause of NL-MGGA-DTR severely underbinding the larger alkali/alkaline earth metals (even compared to HSE06).

\begin{table}[]
\scriptsize
\captionsetup{font=scriptsize}
\caption{Bond lengths for diatomic molecules in \AA. Note that these molecules are contained in the CIDER training set. All ML functionals use the PBE0/CIDER surrogate hybrid form.}
\begin{ruledtabular}
\begin{tabular}{lrrrrrrrr}
{} &   PBE &  r$^2$SCAN &  SL-GGA &  NL-GGA &  SL-MGGA &  NL-MGGA &  NL-MGGA-DTR &  PBE0 \\
\colrule
H2           & 0.750 &   0.741 &          0.750 &          0.747 &           0.746 &           0.742 &           0.744 & 0.745 \\
N2           & 1.102 &   1.093 &          1.104 &          1.096 &           1.095 &           1.089 &           1.094 & 1.089 \\
O2           & 1.218 &   1.206 &          1.219 &          1.206 &           1.205 &           1.194 &           1.200 & 1.192 \\
F2           & 1.413 &   1.397 &          1.417 &          1.403 &           1.390 &           1.385 &           1.385 & 1.375 \\
HF           & 0.930 &   0.921 &          0.930 &          0.922 &           0.919 &           0.916 &           0.918 & 0.918 \\
CO           & 1.135 &   1.126 &          1.139 &          1.129 &           1.130 &           1.123 &           1.127 & 1.122 \\
MAD vs. PBE0 & 0.018 &   0.008 &          0.020 &          0.010 &           0.007 &           0.003 &           0.005 & --- \\
\end{tabular}
\end{ruledtabular}
\label{tab:bond_lengths}
\end{table}

\begin{table}[]
\scriptsize
\captionsetup{font=scriptsize}
\caption{Atomization energies for diatomic molecules in kcal/mol. Note that these molecules are contained in the CIDER training set. All ML functionals use the PBE0/CIDER surrogate hybrid form.}
\begin{ruledtabular}
\begin{tabular}{lrrrrrrrr}
{} &   PBE &  r$^2$SCAN &  SL-GGA &  NL-GGA &  SL-MGGA &  NL-MGGA &  NL-MGGA-DTR &  PBE0 \\
\colrule
H2           & 104.7 &   107.7 &          104.1 &          105.5 &           104.7 &           104.3 &           104.6 & 104.4 \\
N2           & 243.6 &   220.3 &          223.8 &          222.5 &           221.3 &           223.1 &           223.2 & 225.6 \\
O2           & 143.7 &   129.9 &          129.5 &          121.8 &           125.7 &           122.8 &           126.8 & 124.6 \\
F2           &  52.8 &    38.7 &           45.7 &           35.8 &            34.7 &            34.9 &            38.1 &  35.1 \\
HF           & 142.1 &   138.3 &          138.0 &          139.4 &           136.6 &           137.0 &           138.2 & 136.9 \\
CO           & 268.9 &   255.6 &          256.0 &          253.0 &           254.8 &           254.8 &           255.6 & 255.6 \\
MAD vs. PBE0 &  12.3 &     3.1 &            3.2 &            2.1 &             1.2 &             0.9 &             1.5 &   --- \\
\end{tabular}
\end{ruledtabular}
\label{tab:atomization_energies}
\end{table}

\begin{table}[]
\captionsetup{font=scriptsize}
\scriptsize
\caption{Lattice constant predictions for the SOL62 database~\cite{Zhang2018,Trepte2022} using functionals including PBE0/NL-MGGA-DTR, in units of \AA. The midrule bars separate the training, validation, and test sets, in that order from top to bottom. Except for the mean absolute error reported separately for the train, validation, and test sets, error metrics are for the entire dataset.}
\begin{ruledtabular}
\begin{tabular}{lrrrrr}
{} &   PBE &  r$^2$SCAN &  PBE0/NL-MGGA-DTR &  HSE06~\cite{Zhang2018} &  Experimental \\
\colrule
Li   & 3.437 &   3.484 &             3.473 &  3.466 &         3.477 \\
C    & 3.572 &   3.560 &             3.567 &  3.549 &         3.567 \\
GaAs & 5.769 &   5.680 &             5.717 &  5.692 &         5.641 \\
LiF  & 4.096 &   4.032 &             4.059 &  4.003 &         4.010 \\
MgO  & 4.267 &   4.223 &             4.213 &  4.208 &         4.207 \\
Pt   & 3.968 &   3.950 &             3.952 &  3.949 &         3.916 \\
Pd   & 3.941 &   3.919 &             3.956 &  3.929 &         3.881 \\
TiC  & 4.340 &   4.337 &             4.350 &  4.307 &         4.330 \\
ZrC  & 4.703 &   4.709 &             4.707 &  4.700 &         4.696 \\
\colrule
Rb   & 5.672 &   5.745 &             5.834 &  5.756 &         5.585 \\
Na   & 4.200 &   4.225 &             4.242 &  4.215 &         4.225 \\
Sr   & 6.031 &   6.077 &             6.108 &  6.092 &         6.048 \\
MgS  & 5.228 &   5.207 &             5.192 &  5.207 &         5.202 \\
InP  & 5.954 &   5.906 &             5.923 &  5.907 &         5.861 \\
AlSb & 6.221 &   6.175 &             6.182 &  6.184 &         6.128 \\
LiCl & 5.156 &   5.133 &             5.096 &  5.107 &         5.106 \\
GaSb & 6.213 &   6.127 &             6.159 &  6.152 &         6.082 \\
Ge   & 5.767 &   5.679 &             5.708 &  5.702 &         5.652 \\
AlP  & 5.512 &   5.482 &             5.484 &  5.482 &         5.458 \\
InAs & 6.180 &   6.106 &             6.133 &  6.125 &         6.036 \\
Cu   & 3.644 &   3.592 &             3.669 &  3.633 &         3.603 \\
Ir   & 3.867 &   3.848 &             3.848 &  3.840 &         3.835 \\
Nb   & 3.305 &   3.303 &             3.324 &  3.305 &         3.296 \\
Rh   & 3.839 &   3.821 &             3.839 &  3.797 &         3.798 \\
Ta   & 3.319 &   3.311 &             3.318 &  3.328 &         3.301 \\
ZnSe & 5.732 &   5.656 &             5.712 &  5.700 &         5.668 \\
CdTe & 6.599 &   6.545 &             6.585 &  6.574 &         6.480 \\
HfC  & 4.659 &   4.649 &             4.645 &  4.647 &         4.638 \\
TiN  & 4.255 &   4.244 &             4.250 &  4.213 &         4.239 \\
ZnTe & 6.169 &   6.097 &             6.149 &  6.152 &         6.104 \\
ZrN  & 4.587 &   4.584 &             4.572 &  4.575 &         4.585 \\
\colrule
Ba   & 5.108 &   5.172 &             5.221 &  5.087 &         5.007 \\
Ca   & 5.512 &   5.549 &             5.580 &  5.582 &         5.565 \\
Al   & 4.041 &   3.991 &             4.016 &  4.020 &         4.032 \\
K    & 5.286 &   5.351 &             5.406 &  5.313 &         5.225 \\
GaN  & 4.603 &   4.542 &             4.561 &  4.504 &         4.520 \\
GaP  & 5.545 &   5.479 &             5.509 &  5.467 &         5.442 \\
BN   & 3.626 &   3.611 &             3.616 &  3.601 &         3.607 \\
Si   & 5.475 &   5.450 &             5.456 &  5.444 &         5.430 \\
SiC  & 4.389 &   4.365 &             4.372 &  4.353 &         4.358 \\
AlAs & 5.736 &   5.683 &             5.694 &  5.694 &         5.652 \\
BP   & 4.551 &   4.539 &             4.549 &  4.530 &         4.538 \\
NaF  & 4.701 &   4.611 &             4.646 &  4.635 &         4.609 \\
InSb & 6.620 &   6.548 &             6.577 &  6.561 &         6.469 \\
NaCl & 5.696 &   5.616 &             5.622 &  5.645 &         5.595 \\
BAs  & 4.815 &   4.787 &             4.801 &  4.781 &         4.777 \\
Sn   & 6.641 &   6.560 &             6.591 &  6.580 &         6.482 \\
Ni   & 3.530 &   3.498 &             3.554 &  3.512 &         3.513 \\
Fe   & 2.844 &   2.874 &             2.926 &  2.897 &         2.861 \\
Ag   & 4.147 &   4.111 &             4.192 &  4.153 &         4.069 \\
Mo   & 3.160 &   3.155 &             3.165 &  3.145 &         3.144 \\
W    & 3.183 &   3.173 &             3.175 &  3.173 &         3.162 \\
Au   & 4.177 &   4.148 &             4.169 &  4.153 &         4.065 \\
V    & 3.003 &   2.998 &             3.022 &  2.970 &         3.028 \\
HfN  & 4.552 &   4.532 &             4.522 &  4.531 &         4.519 \\
VN   & 4.132 &   4.115 &             4.120 &  4.073 &         4.135 \\
VC   & 4.168 &   4.156 &             4.164 &  4.112 &         4.160 \\
CdS  & 5.920 &   5.877 &             5.914 &  5.885 &         5.818 \\
NbN  & 4.412 &   4.408 &             4.397 &  4.392 &         4.379 \\
CdSe & 6.188 &   6.130 &             6.173 &  6.147 &         6.050 \\
NbC  & 4.478 &   4.478 &             4.472 &  4.461 &         4.470 \\
ZnS  & 5.445 &   5.380 &             5.435 &  5.419 &         5.404 \\
\colrule
Train MAE  & 0.050 &  0.020 &            0.031 & 0.022 &        \multicolumn{1}{c}{-} \\
Val. MAE   & 0.057 &  0.029 &            0.050 & 0.039 &        \multicolumn{1}{c}{-} \\
Test MAE   & 0.058 &  0.036 &            0.052 & 0.038 &        \multicolumn{1}{c}{-} \\
ME         & 0.051 &  0.025 &            0.046 & 0.025 &        \multicolumn{1}{c}{-} \\
MAE        & 0.057 &  0.031 &            0.048 & 0.036 &        \multicolumn{1}{c}{-} \\
RMSE       & 0.071 &  0.046 &            0.070 & 0.050 &        \multicolumn{1}{c}{-} \\
MAXE\footnote{Maximum absolute error.}       & 0.159 &  0.165 &            0.249 & 0.171 &    \multicolumn{1}{c}{-} \\
MAXE Syst.\footnote{System with the maximum absolute error.} &    Sn &     Ba &               Rb &    Rb &    \multicolumn{1}{c}{-} \\
\end{tabular}
\end{ruledtabular}
\label{tab:lattice_constants}
\end{table}

\section{More Detailed Defect Transition Level Data}\label{sec:si_levels}

This section provides more details on the defect transition level data. The bulk lattice constant and band gap, determined by the methods in Section A5, are tabulated in Table \ref{tab:si_bulk_si}. The defect transition levels themselves are shown in Table \ref{tab:defect_level_table}.

\begin{table}[]
    \caption{Bulk data for silicon for different functionals. The silicon experimental lattice constant was taken from Ref.~\cite{Zhang2018}.}
    \begin{ruledtabular}
    \begin{tabular}{cdd}
        Functional&
        \multicolumn{1}{c}{Lattice Constant (\AA)}&
        \multicolumn{1}{c}{Band Gap (eV)} \\
        \colrule
        PBE & 5.475 & 0.61\\
        PBE0/NL-MGGA-DTR & 5.457 & 1.08\\
        PBE0(0.3)/NL-MGGA-DTR & 5.454 & 1.17\\
        Experimental & 5.430 & 1.17\\
    \end{tabular}
    \end{ruledtabular}
    \label{tab:si_bulk_si}
\end{table}

\begin{table}[]
\caption{Transition levels for several defects in silicon for different functionals and correction schemes. ``Fr.'' is the Freysoldt correction scheme, while ``Pot.'' is the potential alignment-only scheme. All PBE0($\alpha$)/CIDER calculations use the NL-MGGA-DTR exchange functional. CIDER refers to PBE0(0.3)/NL-MGGA-DTR. The experimental references are from the following sources: vacancy~\cite{Watkins1980}, boron~\cite{Fischer1983}, phosphorus~\cite{Jagannath1981}, copper~\cite{PhysRevB.65.165203}, and sulfur~\cite{Schulz2002,Deak2010}.}
\begin{ruledtabular}
\begin{tabular}{ccrrrrrrr}
Defect & Level &  \multicolumn{2}{c}{PBE} & \multicolumn{4}{c}{PBE0($\alpha$)/CIDER} & Expt. \\
{} & {} & {}  & {}   & \multicolumn{2}{c}{$\alpha=0.25$} & \multicolumn{2}{c}{$\alpha=0.30$} & {} \\
{} & {} & Fr. & Pot. & Fr. & Pot. & Fr. & Pot. & {} \\
\colrule
   vac &  ++/+ &   -0.09 &    0.06 &       -0.01 &        0.14 &       -0.01 &        0.14 & 0.13 \\
   vac &   +/0 &   -0.04 &    0.01 &        0.05 &        0.10 &        0.08 &        0.13 & 0.05 \\
   vac &  ++/0 &   -0.06 &    0.04 &        0.02 &        0.12 &        0.03 &        0.13 & 0.09 \\
   vac &   0/- &    0.46 &    0.41 &        0.71 &        0.66 &        0.76 &        0.71 &  --- \\
   vac &  -/-- &    0.50 &    0.35 &        0.84 &        0.69 &        0.92 &        0.77 &  --- \\
   vac &  0/-- &    0.48 &    0.38 &        0.77 &        0.68 &        0.84 &        0.74 &  --- \\
     P &   +/0 &    0.53 &    0.58 &        0.99 &        1.03 &        1.08 &        1.13 & 1.12 \\
     B &   0/- &    0.07 &    0.02 &        0.07 &        0.02 &        0.08 &        0.03 & 0.04 \\
    Cu &   +/0 &    0.04 &    0.09 &        0.03 &        0.07 &        0.02 &        0.07 & 0.21 \\
    Cu &   0/- &    0.24 &    0.19 &        0.53 &        0.48 &        0.58 &        0.53 & 0.48 \\
    Cu &  -/-- &    0.66 &    0.52 &        0.91 &        0.76 &        0.97 &        0.82 & 1.00 \\
     S &  ++/+ &    0.12 &    0.27 &        0.47 &        0.61 &        0.54 &        0.69 & 0.58 \\
     S &   +/0 &    0.41 &    0.46 &        0.83 &        0.88 &        0.92 &        0.97 & 0.85 \\
\end{tabular}
\end{ruledtabular}
\label{tab:defect_level_table}
\end{table}

\subsection{The Cu +/0 Transition Level} \label{sec:si_cu_tl}

\begin{table}[]
    \caption{+/0 transition level of the copper substitutional defect in silicon (excluding finite-size effects), computed using 64-atom supercells and different k-point meshes. CIDER refers to PBE0(0.3)/NL-MGGA-DTR.}
    \begin{ruledtabular}
    \begin{tabular}{ccr}
        k-point mesh & Functional & $E(+/0)$ (eV) \\
        \colrule
        $2 \times 2 \times 2$ $\Gamma$ & HSE06 & 0.19 \\
        $2 \times 2 \times 2$ $\Gamma$ & CIDER & 0.15 \\
        $2 \times 2 \times 2$ MP & HSE06 & 0.25 \\
        $2 \times 2 \times 2$ MP & CIDER & 0.23 \\
        $4 \times 4 \times 4$ $\Gamma$ & HSE06 & 0.10 \\
        $4 \times 4 \times 4$ $\Gamma$ & CIDER & 0.14 \\
    \end{tabular}
    \end{ruledtabular}
    \label{tab:si_cu_conv}
\end{table}

As discussed in the main text, the copper +1/0 transition level computed with PBE0(0.3)/NL-MGGA-DTR is not converged with the supercell size and k-point mesh used with HSE06~\cite{Krukau2006} by Sharan \emph{et al.}~\cite{Sharan2017}. To further illustrate this point, Table~\ref{tab:si_cu_conv} shows the  +1/0 energy transition level for copper $E(+/0)$ for HSE06 and PBE0(0.3)/NL-MGGA-DTR for several k-point meshes. For this comparison, finite-size effects are ignored, so
\begin{equation}
    E(+/0) = E_X^{0}-E_X^{1}-\epsilon_{\text{VBM}} \label{eq:si_cu_tl}
\end{equation}
(with $X$ being the defect supercell). For consistency, all calculations were performed with a geometry relaxed with HSE06. The lattice constant was optimized with HSE06 with an $8\times 8\times 8$ $\Gamma$-centered k-point mesh, and the defect geometries were optimized with HSE06 with the $2\times 2\times 2$ $\Gamma$-centered k-point mesh. No restrictions were placed on the magnetic moment for structural relaxation, but for consistency, all static calculations for computing energy level differences were computed using fixed total magnetic moments of $S=1/2$ for the neutral state and $S=0$ for the +1 state. These are the ground-state spins as determined by Sharan \emph{et al.}~\cite{Sharan2017}, though it is worth noting that for the 512-atom supercell and $2\times 2\times 2$ $\Gamma$-centered k-point mesh, the $S=3/2$ spin state is favored over $S=1/2$. The HSE06 calculations were performed with VASP~\cite{Kresse1993}, while the CIDER calculations were performed with GPAW~\cite{Mortensen2005}.

The primary takeaway from Table \ref{tab:si_cu_conv} is that neither the CIDER nor the HSE06 calculations are converged with the $2 \times 2 \times 2$ k-point meshes. The most significant deviation is between the $2\times 2\times 2$ MP and $4\times 4\times 4$ $\Gamma$-centered mesh for HSE06, for which the difference is 0.15 eV. In addition, PBE0(0.3)/NL-MGGA-DTR and HSE06 are in good agreement for all calculations for a given k-point mesh, with deviations of 0.02-0.04 eV.

\section{Numerical Precision of Nonlocal Feature Implementation}

One of the key details in determining the practical usefulness of a functional is its numerical precision, or how accurately the approximate numerical evaluation of the model XC energy matches the hypothetical ``exact'' evaluation of the model. This section covers a few useful metrics of numerical precision for the NL-MGGA-DTR functional used in the main text.

\subsection{Comparison of Numerical Feature Integration to Fast Feature Evaluation in PySCF} \label{sec:si_num_aux}

Section IIE describes how the evaluation of nonlocal density features can be accelerated within all-electron, localized orbital codes for molecular systems. One might be interested in how accurately this approximation reproduces the simple numerical integration approach, which is much slower:
\begin{align}
    G_i(\mathbf{r}_1) &= \left(\frac{B_i+B_0}{2}\right)^{3/2} \int \dd[3]\mathbf{r}_2 \, \Phi\left(a(\mathbf{r}_2), b_i(\mathbf{r}_1), r_{12}\right) n(\mathbf{r}_2) \\
    &\approx \left(\frac{B_i+B_0}{2}\right)^{3/2} \sum_g w_g \Phi\left(a(\mathbf{r}_g), b_i(\mathbf{r}_1), |\mathbf{r}_1-\mathbf{r}_g|\right) n(\mathbf{r}_g) \label{eq:numerical_nldf}
\end{align}
Table \ref{tab:si_numerical1} displays the deviation in atomization energy obtained for some small molecules with the PBE0/NL-MGGA-DTR functional, when the functional is evaluated from simple numerical integration (Eq.~\ref{eq:numerical_nldf}, ``Numerical'') vs. the auxiliary approximation of Section IIE (``Auxiliary''). The deviations for molecules with first and second-row atoms (the first 8 molecules) are all less than 1 meV, and only 0.3 meV on average. For molecules containing larger atoms (the last 5 molecules, with elements from row 3 or higher), the errors are a bit higher, but still less than 2 meV in all cases. All calculations were performed in PySCF with the def2-QZVPPD basis set, using level 3 grids with the default NWChem pruning for the auxiliary basis approach and level 3 grids \emph{without} pruning (necessary for numerical precision in the core region) for the numerical integration. See the main text Appendix A10 for details on the CIDER auxiliary expansion settings.

\begin{table}[]
    \centering
    \caption{Atomization energies of small molecules predicted using the PBE0/NL-MGGA-DTR functional with numerical integration or auxiliary expansion used to evaluate the nonlocal features (in eV). MAD is mean absolute deviation, and RMSD is root mean square deviation.}
    \begin{ruledtabular}
    \begin{tabular}{lrrr}
     & Numerical & Auxiliary & Difference \\
    \colrule
    H2 & 4.53747 & 4.53747 & 0.00000 \\
    HF & 5.98539 & 5.98537 & -0.00002 \\
    F2 & 1.63125 & 1.63052 & -0.00074 \\
    O2 & 5.42328 & 5.42289 & -0.00040 \\
    N2 & 9.58508 & 9.58557 & 0.00049 \\
    NH3 & 12.87148 & 12.87183 & 0.00035 \\
    CH4 & 18.16628 & 18.16621 & -0.00006 \\
    H2O & 9.93861 & 9.93869 & 0.00008 \\
    \colrule
    SiH4 & 13.70708 & 13.70883 & 0.00175 \\
    H2Se2 & 9.19999 & 9.19967 & -0.00032 \\
    Te2Me2 & 33.99429 & 33.99250 & -0.00179 \\
    As2Me4 & 66.27788 & 66.27955 & 0.00166 \\
    Sb2Me4 & 64.69086 & 64.69097 & 0.00011 \\
    \colrule
    MAD (H2-H2O) & --- & --- & 0.00027 \\
    MAD (SiH4-Sb2Me4) & --- & --- & 0.00113 \\
    MAD (All) & --- & --- & 0.00060 \\
    RMSD (H2-H2O) & --- & --- & 0.00037 \\
    RMSD (SiH4-Sb2Me4) & --- & --- & 0.00135 \\
    RMSD (All) & --- & --- & 0.00089 \\
    \end{tabular}
    \end{ruledtabular}
    \label{tab:si_numerical1}
\end{table}

\subsection{Comparison of Localized-Orbital All-Electron Implementation and PAW Plane-wave Implementation}

Having seen that the auxiliary approximation of Section IIE agrees quite accurately with the direct numerical integration of the nonlocal features, we now turn our attention to the PAW plane-wave implementation of the nonlocal CIDER features in Section IIF. Due to the approximations inherent in PAW~\cite{Blochl1994}, as well as the differences between plane-wave and Gaussian-type orbital basis sets, results from Gaussian-type orbital and plane-wave calculations are not expected to agree completely, but ideally the deviations between these methods should be no larger when using CIDER functionals than when using standard functionals. In Table~\ref{tab:si_numerical2}, we see that this is indeed the case for PBE0/NL-MGGA-DTR. For molecules with smaller atoms (rows 1 and 2), there is an average deviation between PySCF (Gaussian-type orbital DFT) and GPAW (plane-wave DFT with the PAW method) of 0.02 eV for both PBE and PBE0/NL-MGGA-DTR. The PAW potentials for GPAW were generated with the PBE functional, so it is reassuring that using a CIDER functional instead does not increase the deviation between the all-electron and PAW calculations significantly. For molecules with larger atoms, the disagreement is larger (around 0.1 eV on average), but again no larger when using CIDER functionals than when using PBE. There are two possible causes of the larger deviations seen for the molecules with larger atoms. First, these molecules are simply larger and therefore have more bonds being broken during atomization, so the same error per bond will yield a larger total error. Second, the def2-QZVPPD basis uses effective core potentials for atoms larger than Kr rather than performing all-electron calculations, which adds an additional source of disagreement between GPAW and PySCF.

For these calculations, the same PySCF settings were used as in Section~\ref{sec:si_num_aux}. For GPAW, an energy cutoff of 800 eV and grid spacing of 0.108 \AA\ were used to store the wave functions, and the XC energy was integrated on a 2-times denser grid than the wave functions. 5 \AA\ of vacuum was added on all sides of each system to avoid spurious periodic image interactions. The same CIDER-specific settings were used as in the main text, Appendix A10.

\begin{table}[]
    \centering
    \caption{Atomization energies of small molecules predicted using the PBE and PBE0/NL-MGGA-DTR functionals in the PySCF and GPAW codes (in eV). MAD is mean absolute deviation, and RMSD is root mean square deviation.}
    \begin{ruledtabular}
    \begin{tabular}{lrrrrrr}
     & PBE, PySCF & PBE, GPAW & PBE, Diff & CIDER, PySCF & CIDER, GPAW & CIDER, Diff \\
    \colrule
    H2 & 4.539 & 4.529 & -0.010 & 4.537 & 4.527 & -0.010 \\
    HF & 6.165 & 6.139 & -0.026 & 5.985 & 5.956 & -0.029 \\
    F2 & 2.289 & 2.311 & 0.022 & 1.631 & 1.666 & 0.035 \\
    O2 & 6.206 & 6.228 & 0.022 & 5.423 & 5.445 & 0.022 \\
    N2 & 10.514 & 10.539 & 0.026 & 9.586 & 9.595 & 0.009 \\
    NH3 & 13.104 & 13.091 & -0.013 & 12.872 & 12.862 & -0.010 \\
    CH4 & 18.215 & 18.194 & -0.021 & 18.166 & 18.147 & -0.019 \\
    H2O & 10.169 & 10.148 & -0.021 & 9.939 & 9.918 & -0.021 \\
    \colrule
    SiH4 & 13.605 & 13.535 & -0.070 & 13.709 & 13.648 & -0.060 \\
    H2Se2 & 9.474 & 9.369 & -0.105 & 9.200 & 9.132 & -0.068 \\
    Te2Me2 & 34.444 & 34.486 & 0.042 & 33.992 & 34.077 & 0.085 \\
    As2Me4 & 66.791 & 66.525 & -0.266 & 66.280 & 66.063 & -0.216 \\
    Sb2Me4 & 65.278 & 65.247 & -0.031 & 64.691 & 64.692 & 0.001 \\
    \colrule
    MAD (H2-H2O) & --- & --- & 0.020 & --- & --- & 0.020 \\
    MAD (SiH4-Sb2Me4) & --- & --- & 0.103 & --- & --- & 0.086 \\
    MAD (All) & --- & --- & 0.052 & --- & --- & 0.045 \\
    RMSD (H2-H2O) & --- & --- & 0.021 & --- & --- & 0.021 \\
    RMSD (SiH4-Sb2Me4) & --- & --- & 0.134 & --- & --- & 0.111 \\
    RMSD (All) & --- & --- & 0.085 & --- & --- & 0.071 \\
    \end{tabular}
    \end{ruledtabular}
    \label{tab:si_numerical2}
\end{table}

\subsection{Convergence of Total Energy with Respect to Grid Size}

Having verified that both implementations of the nonlocal density features are numerically accurate, we now turn to the question of energy convergence with respect to the grid size used to integrate the XC energy. Ideally, an XC functional should be sufficiently smooth that numerical integrals of the functional over coarse grids accurately match more precise numerical integrals using denser grids. Being able to use a coarser grid can significantly decrease the cost of a DFT calculation. As such, we are interested in how quickly CIDER functionals converge with respect to grid size compared to other functionals.

\begin{figure}
    \centering
    \includegraphics[width=0.48\textwidth]{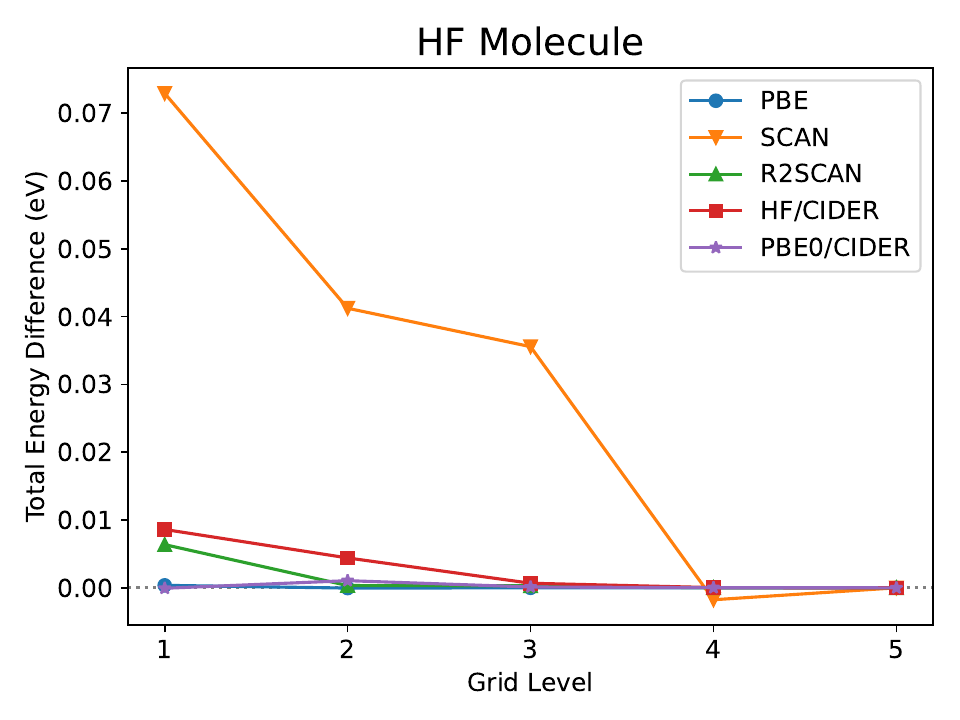}
    \includegraphics[width=0.48\textwidth]{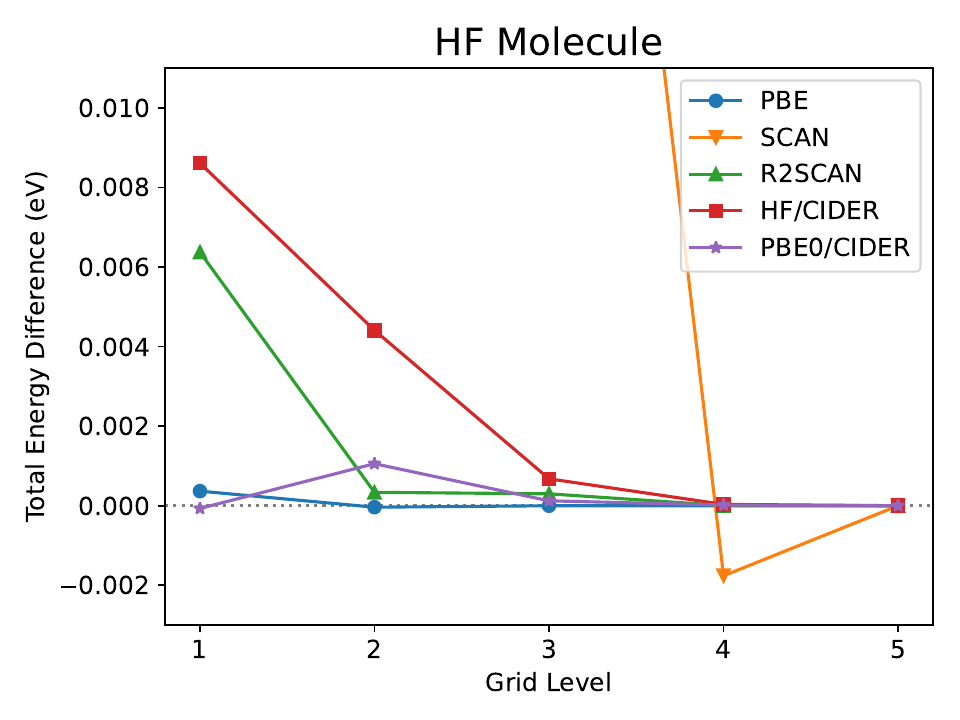}\\
    \includegraphics[width=0.48\textwidth]{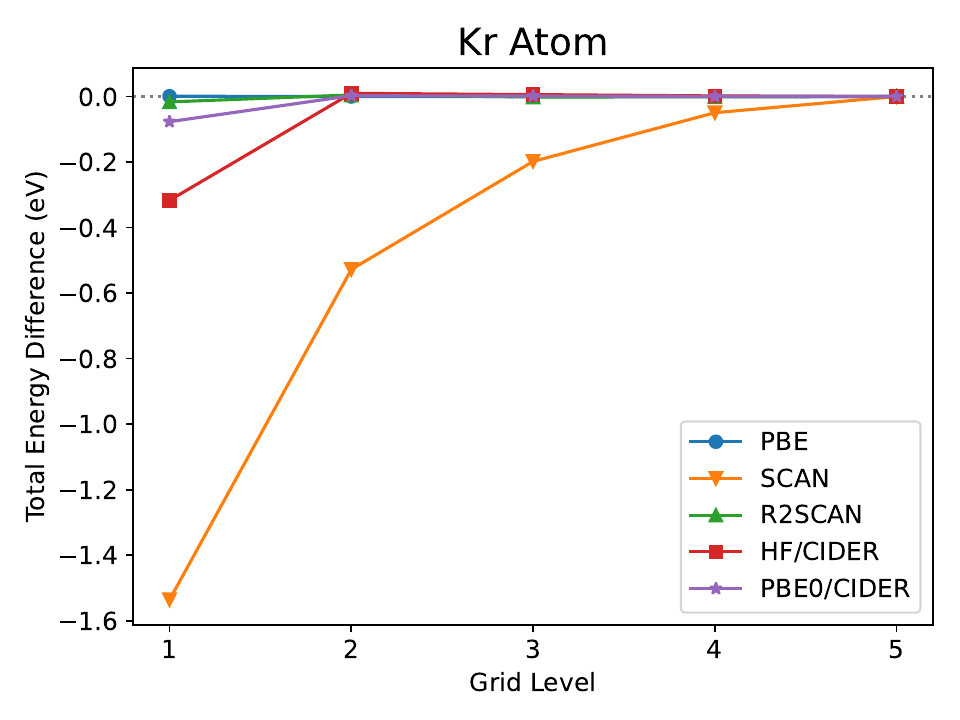}
    \includegraphics[width=0.48\textwidth]{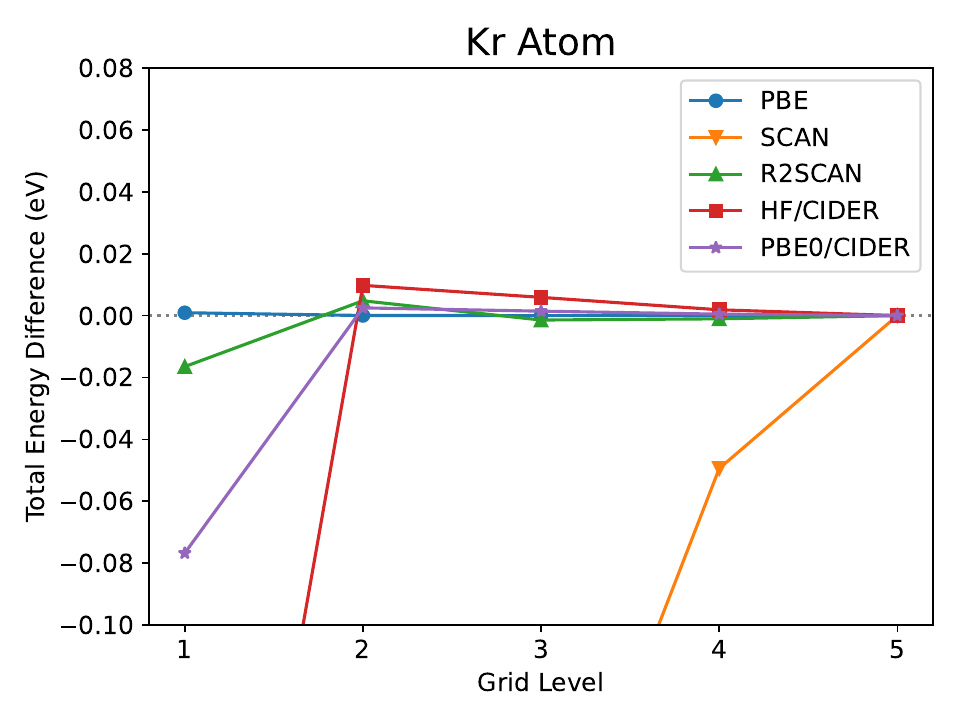}
    \caption{Convergence of the total energy of the \ce{HF} molecule and \ce{Kr} atom for different functionals with respect to the PySCF grid level. A larger grid level indicates a larger grid, and the energy for each functional is referenced to the total energy for grid level 5. HF/CIDER corresponds to the CIDER exchange functional with no correlation. The NL-MGGA-DTR model is used for the CIDER functionals. The plots on the left side show the full convergence plots, while the plots on the right provide a zoomed-in view to present a clearer picture of the behavior of the functionals for larger grid sizes.}
    \label{fig:pyscf_grid_convergence}
\end{figure}

Figure~\ref{fig:pyscf_grid_convergence} depicts the convergence of total energy of the \ce{HF} molecule and \ce{Kr} atom with respect to grid size for different functionals, as computed with PySCF. The calculations were performed with a def2-QZVPPD basis set. The most immediately obvious observation is that PBE (a GGA functional) has negligible numerical integration error compared to the SCAN and r$^2$SCAN meta-GGAs and the CIDER functionals, which use the NL-MGGA-DTR model from the main text. The original SCAN functional is known to have significant numerical stability issues, and this is illustrated in Fig.~\ref{fig:pyscf_grid_convergence} by the large integration errors for smaller grid sizes. The r$^2$SCAN functional fixes these issues. Reassuringly, the errors for the CIDER exchange functional (HF/CIDER) are not much larger than for r$^2$SCAN for most grid sizes for both systems. In practice, it will be more common to use functionals with a fraction of CIDER like PBE0/CIDER. Since these functionals only use a small fraction of the CIDER functional and otherwise rely on GGA-level information, the numerical error is smaller than for the full CIDER functional.

\begin{figure}
    \centering
    \includegraphics[width=0.48\textwidth]{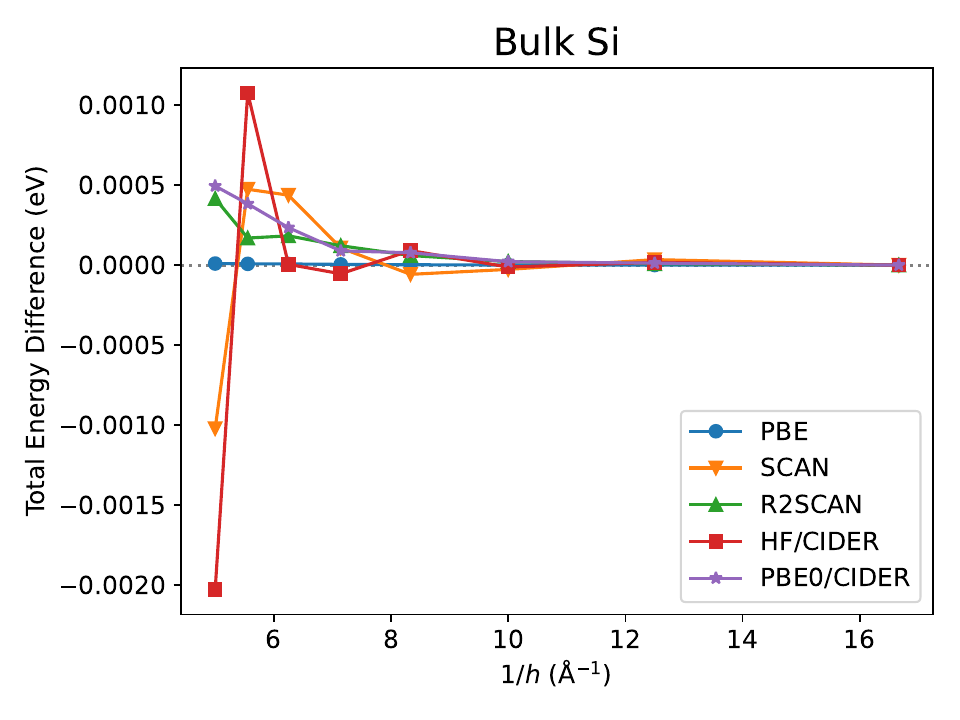}
    \includegraphics[width=0.48\textwidth]{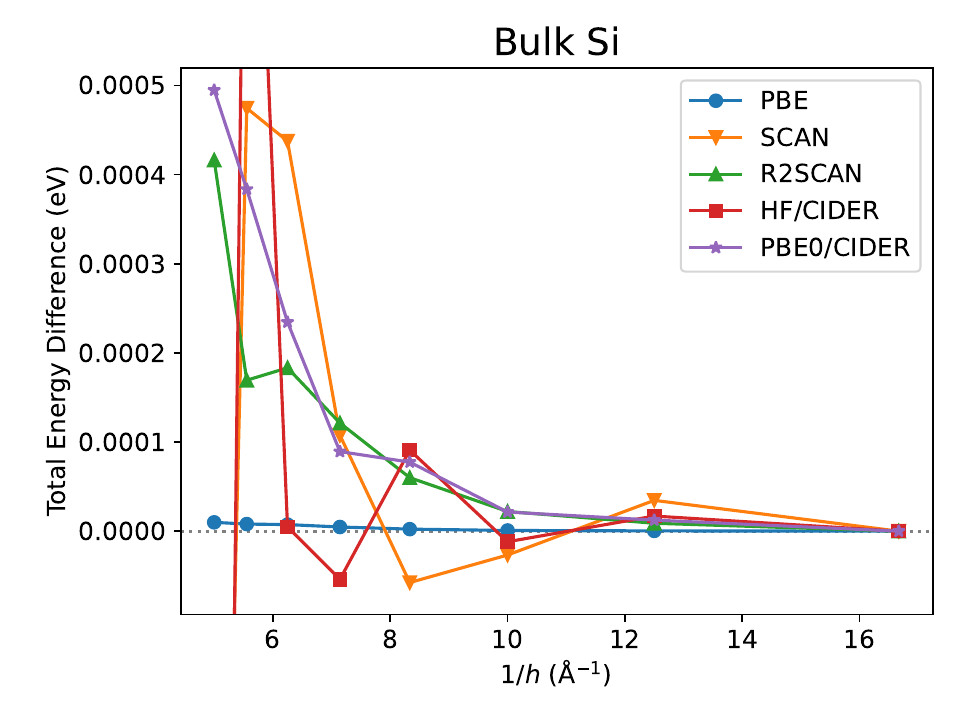}\\
    \includegraphics[width=0.48\textwidth]{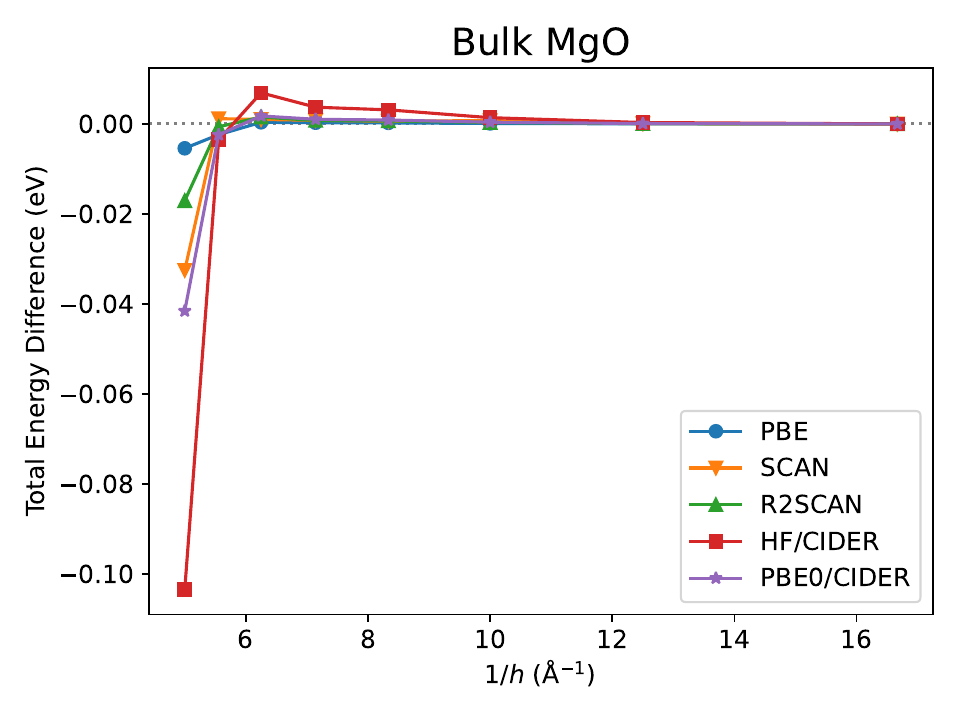}
    \includegraphics[width=0.48\textwidth]{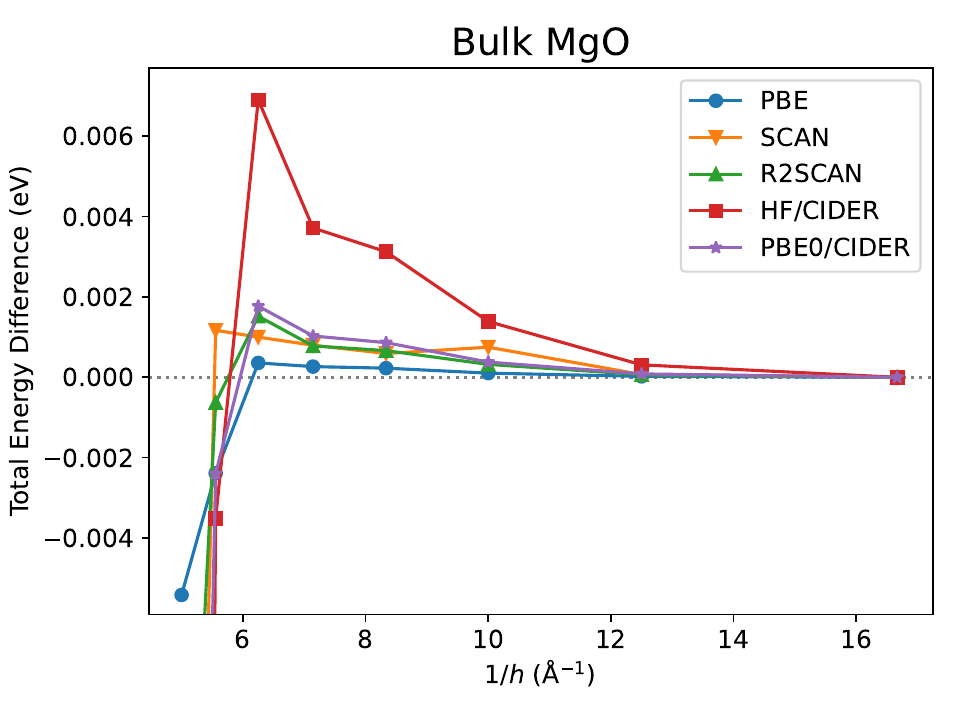}
    \caption{Convergence of the total energy of bulk \ce{Si} and \ce{MgO} for different functionals with respect to the grid spacing setting $h$ in GPAW. (Note: the XC integration grid has spacing $h/2$.) The energy for each functional is referenced to the total energy for $h=0.06$ \AA. HF/CIDER corresponds to the CIDER exchange functional with no correlation. The NL-MGGA-DTR model is used for the CIDER functionals. The plots on the left side show the full convergence plots, while the plots on the right provide a zoomed-in view to present a clearer picture of the behavior of the functionals for larger grid sizes.}
    \label{fig:gpaw_grid_convergence}
\end{figure}

Figure~\ref{fig:gpaw_grid_convergence} shows the convergence of total energy of bulk \ce{Si} and \ce{MgO} with respect to grid size for different functionals, as computed with GPAW. The calculations used a 520 eV energy cutoff for both systems, an $8\times 8\times 8$ k-point mesh for \ce{Si}, and a $4 \times 4 \times 4$ k-point mesh for \ce{MgO}. For these systems, SCAN and r$^2$SCAN have similar convergence behavior with respect to grid size (though SCAN is still a bit less stable), and PBE0/CIDER also matches the convergence behavior of these functionals fairly closely. However, HF/CIDER has slower convergence due to the larger fraction of CIDER exchange, especially for \ce{MgO}. Therefore, improving the numerical precision of CIDER functionals for solid-state systems could be important for future applications.

\section{Accuracy of Nystr\"om Approximation for Gaussian Process}

In the main text (Eqs. 37 and 38), the Nystr\"om approximation was used to reduce the computational complexity of fitting the Gaussian process models. The control points $\tilde{\mathbf{x}}_a$ for this method were selected using pivoted Cholesky decomposition as described in Appendix A6. This method requires selecting a tolerance parameter, for which $10^{-5}$ was chosen in the main text. The Nystr\"om approximation becomes exact if all feature vectors in the training set are included as control points, though this negates the computational efficiency of the approximation.

It is not computationally feasible to evaluate the covariance kernel exactly via Eq. 35 for the full training set, but we can test the effect of changing the number of control points used in the Nystr\"om approximation, which becomes more accurate as more control points are added. To change the number of control points, we test a larger tolerance of $10^{-3}$ (which decreases the number of control points) and a smaller tolerance of $10^{-8}$ (which increases the number of control points) in the Cholesky factorization step described in Appendix A6. Table~\ref{tab:bench_nystrom_approx} shows the deviations between models trained using these different tolerance parameters, with the NL-MGGA-DTR functional form and training set used as an example. The deviations are quite small, with the $10^{-3}$ and $10^{-5}$ tolerances disagreeing by a few meV on average and the $10^{-8}$ and $10^{-5}$ tolerances disagreeing by a few tenths of an meV on average.

\begin{table}[]
    \centering
    \caption{Mean absolute deviations between the NL-MGGA-DTR model trained with a tolerance factor for the Nystr\"om approximation of $10^{-5}$ (as used in the main text) and models trained with tolerances of $10^{-3}$ (``large tol'') and $10^{-8}$ (``small tol''), in meV. SOL62 corresponds to the solid-state cohesive energy database, and all the other datasets constitute the small molecule category of the GMTKN55 database. MoM indicates the mean-of-means error for the 18 sub-databases of GMTKN55 in this table.}
    \begin{ruledtabular}
    \begin{tabular}{lrr}
     & large tol (meV) & small tol (meV) \\
    \colrule
    SOL62 & 3.92 & 0.24 \\
    \colrule
    W4-11 & 3.64 & 0.29 \\
    G21EA & 1.87 & 0.14 \\
    G21IP & 2.44 & 0.16 \\
    DIPCS10 & 3.25 & 0.16 \\
    PA26 & 2.14 & 0.07 \\
    SIE4x4 & 3.10 & 0.14 \\
    ALKBDE10 & 2.05 & 0.20 \\
    YBDE18 & 2.50 & 0.08 \\
    AL2X6 & 2.79 & 0.21 \\
    HEAVYSB11 & 3.24 & 0.17 \\
    NBPRC & 2.27 & 0.04 \\
    ALK8 & 4.27 & 0.22 \\
    RC21 & 3.04 & 0.16 \\
    G2RC & 4.72 & 0.12 \\
    BH76RC & 1.60 & 0.10 \\
    FH51 & 2.92 & 0.08 \\
    TAUT15 & 0.67 & 0.02 \\
    DC13 & 6.08 & 0.36 \\
    \colrule
    MoM & 2.92 & 0.15 \\
    \end{tabular}
    \end{ruledtabular}
    \label{tab:bench_nystrom_approx}
\end{table}

\bibliography{references}